\documentclass[11pt,a4paper]{article}
\pdfoutput=1

\usepackage{jheppub}

\usepackage{amssymb}
\usepackage{amsfonts}
\usepackage{graphicx}
\usepackage{epstopdf}
\usepackage{dcolumn}
\usepackage{amsmath}
\usepackage{latexsym,bm}
\usepackage{amsthm}
\usepackage{slashed}
\usepackage{float}
\usepackage{color}
\usepackage{url}
\usepackage{longtable}
\usepackage{subfig}
\usepackage{tikz}
\usepackage{dynkin-diagrams}
\usepackage[all,cmtip]{xy}

\usepackage{multirow}
\usepackage{longtable}

\usepackage[titletoc]{appendix}
\makeatletter
\newtheorem*{rep@theorem}{\rep@title}
\newcommand{\newreptheorem}[2]{%
\newenvironment{rep#1}[1]{%
 \def\rep@title{#2 \ref{##1}}%
 \begin{rep@theorem}}%
 {\end{rep@theorem}}}
\makeatother

\newreptheorem{lemma}{Lemma}

\newreptheorem{conj}{Conjecture}
\theoremstyle{definition}

\newcommand \xoverline[2][0.75]{
    \sbox{\myboxA}{$\m@th#2$}
    \setbox\myboxB\null
    \ht\myboxB=\ht\myboxA
    \dp\myboxB=\dp\myboxA
    \wd\myboxB=#1\wd\myboxA
    \sbox\myboxB{$\m@th\overline{\copy\myboxB}$}
    \setlength\mylenA{\the\wd\myboxA}
    \addtolength\mylenA{-\the\wd\myboxB}
    \ifdim\wd\myboxB<\wd\myboxA
       \rlap{\hskip 0.5\mylenA\usebox\myboxB}{\usebox\myboxA}%
    \else
        \hskip -0.5\mylenA\rlap{\usebox\myboxA}{\hskip 0.5\mylenA\usebox\myboxB}%
    \fi}
\makeatother

\newcommand{\ba}{\begin{aligned}}
\newcommand{\ea}{\end{aligned}}
\def\be{\begin{equation}}
\def\ee{\end{equation}}
\def\bsp{\begin{split}}
\def\esp{\end{split}}
\def\bea{\begin{eqnarray}}
\def\eea{\end{eqnarray}}

\def\mc{\mathcal}

\def\mb{\mathbb}
\def\mbf{\mathbf}
\def \bp{\begin{pmatrix}}
\def\ep{\end{pmatrix}}

\def\min{\mathrm{min}}

\newcommand{\dd}{\mathrm{d}}
\renewcommand{\d}{\partial }

\newcommand{\op}[1]{\operatorname{#1}}
\def\bds{\boldsymbol}
\tikzset{global scale/.style={
    scale=#1,
    every node/.append style={scale=#1}
  }
}

\usepackage{tikz}
\usepackage{diagbox}
\usetikzlibrary{positioning}
\usetikzlibrary{calc}
\usetikzlibrary{decorations.pathreplacing,calligraphy}

\usepackage{xstring}
\usetikzlibrary{decorations.pathmorphing} 
\usetikzlibrary{decorations.markings} 
\usetikzlibrary{snakes}
\usetikzlibrary{arrows} 
\usetikzlibrary{shapes} 
\usetikzlibrary{matrix} 
\usetikzlibrary{positioning} 
\usepackage[english]{babel} 
\usepackage[autostyle]{csquotes}

\usepackage{tikz}
\usetikzlibrary{arrows}
\usetikzlibrary{arrows.meta}
\usetikzlibrary{shapes.geometric,calc,arrows, positioning,shapes.misc,decorations.markings}
\tikzset{
  big arrow/.style={
    decoration={markings,mark=at position 1 with {\arrow[scale=2,#1]{>}}},
    postaction={decorate},
    shorten >=0.4pt},
  big arrow/.default=black}
  
\tikzstyle{none}=[inner sep=0pt] 

\tikzstyle{NodeCross}=[draw, shape=circle, cross out, inner sep=0pt, minimum size=6pt,line width=0.25mm]
\tikzstyle{Circle}=[draw, shape=circle, black,  fill=black, inner sep=0pt, minimum size=6pt]
\tikzstyle{Star}=[draw, shape=star, fill=black, star points=8, inner sep=0pt, minimum size=8pt]

\tikzstyle{DashedLine}=[-, densely dashed, line width=0.25mm]
\tikzstyle{DottedLine}=[-, dotted, line width=0.25mm]
\tikzstyle{ThickLine}=[-, line width=0.25mm]
\tikzstyle{ArrowLineRight}=[-, -{Stealth[scale=1.75]}, line width=0.1mm, scale=5]
\tikzstyle{RedLine}=[-, draw={rgb,255: red,191; green,0; blue,0}, fill=none, line width=0.25mm]
\tikzstyle{DottedRed}=[-, dotted, draw={rgb,255: red,191; green,0; blue,0}, fill=none, line width=0.25mm]
\tikzstyle{DashedLineThin}=[-, densely dashed, line width=0.125mm, fill=none, draw=black]
\tikzstyle{ArrowLineRed}=[-, -{Stealth[scale=1.75]}, draw={rgb,255: red,191; green,0; blue,0}, line width=0.1mm, scale=5]
\tikzstyle{brane}=[draw]

\graphicspath{ {figs/} }

\title{Massive Spectrum in F-theory and the Distance Conjecture}

\preprint{\today \hspace*{0.1in} }

\author[a]{Keren Chen}
\author[a]{Qinjian Lou}
\author[a,b]{Yi-Nan Wang}

\affiliation[a]{School of Physics, \\
Peking University, Beijing 100871, China}
\affiliation[b]{Center for High Energy Physics, Peking University, \\
Beijing 100871, China}

\emailAdd{keren.chen246@gmail.com,qinjian.lou@pku.edu.cn,ynwang@pku.edu.cn}

\abstract{We investigate the massive states in F-theory compactification models, including BPS string junctions stretching between parallel 7-branes and KK modes. We obtain analytical results when there are two colliding bunches of 7-branes with a locally constant axiodilaton profile. In particular, in 8D F-theory setups when the 7-branes collide into a codimension-one $(4,6,12)$ singularity, we found an infinite light tower of BPS string junctions, which should match the light KK tower in the dual heterotic description. To exactly match with the results in the distance conjecture, we propose that the definition of 8D Planck mass should receive a large correction. We have also computed parts of KK modes in 8D F-theory in a simplified setup, as well as the BPS string junction spectrum in specific setups of 6D and 4D F-theory.}

\begin{document}

\maketitle

\section{Introduction}

F-theory is a powerful geometric framework to describe strongly coupled IIB superstring theory in presence of background 7-branes and a varying axiodilaton profile~\cite{Vafa:1996xn,Morrison:1996na,Morrison:1996pp,Weigand:2018rez}. Compactifying F-theory on elliptically fibered Calabi-Yau manifolds leads to a vast landscape of supergravity models in even space-time dimensions~\cite{Morrison:2012np,Morrison:2012js,Taylor:2015isa,Halverson:2015jua,Taylor:2015ppa,Watari:2015ysa,Taylor:2015xtz,Halverson:2017ffz,Taylor:2017yqr,Tian:2020yex,Morrison:2023hqx}, including different ways to realize standard model-like physics~\cite{Donagi:2008ca,Beasley:2008dc,Beasley:2008kw,Donagi:2008kj,Marsano:2009gv,Blumenhagen:2009yv,Heckman:2010bq,Braun:2013yti,Lin:2014qga,Grassi:2014zxa,Cvetic:2018ryq,Cvetic:2019gnh,Taylor:2019wnm,Raghuram:2019efb,Li:2021eyn,Jefferson:2022yya,Marchesano:2022qbx,Li:2023dya}. 

Such lower dimensional supergravity effective actions are typically built with massless fields, including the graviton, closed string sector gauge fields, moduli scalars (before the moduli stabilization), D-brane gauge fields and their superpartners. On the other hand, the massive spectrum in F-theory is largely unexplored, despite their significance in the swampland conjectures~\cite{Vafa:2005ui,Brennan:2017rbf,Palti:2019pca,vanBeest:2021lhn,Agmon:2022thq} and model building. 

In this paper, we present a preliminary exploration of the massive spectrum in F-theory compactifications. We consider the massive states in lower dimensional supergravity models from two sources: the massive BPS string junction states, and the Kaluza-Klein (KK) modes.

In F-theory, segments of $(p,q)$-strings can join together and form string junctions connecting multiple bunches of 7-branes~\cite{Gaberdiel:1998mv,Mikhailov:1998bx,DeWolfe:1998zf,DeWolfe:1998eu,DeWolfe:1998pr,Grassi:2013kha,Grassi:2014ffa,Grassi:2018wfy,Grassi:2021ptc}. In presence of parallel bunches of 7-branes $B_i$, i.e. they do not intersect each other on the internal base manifold, there would be massive BPS string junction stretching between them. The mass of such states can be computed by the minimization of string tension $T_{p,q}$ integrated over the string segments:
\be
m_{\op{junction}}=\min\sum_i\left|\int_{L_i}T_{p_i,q_i}ds\right|\,.
\ee
In particular, the lightest ones are given by strings connecting a pair of parallel 7-branes. Such configurations of parallel 7-branes are commonly occurring not only in the case of 8D F-theory where the bunches of 7-branes are distinct points on the base $\mb{P}^1$, but also on higher-dimensional bases where the divisors supporting different bunches of 7-branes do not intersect, see e.g. \cite{Morrison:2014lca,Halverson:2015jua,Taylor:2015ppa,Taylor:2015xtz,Halverson:2017ffz,Taylor:2017yqr}.

We provide a detailed analysis of massive BPS string junctions in the case of 8D F-theory, with in total 24 7-branes on the base manifold $\mb{P}^1$. 
Denoting the local base coordinate of $\mb{P}^1$ by $t$, and there are $n_i$ 7-branes located at the point $t=t_i$ (we set $t_0=0$), then the metric of such base $\mb{P}^1$ is given by~\cite{Greene:1989ya}\footnote{$C_g$ is a numerical constant depending on the volume of base $\mb{P}^1$.}
\begin{equation}
\dd s^2= C_g^2\tau_2 \eta(\tau)^2 \bar{\eta}(\bar{\tau})^2 \prod_i\left(t-t_i\right)^{-n_i / 12}\left(\bar{t}-\bar{t}_i\right)^{-n_i / 12} \dd t \dd \bar{t}\,.
\end{equation}
In particular, when there is a bunch of $n_1$ 7-branes at $t=t_1$ approaching $n_0$ 7-branes at $t=0$ with a ``locally constant'' axiodilaton profile\footnote{The value of $\tau$ for the two bunches of parallel 7-branes are compatible (the same).}, we obtain analytical results for the mass of the BPS string junction between them. Denoting the asymptotic charge connecting the two bunches of 7-branes by $(P,Q)$, if $n_0+n_1<12$ we have 
\begin{equation}
    \begin{aligned}
        m_{\text{junction}}=&C_g\left|(P-Q\tau) \eta^2(\tau)\prod_{i=2} t_i^{-\frac{n_i}{12}}B\left(1-\frac{n_0}{12},1-\frac{n_1}{12}\right)\right|\\
        &\times \left|1+\left(\frac{12-n_0}{24-n_0-n_1}\sum_{i=2} \frac{n_i}{12t_i}\right)t_1\right| |t_1|^{1-\frac{n_0+n_1}{12}}+\mc{O}(t_1^{3-\frac{n_0+n_1}{12}})\,.
    \end{aligned}
\end{equation}

More interestingly, if we have $n_0+n_1=12$, which means that there would be a codimension-one $(4,6,12)$ singularity after $t_1\rightarrow 0$ and the two bunches of 7-branes collide, we get the analytical formula for the mass of the BPS string junction stretching between these bunches of 7-branes:
\begin{equation}
        \begin{aligned}
        m_{\text{junction}}=&\left|\frac{P-Q\tau}{\sqrt{2\pi\tau_2}} B\left(1-\frac{n_0}{12},1-\frac{n_1}{12}\right)  \left[1+\left(\frac{12-n_0}{24-n_0-n_1}\sum_{i=2} \frac{n_i}{12t_i}\right)\right]\right|\sqrt{\left|\frac{V_{\mb{P}^1}}{\log t_1}\right|}\\&+\mc{O}(|t_1|^2|\log t_1|^{-\frac{1}{2}}) \,.
    \end{aligned} 
\end{equation}

It is known that the codimension-one $(4,6,12)$ singularity in F-theory is an infinite distance limit in the complex structure moduli space~\cite{Lee:2021usk,Alvarez-Garcia:2023qqj}. More precisely, when the order of vanishing of $(f,g,\Delta)=(4,6,12)$, it describes a type II.a Kulikov degeneration of K3 surface, which should have a decompactified 10D heterotic string description (on a $T^2$ with $V_{T^2}\rightarrow\infty$)~\cite{Lee:2021qkx,Lee:2021usk}.

In our setup when $t_1\rightarrow 0$, we exactly obtain an infinite light tower from the string junction connecting the two bunches of 7-branes that collide into a codimension-one $(4,6,12)$ singularity. After comparing with the properly normalized metric of complex structure moduli space on the heterotic side, we find that the mass of such light tower scales as
\be
\label{junction-scales}
m\propto e^{-\sqrt{\frac{3}{8}}\Delta s}\,.
\ee
Interestingly, the numerical factor $\sqrt{\frac{3}{8}}$ does not match the expected number
\be
\sqrt{\frac{(D-2)}{(D-d)(d-2)}}=\sqrt{\frac{2}{3}}
\ee
from the 10D to 8D KK reduction on the heterotic side. We claim that the mismatch should be due to the fact that our F-theory description is strongly coupled, and the definition of 8D Planck mass in the F-theory description has the following discrepancy from the naive result related to the volume of base $\mb{P}^1$ ($M_{\op{P},10}$ is the 10D Planck mass):
\be
\label{Planck-mass-conj}
M_{\op{P},F,8}\sim \left(M_{\op{P},10}^{8}V_{\mb{P}^1}\right)^{\frac{1}{6}}e^{\sqrt{\frac{1}{24}}\Delta s}\,.
\ee
Thus we provided a non-trivial example of infinite distance limit in a strongly coupled string theory setup, as well as a new quantitative prediction of F/heterotic duality in the strongly coupled regime of F-theory.

We have also attempted to extend the discussion to the lower dimensional cases. For general 6D and 4D F-theory setups, it is hard to write down the base metric \cite{Katz:2022vwe}. Nonetheless, if we limit ourselves to the specific configurations with sets of parallel 7-branes, we can obtain massive spectrum of string junctions stretching between them analogous to the 8D cases.

 Besides the massive spectrum in F-theory from the BPS string junctions, we have also attacked the problem of computing massive KK modes in 8D F-theory, which originates from the reduction of the $B_2^I=(B_2,C_2)$ doublet of 10D IIB superstring theory. We consider the simplification which only keeps the leading order kinetic term of $B_2^I$ in the F-theory action\footnote{Such computations could be improved by taking into account the higher order corrections in the future~\cite{Grimm:2013bha,Grimm:2013gma,Weissenbacher:2019mef,Cicoli:2021rub}.} 
\be
\ba
    S_{2-form}=\int M_{IJ}dB^I\wedge*dB^J\quad,\ 
    M_{IJ}=\frac{1}{\operatorname{Im} \tau}\left( \begin{array}{cc}
        |\tau^2| & -\operatorname{Re} \tau \cr
        -\operatorname{Re} \tau & 1
    \end{array}\right)\,.
\ea
\ee
For simplicity, we only study in detail the reduction of 10D  2-form doublet $B_2^I$ to 8D massive tensor fields $c_{\mu\nu}$:
\be
B_2^I\sim b^I c_{\mu\nu}dx^\mu\wedge dx^\nu\quad,\ (\mu,\nu=0,\dots,7)\,.
\ee
As explicit examples, we consider 8D F-theory setups with a constant axiodilaton $\tau$ profile, for instance when there are four bunches of type $I_0^*$ $D_4$ branes located at the points $t=t_1/2,-t_1/2,T/2,-T/2$, with $t_1\ll T$. After numerical computations, we find that the lowest massive KK mode has a mass of
\be
m_{\rm KK}=\sqrt{2\pi\lambda\log(T/t_1)/{V_{\mb{P}^1}}}\ ,\ \lambda\approx 0.06\,.
\ee
In particular, in the infinite complex structure distance limit of $t_1\rightarrow 0$, the KK modes would become infinitely heavy, and scale as
\be
m_{\rm KK}\propto e^{\sqrt{\frac{1}{6}}\Delta s}\,,
\ee
after taking into account the conjectured asymptotic behaviour of 8d Planck mass in F-theory (\ref{Planck-mass-conj}). Note that this result is obtained after we fix the total volume $V_{\mb{P}^1}$ of the base manifold, hence it is not inconsistent with the existing results in the literature~\cite{Blumenhagen:2018nts,Joshi:2019nzi,Ashmore:2021qdf}. For other configurations of constant axiodilaton $\tau$, we get qualitatively the same results.

The structure of this paper is as follows: in section \ref{sec:string-junction} we first review the setups of 8D F-theory in section \ref{subsec_field_config_8d_F_theory} and the definition of string junctions in 8D F-theory in section \ref{subsec_basic_string_junction}. In section \ref{subsec_massive_junction_two_bunches} and section \ref{subsec_massive_junction_n_bunches} we discuss the configuration of BPS string junctions connecting two bunches and $n\geq 3$ bunches of 7-branes, respectively. In section \ref{sec:stringjunc-constant-tau} and \ref{sec:more-constant-tau} we study the mass of BPS string junctions in configurations with a constant $\tau$. 

In section \ref{sec:KK} we study the KK modes of 8D F-theory. We derive the dimensional reduction of 2-form field doublet in IIB to 8D supergravity in section \ref{sec:dim-red-2-form}. Then we discuss the spectrum of certain KK modes for the case of four $D_4$ bunches in section \ref{sec:KK-4-D4} and the other cases of constant $\tau$ in section \ref{sec:KK-other}. 

In section \ref{sec:infinite-distance} we discuss the relation of our results with the distance conjectures, which is briefly reviewed in section \ref{sec:infinite-conj}. In section \ref{sec:F-heterotic} we review the map between F-theory moduli space and the properly normalized moduli space in heterotic string, with no Wilson lines. In section \ref{sec:codimension-one-46-limit} we compare the achieved massless string junction tower in the codimension-one $(4,6,12)$ limit with the moduli space on the heterotic side, and propose the correction to the 8D F-theory Planck mass to fix the discrepancy on the exponent of (\ref{junction-scales}). 

In section \ref{sec:massive-lower} we discuss the massive tower of BPS string junctions in the cases of lower dimensional F-theory. In section \ref{subsec_field_config_lower_d_F_theory} we review the difficulty in the computation of base metric in these cases, and propose a number of local and global forms for certain types of brane configurations, that can be applied to 6D or 4D F-theory. In section \ref{sec:massive-6D} we provide a general discussion of massive BPS string junctions in 6D F-theory, while in section \ref{sec:4D} we provide a more detailed computation of such spectrum in specific 4D F-theory setups with $E_6$ or $E_7$ gauge groups. We conclude the paper by outlooks in section \ref{sec:discussions}.

In appendix \ref{app:special-function} we summarized the special functions used in the paper. In appendix \ref{sec_mass_string_junction} we provide a more detailed calculation of the mass of BPS string junctions in a constant $\tau$ background that is omitted in the main text. In appendix \ref{sec_metric_moduli_space} we review the computation of the exponent of the KK tower and the normalized metric of the heterotic moduli space with no Wilson line. In appendix \ref{sec_junction_spec} we present the detailed computation of BPS string junctions for the colliding bunches of branes mentioned in this paper.

\section{8D F-theory and string junctions}
\label{sec:string-junction}

\subsection{Field configurations of 8D F-theory}
\label{subsec_field_config_8d_F_theory}

In this section we review the setup of 8D F-theory, which is  type IIB superstring theory on $X_{10}=\mb{R}^{7,1}\times \mb{P}^1$ with 24 7-branes. The base manifold $\mb{P}^1$ has the topology of a 2-sphere $S^2$. The worldvolume of 7-branes are $\mb{R}^{7,1}\times p_i$, where $p_i$ are points on the base manifold $\mb{P}^1$. It is known that one can solve the field configurations of such 7-brane solutions through the equation of motion of type IIB supergravity \cite{Greene:1989ya}.

The action of 10D IIB superstring theory is (see e.g. \cite{Blumenhagen:2013fgp})
\begin{equation}
\begin{aligned}
S_{\text {IIB }}= & \frac{1}{2 \tilde{\kappa}_{10}^2} \int d^{10} x \sqrt{-g}\left[R-\frac{\partial_M \tau \partial^M \bar{\tau}}{2(\operatorname{Im} \tau)^2}-\frac{1}{2} \frac{\left|G_3\right|^2}{\operatorname{Im} \tau}-\frac{1}{4}\left|F_5\right|^2\right] \\
& +\frac{1}{8 i \tilde{\kappa}_{10}^2} \int \frac{1}{\operatorname{Im} \tau} C_4 \wedge G_3 \wedge \bar{G}_3 + S_{\op{7-branes}} \,.
\end{aligned}
\label{type_IIB_sugra}
\end{equation}
With no 3-branes and 5-branes in 8D F-theory, we can set $G_3$ and $F_5$ to be vanishing in the supergravity solution. Then the only non-trivial field configurations involved are the complex scalar $\tau$ and the space-time metric $g_{MN}$.

The 7-branes are the sources of non-trivial monodromy of $\tau$. Although it is difficult to write down the 7-brane action, we can read out the configuration of $\tau$ from the Weierstrass model
\begin{equation}
    y^2=x^3+f(t) x +g(t)\,,
\end{equation}
where $t$ is the local coordinate of $\mb{P}^1$.

From the discriminant
\begin{equation}
    \Delta=4 f^3+27 g^2=\Delta_0\prod_i(t-t_i)^{n_i}\,,
\end{equation}
we see that there are a bunch of $n_i$ 7-branes at the location $t=t_i$. The value of axiodilaton $\tau$ can be computed up to $\op{SL}(2,\mb{Z})$ monodromy by the inverse of Jacobi $j$-function
\begin{equation}
    j(\tau)=4 \frac{12^3 f^3}{\Delta}\,.
    \label{j_tau_fg}
\end{equation}

It is known that the space-time metric can be determined by the Einstein equation~\cite{Greene:1989ya}
\begin{equation}
\dd s^2= C_g^2\tau_2 \eta(\tau)^2 \bar{\eta}(\bar{\tau})^2 \prod_i\left(t-t_i\right)^{-n_i / 12}\left(\bar{t}-\bar{t}_i\right)^{-n_i / 12} \dd t \dd \bar{t}\,.
\label{metric_8d_F-theory}
\end{equation}
Here $C_g$ is a real positive constant, which can be fixed by the overall volume of $\mb{P}^1$. 

\subsection{Basic concepts of string junctions}
\label{subsec_basic_string_junction}
In this section we review the basics of string junctions connecting different bunches of 7-branes in 8D F-theory, which are candidates of massive BPS states~\cite{Schwarz:1995dk}. 

Apart from a single open string connecting two bunches of 7-branes, we also allow $n$-junctions consisting of different types of $(p,q)$-string under the constraint of charge conservation. For example, a 3-junction with a $(1,0)$-string and a $(0,1)$-string as two ends can only have a $(-1,-1)$-string as the third end. For a general $n$-junction ending with  $(p_i,q_i)$-strings, we have the conservation law
\begin{equation}
    \sum_i p_i=\sum_i q_i=0\,.
\end{equation}
For a BPS string junction, the relative angles between different ends are fixed. Near the junction point, we denote the tangent vector of each end is $\hat{n}_i$, and one of the BPS conditions is \cite{Dasgupta:1997pu}
\begin{equation}
\label{rel-angle-cond}
    \sum_i T_{p_i,q_i}\hat{n}_i=0\,.
\end{equation}
%And because of conformal property of holomorphic maps, such BPS condition is valid up to complex diffeomorphism. 

Here the tension of $(p,q)$-string is \cite{Schwarz:1995dk}
\begin{equation}
    T_{p, q}=\frac{1}{\sqrt{\tau_2}}|p-q \tau|\,.
    \label{pq_string_tension}
\end{equation}
Let us consider a BPS string junction, which consists of $N$ segments of strings with charges $(p_i,q_i)$ along different paths $L_i$. The mass of the BPS string junction is
\begin{equation}
    m_{\text{junction}}=\sum_i \int_{L_i} T_{p_i,q_i}\dd s\,.
\end{equation}
Here we need to integrate along the paths $L_i$ because in general $T_{p,q}$ is not constant in F-theory. The other BPS condition is that $m_{\text{junction}}$ should be minimized among all the choices of the paths $L_i$.

The charges of the $(p,q)$-string can be transformed under the $\op{SL}(2,\mb{Z})$ monodromy on $\mb{P}^1$, nonetheless we can always use the standard presentation in \cite{DeWolfe:1998zf} to avoid such complications. 

With the convention of \cite{DeWolfe:1998zf}, one $(p,q)$ 7-brane can be written as $X_{p,q}$.
The monodromy around a $(p,q)$ 7-brane is 
\begin{equation}
K_{p,q}=\left(
\begin{array}{cc}
    1+pq & -p^2 \\
    q^2 & 1-pq
\end{array}
\right).
\end{equation}
There are three special $(p,q)$ 7-branes that can be used as basis to construct a stack of 7-branes
\begin{equation}
    \mbf{A}=X_{1,0}\ ,\  \mbf{B}=X_{1,-1}\ ,\  \mbf{C}=X_{1,1}\,.
\end{equation}

In this paper, we do not consider D3-branes in the background, and the string junction can only end on 7-branes. In particular, only $(p,q)$-strings can end on the $(p,q)$ 7-branes, hence all possible string junctions can be drawn in the form of Figure~\ref{fig:string_junc_bet_bunch}, where all branch cuts are pointed downward. For a bunch of 7-branes $X_{p_1,q_1}X_{p_2,q_2}\dots X_{p_i,q_i}\dots X_{p_n,q_n}$, with index $i$ increasing from left to right, we can get the total monodromy around it:
\begin{equation}
    K=K_{p_n,q_n}\dots K_{p_i,q_i}\dots K_{p_2,q_2}K_{p_1,q_1}\,.
\end{equation}
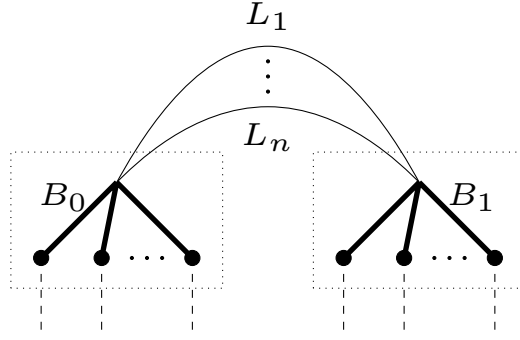
\begin{figure}
    \centering
    \begin{tikzpicture}[global scale =2]
        \filldraw (0,0) circle (.05)
        (.4,0) circle (.05)
        (.6,0) circle (.01)
        (.7,0) circle (.01)
        (.8,0) circle (.01)
        (1.,0) circle (.05); 
        \draw[line width =2pt] (0,0)--(.5,.5)
        (.4,0)--(.5,.5)
        (1,0)--(.5,.5);
        \draw[dotted] (-.2,-.2) rectangle (1.2,.7);
        \draw[dashed] (0,0)--(0,-0.5)
        (.4,0)--(.4,-0.5)
        (1.,0)--(1.,-0.5);
        \node[font=\fontsize{6}{6}\selectfont] at(0.15,0.4) {$B_0$};

        \filldraw[shift={(2,0)}] (0,0) circle (.05)
        (.4,0) circle (.05)
        (.6,0) circle (.01)
        (.7,0) circle (.01)
        (.8,0) circle (.01)
        (1.,0) circle (.05); 
        \draw[shift={(2,0)},line width =2pt] (0,0)--(.5,.5)
        (.4,0)--(.5,.5)
        (1,0)--(.5,.5);
        \draw[dotted,shift={(2,0)}] (-.2,-.2) rectangle (1.2,.7);
        \draw[dashed,shift={(2,0)}] (0,0)--(0,-0.5)
        (.4,0)--(.4,-0.5)
        (1.,0)--(1.,-0.5);
        \node[font=\fontsize{6}{6}\selectfont] at(2.85,0.4) {$B_1$};

        \node[font=\fontsize{6}{6}\selectfont] at(1.5,1.6) {$L_1$};
        \node[font=\fontsize{6}{6}\selectfont] at(1.5,.8) {$L_n$};
        \draw (.5,.5) parabola bend (1.5,1) (2.5,.5);
        \draw (.5,.5) parabola bend (1.5,1.4) (2.5,.5);
        \filldraw (1.5,1.1) circle (.01)
        (1.5,1.2) circle (.01)
        (1.5,1.3) circle (.01);
    \end{tikzpicture}
    \caption{The standard presentation of the string junction connecting two bunches of 7-branes. The black circles in the square $B_0,B_1$ denotes the 7-branes in the bunch $B_0,B_1$. The dashed lines under the circles denote the branch cut in standard presentation \cite{DeWolfe:1998zf}. $L_j$ represents the strings connecting two bunches. The thick lines in the square could represent multiple $(p,q)$-strings while a thin line $L_j$ denotes a single $(p,q)$-string.}
    \label{fig:string_junc_bet_bunch}
\end{figure}
For massless string junctions, all 7-branes collide, and all string segments shrink to a point. In Figure \ref{fig:string_junc_bet_bunch}, when all of the points in the two squares collide and all thick and thin lines contract to a point, the string junction is massless. In this case, the configurations of all lines are fixed, and the only non-trivial information is the number of strings ending on each 7-brane, i.e. the number of thick lines in Figure \ref{fig:string_junc_bet_bunch}. Hence massless string junctions can be represented as a linear combination
\begin{equation}
    \mathbf{J}=\sum_i n_i(\mathbf{J}) \mathbf{s}_i\,,
    \label{standard_rep_massless_junction}
\end{equation}
where $\mbf{s}_i$ denotes one $(p_i,q_i)$-string ending on the $i$-th 7-brane of type $(p_i,q_i)$. $n_i$ denotes the number of strings ending on the $i$-th 7-brane. We can define the asymptotic charge $(P,Q)$ of the string junction, which is independent of the choice of branch cuts:
\begin{equation}
    P=\sum_i n_i p_i, \quad Q=\sum_i n_i q_i \,.
\end{equation}
For massive string junctions connecting 7-branes at different locations, it can also be written in the form (\ref{standard_rep_massless_junction}) even if they have non-trivial internal string configurations. We can also interpret the process that pulls 7-branes apart as a process of Higgs mechanism, in which the BPS string junctions obtain a non-zero mass.

As discussed in \cite{DeWolfe:1998zf}, through the F/M-duality, we can define the inner product of the string junctions using the intersection numbers of M2-branes
\be
(\mbf{J}_1,\mbf{J}_2)=(\mbf{J}_1,\mbf{J}_2)_1+(\mbf{J}_1,\mbf{J}_2)_2\,.
\ee
Here $(\mbf{J}_1,\mbf{J}_2)_1$ denotes the self-intersection part $(\mbf{s}_i,\mbf{s}_i)=-1$, and $(\mbf{J}_1,\mbf{J}_2)_2$ denotes the intersection number at the 3-junction points. In principle, all of the junction points of the string junction can be decomposed into such 3-junction points. At this point we have
\be
(\mbf{s}_i,\mbf{s}_{i+1})=(\mbf{s}_{i+1},\mbf{s}_i)=\frac{1}{2}\left|\begin{array}{cc}
    p_i & p_{i+1} \\
    q_i & q_{i+1}
\end{array}\right|\,.
\ee 
The subscript $i$ increases as we go around the intersection point in the counterclockwise direction. As for the junction point of $n>3$ strings, we can decompose it into a number of 3-junction points, and then calculate the total inner product.

From the inner product, we arrive at the BPS condition for the string junction~\cite{DeWolfe:1998bi,Mikhailov:1998bx}
\begin{equation}
    (\mbf{J},\mbf{J})-\op{gcd}(P,Q)\geq -2\,.
    \label{string_junction_BPS_condition}
\end{equation}
Here $(P,Q)$ denotes the total asymptotic charge of the string junction. 
%If there is no D3-brane, then $P=Q=0$. 

At the bunch of 7-branes with non-abelian gauge symmetry $G$, the BPS string junction can be written as 
\begin{equation}
    \mbf{J}=\bds{\lambda}+P\bds{\omega}^p+Q\bds{\omega}^q\,.
\end{equation}
$\bds{\lambda}$ denotes all string junctions with zero asymptotic charge, which comprise the weights of the adjoint representation of $G$. $\bds{\omega}^p,\bds{\omega}^q$ are the extended weights with asymptotic charge $(1,0),(0,1)$ respectively, which satisfy
\be
(\bds{\omega}^p,\bds{\lambda})=(\bds{\omega}^q,\bds{\lambda})=0\,.
\ee
We can read off the asymptotic charge $(p,q)$ of the string junction from the coefficients of $(\bds{\omega}^p,\bds{\omega}^q)$. In particular, if the asymptotic charge of the string junction $(P,Q)\neq (0,0)$, it means that the string junction is incomplete, and it should end on some bunches of 7-branes or D3-branes. 

\subsection{Massive BPS string junctions connecting two bunches of 7-branes}

\label{subsec_massive_junction_two_bunches}

For a general massive string junction,  (\ref{standard_rep_massless_junction}) does not include all the information of the string junction as the configuration of internal strings are omitted. We would first consider the simplified cases with only two bunches of 7-branes. 

Before calculating the mass of the BPS string junctions, we can first consider a single $(p,q)$-string connecting two $(p,q)$ 7-branes. As discussed in \cite{Gaberdiel:1998mv}, independent of the path, the mass of such $(p,q)$-string is
\begin{equation}
    m_{\op{BPS}}= C_g \left|\int h_{p,q}\dd t\right|\,.
\end{equation}
Here 
\begin{equation}
    h_{p,q}=(p-q\tau)\eta^2 \prod_{i}(t-t_i)^{-\frac{n_i}{12}}\,.
\end{equation}

We can prove the path independence property of the BPS mass of the string connecting two 7-branes as follows. Defining the new coordinate 
\begin{equation}
    w=\int h_{p,q}\dd t\,,
\end{equation}
due to the holomorphicity of $h_{p,q}$, $w$ is well-defined away from the branch cuts, and the line integration is independent of the path if the path does not cross any branch cut. 

Then we can find 
\begin{equation}
    |T_{p,q}\dd s|^2=C_g^2\dd w \dd \Bar{w}\,,
\end{equation}
which means that in the new coordinate $w$, the base $\mb{P}^1$ is mapped to a 1D complex plane with a flat Euclidean metric, as shown in \cite{Gaberdiel:1998mv}. 

The BPS string state corresponds to the geodesic in the complex space with metric $\dd s_{p,q}^2=|T_{p,q}\dd s|^2$, i.e. the path with minimal length. So in the $w$ space, it is precisely the straight line connecting two points, and the length is 
\begin{equation}
    l_{\text{geodesic}}=C_g |\Delta w|= C_g |\int h_{p,q} dt| \,.
    \label{length_geodesic}
\end{equation} 

Hence we obtain the BPS mass of the $(p,q)$-string connecting two 7-branes
\begin{equation}
    m_{\op{BPS}}=\op{min}\left(\int| T_{p,q}\dd s| \right)=\op{min}\left(\int C_g| \dd w| \right)=l_{\text{geodesic}}=C_g |\Delta w|= C_g |\int h_{p,q} dt|
\end{equation}
by a line integration over holomorphic function instead over real function. If there is no singularity in the line, i.e. the string does not cross the branch cut, then the integration is independent of the path. As discussed in \cite{Gaberdiel:1998mv}, for any string junction, we can find the configuration with the same mass independent of the way it crosses the branch cuts. Hence we can assume that the string junction does not cross any branch cut for simplicity.

In the limit of enhanced non-abelian gauge symmetry, as shown in Figure \ref{fig:string_junc_bet_bunch}, the branes in each bunch (square) collide while the distance between the two bunches (squares) is non-zero. In this case, we can argue that in the BPS string junction, only the string segments $L_j$ have non-vanishing length, and all strings in the square are contracted to a point. 

\begin{figure}
    \centering
    \subfloat[]{
        \begin{tikzpicture}
        \filldraw (0,0) circle (.1)
        (1,0) circle (.05)
        (3,0) circle (.1);
        \draw (0,0) parabola bend (0.5,0.3) (1,0);
        \draw (0,0) parabola bend (0.5,-0.3) (1,0);
        \draw (1,0)--(3,0);
        \node[font=\fontsize{4}{6}\selectfont] at(-0.2,-0.2) {$B_0$};
        \node[font=\fontsize{4}{6}\selectfont] at(3.2,-0.2) {$B_1$};
        \node[font=\fontsize{4}{6}\selectfont] at(1.2,0.2) {$N$};
        \node[font=\fontsize{4}{6}\selectfont] at(0.5,0.5) {$(p_1,q_1)$};
        \node[font=\fontsize{4}{6}\selectfont] at(0.5,-0.5) {$(p_2,q_2)$};
        \node[font=\fontsize{4}{6}\selectfont] at(2,-0.5) {$(p_1+p_2,q_1+q_2)$};
    \end{tikzpicture}
    }
    \subfloat[]{
        \begin{tikzpicture}
        \filldraw (0,0) circle (.1)
        (1,0) circle (.05)
        (3,0) circle (.1);
        \draw (0,0)--(3,0);
        \node[font=\fontsize{4}{6}\selectfont] at(-0.2,-0.2) {$B_0$};
        \node[font=\fontsize{4}{6}\selectfont] at(3.2,-0.2) {$B_1$};
        \node[font=\fontsize{4}{6}\selectfont] at(1.2,0.2) {$N$};
        \node[font=\fontsize{4}{6}\selectfont] at(0.5,0.3) {$(p_1,q_1)$};
        \node[font=\fontsize{4}{6}\selectfont] at(0.5,-0.5) {$(p_2,q_2)$};
        \node[font=\fontsize{4}{6}\selectfont] at(2,-0.5) {$(p_1+p_2,q_1+q_2)$};
    \end{tikzpicture}
    }
    \subfloat[]{
    \begin{tikzpicture}
        \filldraw (0,0) circle (.1)
        (3,0) circle (.1);
        \draw (0,0)--(3,0);
        \node[font=\fontsize{4}{6}\selectfont] at(-0.2,-0.2) {$B_0$};
        \node[font=\fontsize{4}{6}\selectfont] at(3.2,-0.2) {$B_1$};
        \node[font=\fontsize{4}{6}\selectfont] at(0.2,0.2) {$N$};
        \node[font=\fontsize{4}{6}\selectfont] at(1.5,-0.5) {$(p_1+p_2,q_1+q_2)$};
    \end{tikzpicture}
    }
    \caption{Visualization of different 3-junction configurations.}
    \label{fig:diff_3_junc_config}
\end{figure}
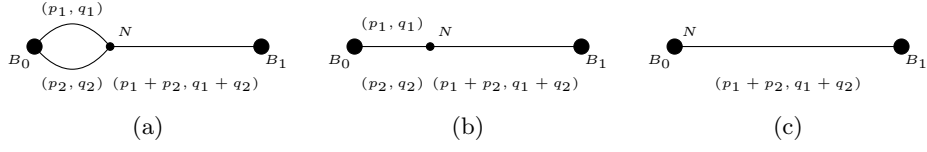

First, the relative angle condition (\ref{rel-angle-cond}) is obviously satisfied if the strings in the square are contracted to a point. Furthermore, let us see Figure \ref{fig:diff_3_junc_config} with different configurations of 3-string junctions. The lower bound of total length of the all three strings can be calculated by the holomorphic expression (\ref{length_geodesic}). By the path independence property of BPS mass, the configuration 2.(a) has the same lower bound of the total length of all three string segments as the configuration 2.(b). Then from the triangular inequality
\begin{equation}
    T_{p_1,q_1}+T_{p_2,q_2}\geq T_{p,q}\,,
    \label{tri_inequ_tension}
\end{equation}
we can see that the lower bound of total length of the all three string segments in the configuration 2.(c) is lower than that in 2.(b). Thus such lower bound of total length of the all three string segments can be reached by choosing the straight line in the holomorphic coordinate. Hence we can see that the BPS string junction, which has the minimal total length, should have the configuration where only $L_j$ has a non-vanishing length.

We conclude that the BPS string junction connecting two bunches takes the form of a number of $(p,q)$-strings connecting the two bunches, and the total mass of these strings should be minimized. To achieve this we should minimize $\sum_j |\int_{L_j} h_{p_j,q_j} \dd t|$, with the fixed asymptotic charges $P=\sum_{j\in B_a} p_j$, $Q=\sum_{j\in B_a} q_j$ in the standard presentation (\ref{standard_rep_massless_junction}). Here $j\in B_a$ denotes all strings ending on the 7-branes in the $B_a$ bunch. $a$ can be chosen as either 0 or 1, which does not affect the final mass as shown below. From the path independence of BPS mass, we can safely set all $L_j$s to be a single straight line connecting the two bunches. Finally, we have the inequality
\begin{equation}
    \sum_j |\int_{L} h_{p_j,q_j} \dd t|\geq |\int_{L} \sum_j h_{p_j,q_j}\dd t|\,.
    \label{lower_bound_string_junction}
\end{equation}
If $P=np, Q=nq$ with $p,q$ relatively prime, then the minimum of l.h.s. above can be achieved when there are $n$ $(p,q)$-strings connecting the two bunches. 

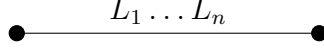
\begin{figure}
    \centering
    \begin{tikzpicture}
    \filldraw (0,0) circle (.1);
    \filldraw (4,0) circle (.1);
    \draw (0,0)--(4,0);
    \node[font=\fontsize{12}{12}\selectfont] at(2,0.3) {$L_1\dots L_n$};
    \end{tikzpicture}
    \caption{The BPS string junction configuration. All $L_i$, $i=1\dots n$ have the same charge $(p,q)$. This configuration is the unique choice up to homology.}
    \label{fig:BPS_string_junction_configuration}
\end{figure}

In summary, the BPS configuration of the string junction connecting two bunches of 7-branes is shown in the Figure \ref{fig:BPS_string_junction_configuration}, and we have
\begin{equation}
    m_{\text{junction}}=C_g n|\int_{L}  h_{p,q}\dd t|=C_g |\int_{L}  h_{P,Q}\dd t|=C_g|\int_L (P-Q\tau)\eta^2 \prod_i (t-t_i)^{-\frac{n_i}{12}}\dd t|\,.
    \label{string_junction_mass_PQ}
\end{equation}
We can see that the mass of such string junction only depends on its asymptotic charge $(P,Q)$. 

\subsection{Massive BPS string junctions connecting $n$ bunches of 7-branes}

\label{subsec_massive_junction_n_bunches}

Now we discuss the case with $n>2$ bunches of 7-branes, which is more involved than the cases with two bunches. 

\begin{figure}
    \centering
    \begin{tikzpicture}
        \draw[fill=gray!20] (0,0) circle (1);
        \filldraw (0,3) circle (.1);
        \draw plot [smooth] coordinates{(0.,3) (0.5,2) (0.707,0.707)};
        \draw plot [smooth] coordinates{(0.,3) (-0.5,2) (-0.707,0.707)};
        \draw (-0.1,2) circle (.01);
        \draw (0,2) circle (.01);
        \draw (0.1,2) circle (.01);
        \node at (0.5,3) {$B_0$};
        \node at (1.2,2) {$(P_0,Q_0)$};
        
        \filldraw[rotate=120] (0,3) circle (.1);
        \draw[rotate=120] plot [smooth] coordinates{(0.,3) (0.5,2) (0.707,0.707)};
        \draw[rotate=120] plot [smooth] coordinates{(0.,3) (-0.5,2) (-0.707,0.707)};
        \draw[rotate=120] (-0.1,2) circle (.01);
        \draw[rotate=120] (0,2) circle (.01);
        \draw[rotate=120] (0.1,2) circle (.01);
        \node at (2.5,-2) {$B_1$};
        \node at (1,-1.7) {$(P_1,Q_1)$};
        
        \filldraw[rotate=240] (0,3) circle (.1);
        \draw[rotate=120] (-0.1,2) circle (.01);
        \draw[rotate=120] (0,2) circle (.01);
        \draw[rotate=120] (0.1,2) circle (.01);
        \draw[rotate=240] plot [smooth] coordinates{(0.,3) (0.5,2) (0.707,0.707)};
        \draw[rotate=240] plot [smooth] coordinates{(0.,3) (-0.5,2) (-0.707,0.707)};
        \draw[rotate=240] (-0.1,2) circle (.01);
        \draw[rotate=240](0,2) circle (.01);
        \draw [rotate=240](0.1,2) circle (.01);
        \node at (-2.5,-2) {$B_2$};
        \node at (-1,-1.7) {$(P_2,Q_2)$};
    \end{tikzpicture}
    \caption{The figure shows different possible configurations of string junctions connecting three bunches of 7-branes $B_i$, which are denoted by the three black dots. The large gray circle denotes a possible complicated web of strings. The lines connecting the bunches and the web denote the strings that end on the bunch and contribute to the asymptotic charge of each bunch. $(P_i,Q_i)$ denotes the total asymptotic charge of each bunch.}
    \label{fig:3-bunch-junction}
\end{figure}
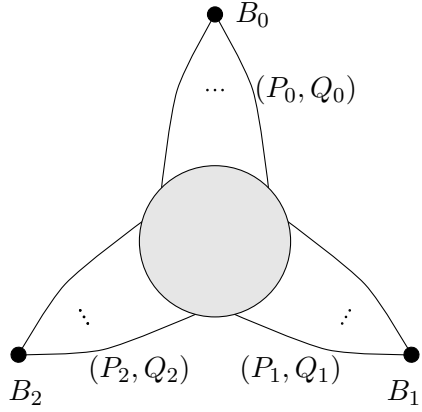

Let us start with the 3-bunch case. With an analogous discussion as in section \ref{subsec_massive_junction_two_bunches}, for BPS string junctions, only the strings that contribute to the asymptotic charge at each bunch of 7-branes can have non-trivial configurations. Hence we can obtain the general form of the 3-bunch string junction as in Figure \ref{fig:3-bunch-junction}.

We can consider the lower bound of the mass among all possible string junction configurations. The mass of the string junction configuration is
\begin{equation}
\label{n-junction-inequal}
    m_{\text{junction}}=\sum_i \left|\int_{L_i} h_{p_i,q_i} \dd t\right|\geq \left|\sum_i\int_{L_i} h_{p_i,q_i} \dd t\right|\,.
\end{equation}
Now we focus the discussion on the lower bound of the r.h.s. of (\ref{n-junction-inequal}). Because $h_{p_1,q_1}+h_{p_2,q_2}=h_{p_1+p_2,q_1+q_2}$, we can separate all $h_{p_i,q_i}$s at the r.h.s. of (\ref{n-junction-inequal}) into $p_i h_{1,0}+q_i h_{0,1}$. 

\begin{figure}
    \centering
    \subfloat[]{
        \begin{tikzpicture}
        \draw[fill=gray!20] (0,0) circle (1);
        \filldraw (0,3) circle (.1);
        \filldraw (2.60,-1.5) circle (.1);
        \filldraw (-2.60,-1.5) circle (.1);
        \draw plot [smooth] coordinates{(0.,3) (0.5,2) (0.707,0.707)};
        \draw plot [smooth] coordinates{(0.,3) (0.6,2) (0.757,0.657)};
        \draw plot [smooth] coordinates{(0.,3) (-0.5,2) (-0.707,0.707)};
        \draw plot [smooth] coordinates{(0.,3) (-0.6,2) (-0.757,0.657)};
        \node at (1.5,2) {$(P_{01},Q_{01})$};
        \node at (-1.5,2) {$(P_{20},Q_{20})$};

        \draw[rotate=120] plot [smooth] coordinates{(0.,3) (0.5,2) (0.707,0.707)};
        \draw[rotate=120] plot [smooth] coordinates{(0.,3) (0.6,2) (0.757,0.657)};
        \draw[rotate=120] plot [smooth] coordinates{(0.,3) (-0.5,2) (-0.707,0.707)};
        \draw[rotate=120] plot [smooth] coordinates{(0.,3) (-0.6,2) (-0.757,0.657)};
        \node at (2.7,0) {$(P_{01},Q_{01})$};
        \node at (1,-1.8) {$(P_{12},Q_{12})$};

        \draw[rotate=240] plot [smooth] coordinates{(0.,3) (0.5,2) (0.707,0.707)};
        \draw[rotate=240] plot [smooth] coordinates{(0.,3) (0.6,2) (0.757,0.657)};
        \draw[rotate=240] plot [smooth] coordinates{(0.,3) (-0.5,2) (-0.707,0.707)};
        \draw[rotate=240] plot [smooth] coordinates{(0.,3) (-0.6,2) (-0.757,0.657)};
        \node at (-1,-1.8) {$(P_{12},Q_{12})$};
        \node at (-2.7,0) {$(P_{20},Q_{20})$};
        \node at (0.5,3) {$B_0$};
        \node at (2.5,-2) {$B_1$};
        \node at (-2.5,-2) {$B_2$};
        \end{tikzpicture}
    }
    \subfloat[]{
        \begin{tikzpicture}
        \filldraw (0,3) circle (.1);
        \filldraw (2.60,-1.5) circle (.1);
        \filldraw (-2.60,-1.5) circle (.1);
        \draw [line width=2pt] (0,3)--(2.60,-1.5);
        \draw [line width=2pt] (2.60,-1.5)--(-2.60,-1.5);
        \draw [line width=2pt] (-2.60,-1.5)--(0,3);
        \node at (2,2) {$L_{01},(P_{01},Q_{01})$};
        \node at (0,-2) {$L_{12},(P_{12},Q_{12})$};
        \node at (-2,2) {$L_{20},(P_{20},Q_{20})$};
        \draw [rounded corners,dashed] (0,2.7)--(2.338,-1.35)--(-2.338,-1.35)--cycle;
        \node at (0,-1) {$\Tilde{L}$};
        \node at (0.5,3) {$B_0$};
        \node at (2.5,-2) {$B_1$};
        \node at (-2.5,-2) {$B_2$};
        \end{tikzpicture}
    }
    \caption{$(P_{ij},Q_{ij})$ is the total asymptotic charge of the strings from $B_i$ to $B_j$. $L_{ij}$ is some path from $B_i$ to $B_j$. When discussing the lower bound (\ref{lower_bound_string_junction}), the general configuration in Figure \ref{fig:3-bunch-junction} can be deformed into (a). By the path independence of (\ref{lower_bound_string_junction}), (a) has the same lower bound as (b). In (\ref{nP_nQ_lower_bound_contri}), the closed path $L_{01}+L_{12}+L_{20}$ can be deformed smoothly into $\Tilde{L}$.}
    \label{fig:3-bunch-junction-transformation}
\end{figure}
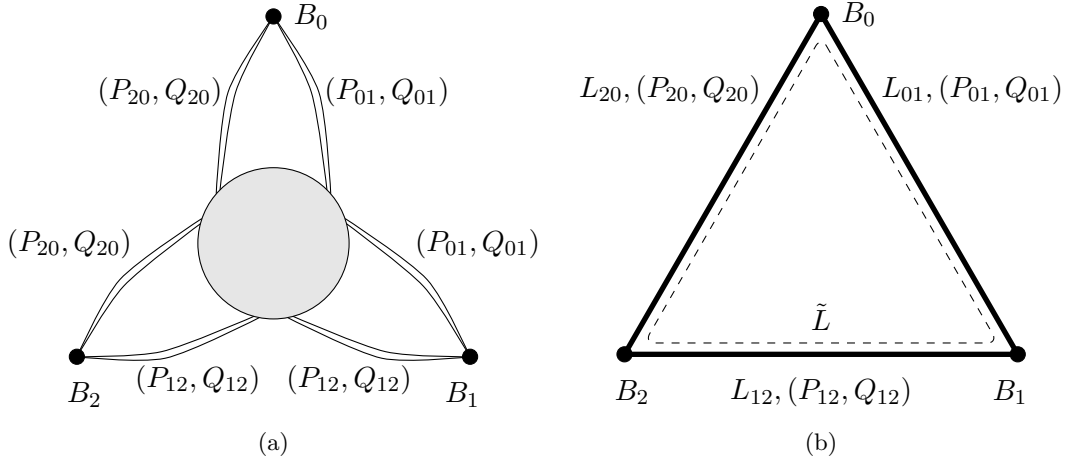

The total asymptotic charge of all bunches are zero, $\sum_i P_i=\sum_i Q_i=0$. Hence with the above separation $h_{p_i,q_i}=p_i h_{1,0}+q_i h_{0,1}$, the configuration in Figure \ref{fig:3-bunch-junction} can be understood as a series $(1,0)$-strings and $(0,1)$-string connecting the 3 bunches. With this interpretation, we can compute the number of $(1,0)$ and $(0,1)$-strings connecting each pair of two bunches of 7-branes, from the total asymptotic charge $(P_{ij},Q_{ij})$ of the strings from $B_i$ to $B_j$, as shown in Figure \ref{fig:3-bunch-junction-transformation}.

In the 3-bunch case above, with the asymptotic charge $(P_i,Q_i)$ of each bunch under the condition $\sum_i P_i=\sum_i Q_i=0$, we can find that the asymptotic charge $(P_{ij},Q_{ij})$ from $B_i$ to $B_j$
\begin{equation}
    \begin{aligned}
        P_{01}-P_{20}&=P_0\ ,\  Q_{01}-Q_{20}=Q_0\,,\\
        P_{12}-P_{01}&=P_1\ ,\ Q_{12}-Q_{01}=Q_1\,,\\
        P_{20}-P_{12}&=P_2\ ,\ Q_{20}-Q_{12}=Q_2\,.\\
    \end{aligned}
\end{equation}
We get the solution
\begin{equation}
    \begin{aligned}
        P_{01}&=P_0+n_P\ ,\ Q_{01}=Q_0+n_Q\,,\\
        P_{12}&=P_0+P_1+n_P\ ,\ Q_{12}=Q_0+Q_1+n_Q\,,\\
        P_{20}&=n_P\ ,\  Q_{20}=n_Q\ ,\ n_P,n_Q\in \mb{Z}\,.\\
    \end{aligned}
\end{equation}

Then we can explicitly express the lower bound of BPS mass
\begin{equation}
    \begin{aligned}
        \left|\sum_{i}\int_{L_i} h_{p_i,q_i} \dd t\right|&=\left|\sum_{i<j,\;0\leq i,j<n=3}\int_{L_{ij}} h_{P_{ij},Q_{ij}}\dd t\right|\\
        &=\left| \int_{L_{01}} h_{P_0,Q_0} \dd t + \int_{L_{12}} h_{P_0+P_1,Q_0+Q_1} \dd t+ \int_{L_{01}+L_{12}+L_{20}}h_{n_P,n_Q}\dd t\right|\,.
    \end{aligned}
    \label{lower_bound_3-junction}
\end{equation}
Here $L_i$ denotes all strings in Figure \ref{fig:3-bunch-junction} and $L_{ij}$ denotes some line connecting $B_i$ and $B_j$ which does not intersect any branch cut. 

Different choices of $n_P,n_Q$ would give the same lower bound, due to the following reason. Because each $L_{ij}$ does not intersect any branch cut, we can transform the closed path $L_{01}+L_{12}+L_{20}$ smoothly into $\Tilde{L}$ in Figure \ref{fig:3-bunch-junction-transformation}. There is no 7-brane in the interior of $\Tilde{L}$, so we have
\begin{equation}
    \int_{L_{01}+L_{12}+L_{20}} h_{01} \dd t=\int_{\Tilde{L}} h_{01} \dd t=\int_{L_{01}+L_{12}+L_{20}} h_{10} \dd t =\int_{\Tilde{L}} h_{10} \dd t=0\,.
    \label{nP_nQ_lower_bound_contri}
\end{equation}
Hench different choices of $n_P,n_Q$ do not affect the lower bound (\ref{lower_bound_3-junction}).

BPS string junctions should have the minimal $m_{\text{junction}}$ among all possible configurations, but it is possible that there exists no configuration that can reach the lower bound (\ref{lower_bound_3-junction}). 

On the other hand, we can also attempt to find some light string junction configuration, which leads to an upper bound of the mass of BPS string junctions. Here we use the special configuration (b) in Figure \ref{fig:3-bunch-junction-transformation} to estimate the upper bound. The string junction in this configuration has the mass
\begin{equation}
    m_{\text{upper}}=\sum_{i<j,\;0\leq i,j<n=3}\left|\int_{L_{ij}} h_{P_{ij},Q_{ij}}\dd t\right|.
    \label{upper_bound_3-junction}
\end{equation}
Notice that this mass (\ref{upper_bound_3-junction}) depends on $n_P,n_Q$, and we need to find the $n_P,n_Q$ giving the minimal value of $m_{\text{upper}}$. Then we can obtain the range of the mass of BPS string junction connecting $n=3$ bunches of 7-branes
\begin{equation}
    \left|\sum_{i<j,\;0\leq i,j<n}\int_{L_{ij}} h_{P_{ij},Q_{ij}}\dd t\right|\leq m_{\text{junction}}\leq\sum_{i<j,\;0\leq i,j<n}\left|\int_{L_{ij}} h_{P_{ij},Q_{ij}}\dd t\right|\,.
    \label{bound_n-junction}
\end{equation}

If the relative phases among $\int_{L_{ij}} h_{P_{ij},Q_{ij}}\dd t$ is not too large, we can expect to have a tight constraint from such inequalities. When all of the bunches locate on a single straight line, obviously the relative phases between different terms in (\ref{bound_n-junction}) are zero and we can get the true mass of the BPS string junction
\begin{equation}
    m_{\text{junction}}=\left|\sum_{i<j,\;0\leq i,j<n}\int_{L_{ij}} h_{P_{ij},Q_{ij}}\dd t\right|\,.
\end{equation}

In fact, for general $n$-bunch cases, we can get the same form as (\ref{bound_n-junction}). Nonetheless for larger $n$, we need more unfixed integers than $n_P,n_Q$ to determine the $(P_{ij},Q_{ij})$ for the upper bound. In general we have the following algorithm to compute the upper bound on the mass of BPS string junction:
\begin{enumerate}
    \item Calculate all $2\binom{n}{2}$ line integrations $\int_{L_{ij}}h_{1,0},\int_{L_{ij}}h_{0,1}$.
    \item Find the general solution of $P_{i,i+1}-P_{i-1,i}=P_i,Q_{i,i+1}-Q_{i-1,i}=Q_i,\;i\in \mb{Z}_n$, with  $2\binom{n}{2}-n+1$ unfixed integers $n_a$.
    \item Find the integral solution of $n_a$ minimize $\sum_{i<j,\;0\leq i,j<n}\left|\int_{L_{ij}} h_{P_{ij},Q_{ij}}\dd t\right|$.
    \item Use (\ref{bound_n-junction}) to determine the upper bound on the mass of BPS string junctions.
\end{enumerate}
For example when $n=5$, we need to calculate $20$ line integrations and minimize a function with 16 integer variables.

\subsection{Mass of BPS string junctions in constant $\tau$ backgrounds}
\label{sec:stringjunc-constant-tau}

With (\ref{string_junction_mass_PQ}), we can calculate the mass of BPS string junctions connecting two bunches through a holomorphic line integration, which is path independent. While it is always possible to get numerical results after choosing the simplest path as possible, it is hard to obtain an analytical result due to the Dedekind $\eta$-function in the expressions. In order to obtain a more intuitive, analytical result, here we consider the 8D F-theory configurations with a constant $\tau$ and $\eta(\tau)$. We provide the details of calculation in Appendix  \ref{sec_mass_string_junction}.

In addition to the requirement of a constant $\tau$, we make another assumption that only two bunches of 7-branes are close to each other. More precisely, we fix two bunches of 7-branes at $t_0=0$ and $t=t_1$, and all the other bunches are far away, i.e. $0<|t_1|\ll |t_i|,i\geq 2$. 

%Here we abuse the notation $t_i$ to denote the location of bunches instead of a single 7-brane.

We can obtain the mass of BPS string junctions up to the first order approximation. Denote the total asymptotic charge connecting the two bunches $B_0,B_1$ as $(P,Q)$, when the total number of 7-branes in the two bunches $n_0+n_1<12$, the BPS mass is computed as
\begin{equation}
    \begin{aligned}
        m_{\text{junction}}=&C_g\left|(P-Q\tau) \eta^2(\tau)\prod_{i=2} t_i^{-\frac{n_i}{12}}B\left(1-\frac{n_0}{12},1-\frac{n_1}{12}\right)\right|\\
        &\times \left|1+\left(\frac{12-n_0}{24-n_0-n_1}\sum_{i=2} \frac{n_i}{12t_i}\right)t_1\right| |t_1|^{1-\frac{n_0+n_1}{12}}+\mc{O}(t_1^{3-\frac{n_0+n_1}{12}})\,.
    \end{aligned}
    \label{2-string_constant_tau}
\end{equation}
Here $B(.,.)$ is the Beta function
\begin{equation}
    B(p, q)=\int_0^1 t^{p-1}(1-t)^{q-1} \dd t \,.
\end{equation}
We find that $m_{\text{junction}}$ has a simple power law expression with respect to $t_1$ while the power depends on the total number of 7-branes in the two approaching bunches.

In this expression, it seems like that the BPS mass decreases when $n_0+n_1<12$, stay constant when $n_0+n_1=12$ and increases when $n_0+n_1>12$ as the distance $t_1$ between two bunches of 7-branes getting smaller. However, the additional subtlety is that the numerical coefficient  $C_g$ would change with the complex structure moduli $t_1$, if we want to keep the volume $V_{\mb{P}^1}$ of $\mb{P}^1$ ( K\"ahler moduli) fixed. Note that since we require that $|t_1|\ll |t_i|$ $(i\geq 2)$, only the effect of varying $t_1$ is significant.

To derive the dependence of $C_g$ on $t_1$, we look at the neighborhood $U=\left\{|t|\leq R\right\}$ with $|t_1|\ll R\ll |t_i| (i\geq 2)$. When $n_0+n_1< 12$, the volume of this neighborhood can be estimated as
\begin{equation}
        V_U=\frac{12\pi}{12-n_0-n_1} C_g^2 A\left|R^{2-\frac{n_0+n_1}{12}}+\mc{O}(|t_1|)+\mc{O}\left(\left|t_1^{2-\frac{n_0+n_1}{6}}\right|\right)\right|\,.
    \label{volume_local_region}
\end{equation}
Here
\begin{equation}
    A(t_2,\dots)=\left|\tau_2 \eta(\tau)^2 \bar{\eta}(\bar{\tau})^2\prod_{i=2}|t_i|^{-\frac{n_i}{6}}\right|
\end{equation}
is independent of $t_1,C_g$, which is considered as a constant in the current discussion.

When $n_0+n_1=12$, we have
\begin{equation}
\label{VU-12}
        V_U= 2\pi C_g^2 A\left|\log t_1+\mc{O}(1)\right|\,.      
\end{equation}
When $n_0+n_1>12$, we have
\begin{equation}
    V_U=C_g^2 A\times \mc{O}\left(\left|t_1^{2-\frac{n_0+n_1}{6}}\right|\right) \,.
\end{equation}
The qualitative feature of (\ref{VU-12}) can be understood as follows. It is known that in 8D F-theory, a single 7-brane contributes $\frac{\pi}{6}$ defect angle to the asymptotic geometry~\cite{Greene:1989ya}. When there are 12 7-branes at a point, the defect angle is $2\pi$, so the asymptotic geometry becomes a cylinder, which leads to the volume tending to infinity at the order $|\log t_1|$. 

As we can see, in the limit of $t_1\rightarrow 0$ with $C_g$ fixed, the volume of the neighborhood $U$ can be ignored when $n_0+n_1<12$ and play the dominant role when $n_0+n_1\geq 12$. Thus when $n_0+n_1\geq 12$ and $|t_1|$ is small enough, we can use $V_U$ as an approximation of $V_{\mb{P}^1}$. 

When $n_0+n_1= 12$, we solve from (\ref{VU-12})
\begin{equation}
    C_g=\left|\frac{V_{\mb{P}_1}}{2\pi A\log t_1}\right|^{\frac{1}{2}}+\mc{O}(|\log t_1|^{-\frac{3}{2}}) .\,.
\end{equation}
Hence the mass of BPS string junction is
\begin{equation}
        \begin{aligned}
        m_{\text{junction}}=&\left|\frac{P-Q\tau}{\sqrt{2\pi\tau_2}} B\left(1-\frac{n_0}{12},1-\frac{n_1}{12}\right)  \left[1+\left(\frac{12-n_0}{24-n_0-n_1}\sum_{i=2} \frac{n_i}{12t_i}\right)\right]\right|\sqrt{\left|\frac{V_{\mb{P}^1}}{\log t_1}\right|}\\&+\mc{O}(|t_1|^2|\log t_1|^{-\frac{1}{2}}) \,.
    \end{aligned}  
    \label{mass_string_junction_12brane}
\end{equation}

In addition, when the two bunches at $t=0$ and $t=t_1$ approach each other, $t_1\rightarrow 0$ and thus $C_g\rightarrow 0$. It  implies that the distances between other bunches at $t_2,t_3\dots$ are also decreasing with the properly normalized base metric. So we can expect that in the limit where $t_1=0$, there are only two bunches on the whole base, each with 12 7-branes. In the language of Weierstrass model, if we have a singular point with $\op{ord}(\Delta)=12$, then we can only have two bunches with vanishing order $(\op{ord}(f),\op{ord}(g),12)$ and $(\geq 8-\op{ord}(f),\geq 12-\op{ord}(g),12)$.

\subsection{More discussions on constant $\tau$}
\label{sec:more-constant-tau}

We have obtained analytical results in a constant $\tau$ background. In this section we discuss the applicability and generalization of such assumption.

Since $\tau$ is related to the Weierstrass model via the Jacobi $j$-function:
\begin{equation}
    j(\tau)=4 \frac{12^3 f^3}{\Delta}\,,
\end{equation}
$\tau$ is constant iff $j(\tau)$ is constant, and there are three cases discussed in the literature.

\begin{enumerate}
\item
$f\equiv 0$, as discussed in \cite{Dasgupta:1996ij}, generically we have 
\begin{equation}
    g(t)=\prod_{i=1}^{12}\left(t-t_i\right)\ , \  \Delta(t)=27 \prod_{i=1}^{12}\left(t-t_i\right)^2\ ,\ \tau \equiv e^{i\pi/3}\,.
\end{equation}
The Weierstrass model describes 12 bunches of type II 7-branes \textbf{CA} without any enhanced gauge symmetry. The constant axiodilaton background $\tau \equiv e^{i\pi/3}$ is the fixed point of the $\op{SL}(2,\mb{Z})$ monodromy. When multiple bunches collide, the singularities can be enhanced to type IV $A_2$ for 2 bunches, type $I_0^*$ $D_4$ for 3 bunches, type $IV^*$ $E_6$ for 4 bunches and type $II^*$ $E_8$ for 5 bunches.

\item
$g\equiv 0$, as discussed in \cite{Dasgupta:1996ij}, generically we have 
\begin{equation}
    f(t)=\prod_{i=1}^8\left(t-t_i\right)\ , \  \Delta(t)=4 \prod_{i=1}^8\left(t-t_i\right)^3\ , \ \tau\equiv i\,.
\end{equation}
The Weierstrass model describes 8 bunches of 7-branes, and each is composed of 3 7-branes \textbf{CAA} and form a type $III$ $A_1$ singularity. The constant axiodilaton background $\tau\equiv i$ is the fixed point of the $\op{SL}(2,\mb{Z})$ monodromy. When multiple bunches collide, the singularities can be enhanced to type $I_0^*$ $D_4$ for 2 bunches and type $III^*$ $E_7$ for 3 bunches.

\item
$f^3/g^2=\text{constant}$, as discussed in \cite{Sen:1996vd}, generically we have 
\begin{equation}
    f(t)=\alpha \phi^2\ ,\ g(t)= \phi^3\ ,\ \Delta(t)=(4\alpha^3+27) \phi^6\ , \  \phi(t)=\prod_{i=1}^4(t-t_i)\ ,\  j(\tau)=\frac{6912\alpha^3}{4\alpha^3+27}\,.
\end{equation}
The Weierstrass model describes 4 bunches of 7-branes, and each is composed of 6 7-branes \textbf{AAAABC} and form a type $I_0^*$ $D_4$ singularity. This configuration is equivalent to type IIB on a $T^2$ orientifold.

\end{enumerate}

The above results (\ref{2-string_constant_tau}) are valid in these cases of constant $\tau$. Nonetheless, we can also apply the result (\ref{2-string_constant_tau}) to estimate the mass for BPS string junctions in more general cases, where $\tau$ is locally constant near some bunches of 7-branes.

More explicitly, we list the fixed values of $\tau$ by the singularity type in Table \ref{tab:tau_value_singu}.
\begin{table}
    \centering
    \begin{tabular}{c|c|c}
         Singularity type& $j(\tau)$ &$\tau$\\\hline
        Type II, type IV $A_2$, type $IV^*$ $E_6$, type $II^*$ $E_8$ & $0$ &$e^{i\pi/3}$\\\hline
        Type III $A_1$, type $III^*$ $E_7$ & $1728$ & i\\\hline
        Type $I_0^*$ $D_4$ &arbitrary & arbitrary\\\hline
        $I_{n}$, type $I_n^*$ $D_{n+4}$& $\infty$ & $i\infty$
    \end{tabular}
    \caption{Value of $j(\tau)$ and $\tau$ for each Kodaira singularity type of 7-branes.}
    \label{tab:tau_value_singu}
\end{table}

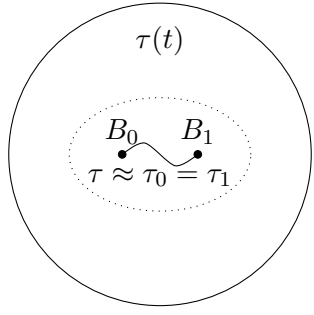
\begin{figure}
    \centering
     \begin{tikzpicture}
        \draw (0,0) circle(2.0);
        \filldraw (-0.5,0) circle (.05)
        (0.5,0) circle (.05);
        \draw plot [smooth] coordinates{(-0.5,0) (-0.2,0.15) (.2,-0.15) (.5,0)};
        \draw[dotted] (0,0) ellipse (1.2 and 0.75);
        \node at (0,1.5) {$\tau(t)$};
        \node at (0,-0.3) {$\tau\approx\tau_0=\tau_1$};
        \node at (-0.5,0.3) {$B_0$};
        \node at (0.5,0.3) {$B_1$};
    \end{tikzpicture}
    \caption{A locally constant $\tau$ configuration.}
    \label{fig:local_constant_tau}
\end{figure}

For instance if $\tau=i$, $e^{i\pi/3}$, locally $j(\tau)$ is continuous and $j(\tau)$, $\tau$ would be slow varying near bunches of 7-branes with the same $\tau$ value, e.g. $E_6$ approaching type II. In this case we can expect that in the neighborhood $U$ of these 7-branes, $\tau$ is nearly constant, as depicted in Figure \ref{fig:local_constant_tau}. So we can use (\ref{2-string_constant_tau}) to estimate the mass of BPS string junctions between these bunches of 7-branes.

If the two approaching bunches have different $\tau$, e.g. $E_7$ approaching type II, then we cannot expect that $\tau$ can be estimated by a constant value. Nonetheless because $j(\tau)$ is continuous, we can still expect that in the neighborhood $U$, there exists some upper bound $C_{\text{upper}}$ for $|(P-Q\tau)\eta^2|$, from which we can obtain an upper bound for the mass of BPS string junctions
\begin{equation}
\begin{aligned}
    &m_{\text{junction}}\leq C_gC_{\text{upper}}\left|\prod_{i=2} t_i^{-\frac{n_i}{12}}B\left(1-\frac{n_0}{12},1-\frac{n_1}{12}\right)\right|\\
    &\quad\quad\quad\times \left|1+\left(\frac{12-n_0}{24-n_0-n_1}\sum_{i=2} \frac{n_i}{12t_i}\right)t_1+\mc{O}(t_1^2)\right| |t_1|^{1-\frac{n_0+n_1}{12}}\,.
\end{aligned}   
\end{equation}

However, this estimation is not valid when there exists type $I_n$ and $D_{n>4}$ singularities. Since in these cases $\tau$ and $j(\tau)$ tend to infinite and have no upper bound, while playing the dominant role in the integration. In fact, the current method of estimating the BPS string junction mass would work better in the strongly coupled regime of F-theory than the weakly coupled IIB superstring theory with type $I_n$ and $D_{n>4}$ 7-branes.

\section{KK modes in 8D F-theory}
\label{sec:KK}

\subsection{Dimensional reduction of IIB 2-form fields}
\label{sec:dim-red-2-form}

In this section we provide a preliminary of analysis of the massive KK modes in 8D F-theory.

The dimensional reduction of 2-form gauge fields creates KK-modes in 8D F-theory on the base $\mb{P}^1$. We turn on the 2-form field in the 10D IIB supergravity action (\ref{type_IIB_sugra}) and only consider the leading order kinetic term: 
\begin{equation}
    S_{\rm 2-form}=\int M_{IJ}dB^I\wedge*dB^J\,.
    \label{SKK}
\end{equation}
In the action, we have
\begin{equation}
    M_{IJ}=\frac{1}{\operatorname{Im} \tau}\left( \begin{array}{cc}
        |\tau^2| & -\operatorname{Re} \tau \\
        -\operatorname{Re} \tau & 1
    \end{array}\right)\\ \,,
\end{equation}
and $B^I$ $(I=1,2)$ is the $\op{SL}(2,\mb{Z})$ 2-form gauge field doublet. From this term we get the equation of motion for $B^I$:
\begin{equation}
    \partial_{j_3}M_{IJ}\sqrt{-g}g^{j_1i_1}g^{j_2i_2}g^{j_3i_3}(\partial_{i_1}B_{i_2i_3}^J+\partial_{i_2}B_{i_3i_1}^J+\partial_{i_3}B_{i_1i_2}^J)=0\,.
    \label{2-form-f-eq}
\end{equation}

Since the 10D space-time manifold is $\mb{R}^{1,7}\times \mb{P}^1$, we can expand the 2-form field $B^I$ as
\begin{equation}
    B^I=b^I\cdot c_{\nu\mu}dx^{\mu}\wedge dx^{\nu}+b^I_i\cdot c_{\nu}dx^i\wedge dx^{\nu}+b^I_{ij}\cdot c dx^i\wedge dx^j\,.
\end{equation}
The Greek letters runs over the Minkowski space indices from 0 to 7 and the lower case Latin letters run over the compact space indices 8, 9. In fact, the equation of motion would mix these three components $b^I$, $b_i^I$ and $b_{ij}^I$. For simplicity, we would assume that only one term in $B^I$ is non-vanishing. We discuss the three different cases separately:

\begin{enumerate}
\item If we only turn on a non-zero $b^I_{ij}$, we get
\begin{equation}
    \partial_{\mu}\partial^{\mu}c=0\ \ ,\  
    b_{89}^I=(M^{-1})^{IJ}\sqrt{g_p}\xi_J\,.
\end{equation}
Here $g_p=g_{88}g_{99}$. The issue is that $\xi_I$ is an invariant doublet on $\mb{P}^1$, hence $b_{89}^I$ would be inconsistent with the monodromy action, except for the cases with trivial monodromy.

\item if only $b_i^I\equiv B_{i\mu}^I$s are nonzero, we have
\begin{equation}
    \frac{\partial_{i'}\left[M_{IJ}\sqrt{g_p}g^{ij}g^{i'j'}(\partial_j b^J_{j'}-\partial_{j'}b^J_{j})\right]}{M_{IJ}\sqrt{g_p}g^{ij}b^J_j}=M^2=-\frac{\partial^{\mu}\partial_{\mu}c_{\nu}}{c_{\nu}}\,.
\end{equation}
$M$ is a constant which is the mass of the 1-form field $c_\nu$ in the 8D Minkowski space-time.

\item If only $b^I$ is nonzero, we arrive at the equation
\begin{equation}
    \frac{\partial_i (M_{IJ}\sqrt{g_p}g^{ij}\partial_jb^J)}{M_{IJ}\sqrt{g_p}b^J}=M^2=-\frac{\partial^{\mu}\partial_{\mu}c_{\nu \rho}}{c_{\nu\rho}}\,.
    \label{2-formKK}
\end{equation}
Similar $M$ is a constant which is the mass of 2-form field $c_{\mu\nu}$ in the 8D Minkowski space-time. This would be the set of KK modes we analyze in detail in this paper.

\end{enumerate}

\subsection{Case with a constant $\tau$: four $D_4$ bunches}
\label{sec:KK-4-D4}

For simplicity, we will only discuss the zero-form $b^I$ on the compact space which leads to the massive 2-form field $c_{\mu\nu}$ in the 8D Minkowski space-time. For constant $\tau$ background, (\ref{2-formKK}) can be simplified to
\begin{equation}
    \frac{4\partial \bar{\partial} b^I}{\sqrt{g_p}b^I}=-M^2=-\frac{\partial^{\mu}\partial_{\mu}c_{\nu \rho}}{c_{\nu\rho}}\,.
\end{equation}
The local holomorphic coordinate is expanded as $t=x+iy$, where $x$, $y$ are the real coordinates of $\mb{P}^1$, and $\partial=\frac{\partial}{\partial t}$. 

As a concrete example, let us consider four type $I_0^*$ $D_4$-bunches of 7-branes located at $(t_1/2,0)$, $(-t_1/2,0)$, $(T/2,0)$, $(-T/2,0)$ and $t_1\ll T$. The monodromy of each bunch should be $-I$. In the following discussions we would keep the volume of base $\mb{P}^1$ fixed and let $t_1$ approaches 0. 

To encode the monodromy behaviours in $b^I$, we use the following ansatz of $b^I$ 
\begin{equation}
   b^1+i b^2=\sqrt{\frac{(t-T/2)(t-t_1/2)}{(t+T/2)(t+t_1/2)}} \beta\,.
   \label{exp1}
\end{equation}
Here $\beta$ has trivial monodromy behaviour, and is smooth away from the location of 7-branes. 

In the codimension-one $(4,6,12)$ limit, from (\ref{ccl_V1}) and (\ref{ccl_V}) we have
\begin{equation}
   \sqrt{g_p}=\frac{V_{\mb{P}^1} T^2}{8\pi|(t^2-t_1^2/4)(t^2-T^2/4)|\log(T/t_1)}\times(1+\mc{O}(\log(T/t_1)^{-1}))\,.
\end{equation}
We do a further coordinate transformation to the spherical coordinate $(\theta,\phi)$:
\begin{equation}
    t=\sqrt{Tt_1}\cot (\theta/2)e^{i\phi}\,.
    \label{ztrf}
\end{equation}
Expanding $\beta$ in (\ref{exp1}) by spherical harmonics, 
\begin{equation}
    \beta=\sum_{l,m} c_{l,m}\mathrm{Y}^{l,m}\,,
\end{equation}
the equation of motion can be rewritten as
\begin{equation}
    \begin{aligned}
     &\sum_{l,m}c_{l,m}\{l(l+1)\sin^2\theta +\sqrt{\frac{t_1}{T}}[(1-\cos\theta)e^{-i\phi}-4(1+\cos\theta)e^{i\phi}]\left(\partial _{\theta}+\frac{m}{\sin \theta}\right)+\mc{O}\left(\frac{t_1}{T}\right)\}\mathrm{Y}^{l,m}(\theta,\phi) \\
     &=\sum_{l,m}\frac{V_{\mb{P}^1}M^2}{2\pi\log(T/t_1)} c_{l,m}\mathrm{Y}^{l,m}(\theta,\phi)\,.
    \end{aligned}
    \label{sphf}
\end{equation}
In the limit of $t_1\rightarrow 0$, we can set $t_1/T=0$ on the l.h.s. of (\ref{sphf}), and the equation can be simplified to 
\begin{equation}
    \begin{aligned}
     &\sum_{l,m}c_{l,m} \sin^2\theta l(l+1)\mathrm{Y}^{l,m}(\theta,\phi)=\sum_{l,m}\frac{VM^2}{2\pi\log(T/t_1)} c_{l,m}\mathrm{Y}^{l,m}(\theta,\phi)\,.
    \end{aligned}
    \label{splf_sphf}
\end{equation}
The problem becomes solving the eigenvalues and eigenvectors of an infinite dimensional matrix, whose elements are
\begin{equation}
    (\mathrm{A})_{(l_1,m_1,l_2,m_2)}= \int^{\pi}_0\sin\theta \mathrm{d}\theta \int^{2\pi}_0\mathrm{d}\phi \overline{\mathrm{Y}}^{l_1,m_1}(\theta,\phi)l_2(l_2+1)\sin^2\theta \mathrm{Y}^{l_2,m_2}(\theta,\phi)\,.
    \label{Alm-matrix}
\end{equation}

\begin{figure}[H]
    \centering
    \includegraphics[width=0.7\linewidth]{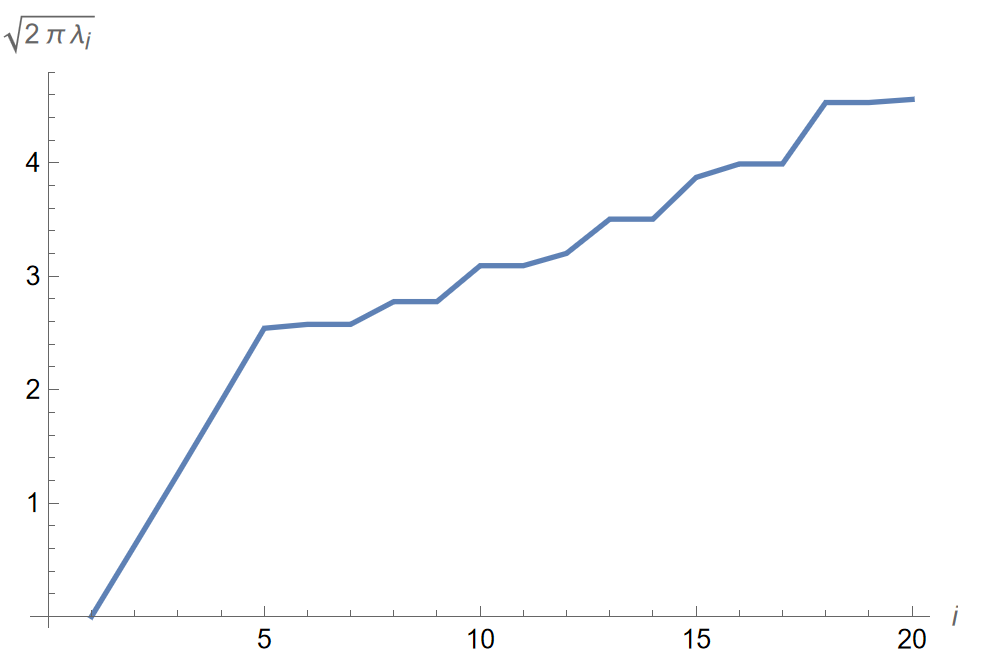}
    \caption{The figure showing the 20 lowest KK modes $\sqrt{2\pi \lambda_i}$.}
    \label{fig_egvtab}
\end{figure}

We set the cut off at $l=300$ to make the matrix (\ref{Alm-matrix}) finite dimensional. Under this approximation, we can solve the 20 smallest eigenvalues $\lambda_i$ of (\ref{splf_sphf}) numerically, which are listed in Table \ref{egvtab}\footnote{Because of the accuracy of the computation program, we do not get precise eigenvalues.} and Figure \ref{fig_egvtab}. 
\begin{table}[h]
    \centering
    \begin{tabular}{|c||c|c|c|c|c|c|c|c|c|c|}
    \hline
       i&1&2&3&4&5&6&7&8&9&10  \\
    \hline
    $\lambda_i$ & 0&0.06&0.25& 0.57& 1.03& 1.06& 1.06& 1.23&1.22&1.52\\
     \hline
     \hline
       i&11&12&13&14&15&16&17&18&19&20  \\
    \hline
    $\lambda_i$ & 1.52& 1.63& 1.95& 1.95& 2.38& 2.53& 2.53&3.27& 3.27& 3.31\\
    \hline
    \end{tabular}
    \caption{The 20 smallest eigenvalue of the system of equations (\ref{splf_sphf}). Each $\lambda_i$ corresponds to a KK mode with mass $m_i=\sqrt{2\pi\lambda_i \log(T/t_1)/V_{\mb{P}^1}}$.}
    \label{egvtab}
\end{table}
The corresponding masses are 
\begin{equation}
    m_i=\sqrt{2\pi\lambda_i \log(T/t_1)/V_{\mb{P}^1}}\,.
    \label{KK_mode_mass_spectrum}
\end{equation}

\begin{figure}[H]
    \centering
    \includegraphics[width=0.8\linewidth]{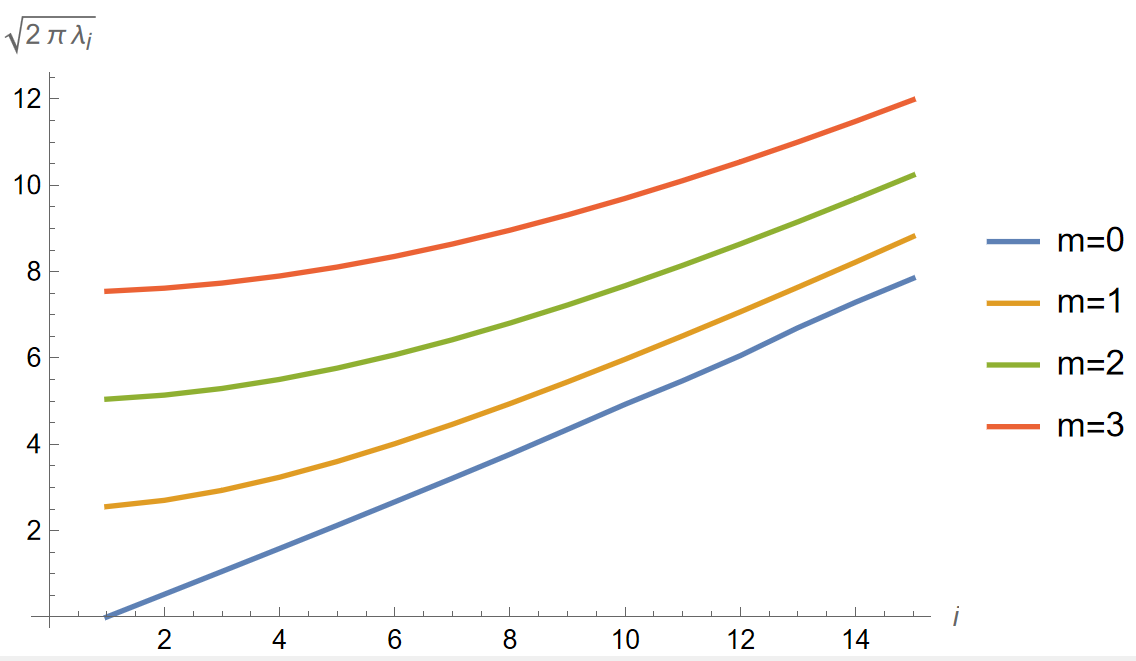}
    \caption{The figures for the eigenvalue $\sqrt{2\pi \lambda_i}$ of different $m$, which is proportional to the KK mass. We can see that each $m$ forms a KK-tower with linearly increasing mass in the large $i$ limit.}\label{fig:fix-m}
\end{figure}

We find an interesting selection rule that in the codimension-one $(4,6,12)$ limit, $A_{(l_1,m_1,l_2,m_2)}$ equals to zero when $m_1 \neq m_2$. Hence the eigenvectors with different $m$ decouple. As a consequence, we can compute the eigenvalues of $A_{(l_1,m_1,l_2,m_2)}$ with different $m$ separately and impose a larger cut off at $l=1000$\footnote{In the Mathematica program, the computation becomes very slow when the number of non-zero matrix elements is bigger than $\mc{O}(10^4)$. So  we choose the cut off at $l=300$ in the previous case. When we calculate the eigenvalue with different $m$ separately, it allows us to choose a larger cut off,  but the error of solving the eigenvalues from the MMA would grow and becomes non-negligible when the cut off reaches some point between l=1000 and l=5000. Hence here we choose the cut off at l=1000.}, and the eigenvalues of $\sqrt{2\pi \lambda_i}$ are shown in Figure \ref{fig:fix-m}.
 
\begin{figure}
    \centering
    \subfloat[$\lambda=0$]{\includegraphics[width=0.3\linewidth]{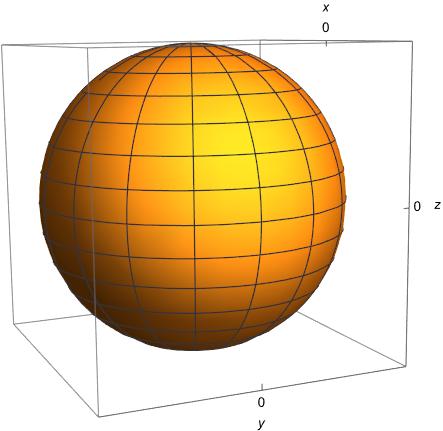}}\hfill
    \subfloat[$\lambda=0.06$]{\includegraphics[width=0.3\linewidth]{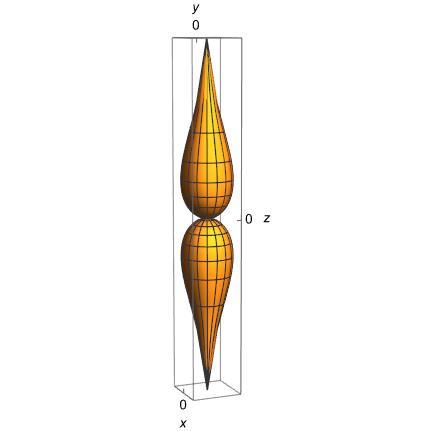}}\hfill
    \subfloat[$\lambda=0.25$]{\includegraphics[width=0.3\linewidth]{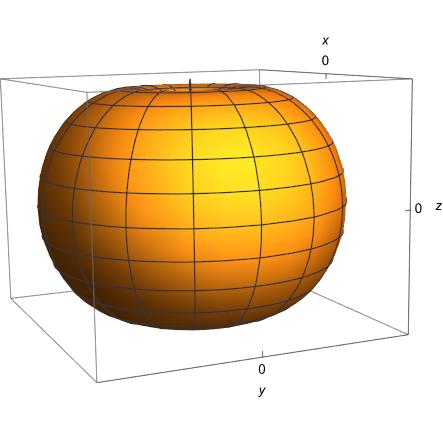}}\hfill
    \subfloat[$\lambda=0.57$]{\includegraphics[width=0.3\linewidth]{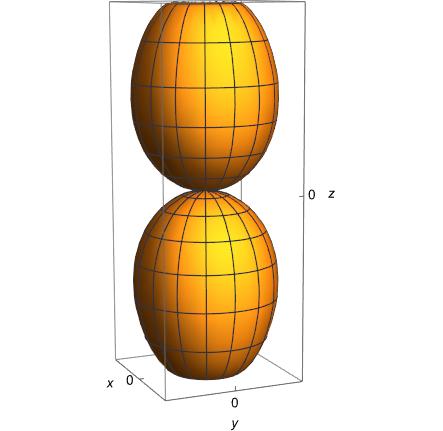}}\hfill
    \subfloat[$\lambda=1.03$]{\includegraphics[width=0.3\linewidth]{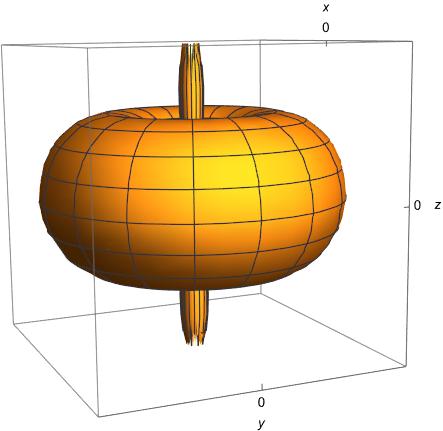}}\hfill
    \caption{The pictures depict the eigenvectors of the 5 lightest KK modes, in the codimension-one $(4,6,12)$ limit with four bunches of type $I_0^*$ $D_4$ branes. We use spherical coordinate to describe the eigenstates. In pictures, $(0,0,0)$ corresponds to the origin of the spherical coordinate and the polar axis is along the positive $z$-axis. The radius is equal to the absolute value of the eigenvector $\beta$ in (\ref{exp1}) at the corresponding angle $(\theta,\phi)$. It is shown that there exists singular behaviour at the location of 7-branes.}
    \label{enigenstate_figure}
\end{figure}

We also depict the eigenvectors of the five lowest eigenvalues in Figure \ref{enigenstate_figure}, which are all in the $m=0$ subset. In fact, these eigenvectors are purely imaginary in the codimension-one $(4,6,12)$ limit.

Each eigenstate corresponds to at most four linearly independent states of two-form field, because we can choose either the compactified $B_2$ and $C_2$ fields to be equal to the real or imaginary parts (if not zero) of $\sqrt{\frac{(t-T/2)(t-t_1/2)}{(t+T/2)(t+t_1/2)}} \beta$. Hence each mass can at most correspond to four KK modes.

In our calculation, we notice that if we do not choose the ansatz (\ref{exp1}) and (\ref{ztrf}), we may get negative eigenvalues, which invalidate the procedure of solving the eigensystem. Hence first, (\ref{exp1}) captures the correct power behaviour of the eigenvector near the 7-branes.

Second, for the justification of (\ref{ztrf}), if we use an alternative map
\begin{equation}
    t=2r\cot(\theta/2)e^{i\phi}\,,
\end{equation}
we can only get a good approximation in the region near the radius $r$. For fix constants $o_1\gg 1$ and $o_2\ll1$. The volume of regions with $|t|<o_1 t_1$ or $|z|>o_2T$ should shrink to zero in the codimension-one (4,6,12) limit, and the most of volume of the base is contributed by the region with $o_1 t_1<|t|<o_2 T$. Hence setting $r=\sqrt{t_1T}/2$ in (\ref{ztrf}) is a good parametrization that can simplify the solving of our eigensystem.

\subsection{Other cases with a constant $\tau$}
\label{sec:KK-other}

Using this method, we can also study the KK spectrum in other cases of codimension-one $(4,6,12)$ limit. 

For instance, we consider 6 type $IV$ $A_2$ bunches locating at $\xi_i t_1$ $(i=1,2,3)$ and $\xi'_i T$ 
 $(i=1,2,3)$ on the $\mb{P}^1$, where $\xi_i$ and $\xi'_i$ are $O(1)$ complex constants, $t_1\ll T$. Then we can keep the volume of base $\mb{P}^1$ fixed and let $t_1\rightarrow 0$. 

To satisfy the monodromy behaviour of $b^I$, we write down an ansatz
 \begin{equation}
     b^*=\beta \frac{[(t-\xi_2 t_1)(t-\xi_3 t_1)(t-\xi'_2 T)(t-\xi'_3 T)]^{1/3}}{[(t-\xi_1 t_1)(t-\xi'_1 T)]^{2/3}} \left( \begin{array}{c}
        1/\sqrt{2}\\
        e^{i\pi/3}/\sqrt{2} 
     \end{array}
     \right)\,.
     \label{exp2}
 \end{equation}
$\beta$ is a function without monodromy behaviour. Similar to the previous case, we expand $\beta$ by spherical harmonics on $\mb{P}^1$ and rewrite the equation of motion to
\begin{equation}
    \begin{aligned}
     &\sum_{l,m}c_{l,m}\{l(l+1)\sin^2\theta+\\
     &\sqrt{\frac{t_1}{T}}[\frac{\xi_2+\xi_3-2\xi_1}{3}(1-\cos\theta)e^{-i\phi}-4\frac{\xi'_1(\xi'_2+\xi'_3)-2\xi'_2\xi'_3}{3\xi'_1\xi'_2\xi'_3}(1+\cos\theta)e^{i\phi}](\partial _{\theta}+\frac{m}{\sin \theta})+\\
     &\mc{O}\left(\frac{t_1}{T}\right)\}\mathrm{Y}^{l,m}(\theta,\phi) \\
     &=\sum_{l,m}\frac{V_{\mb{P}^1}M^2}{2\pi\log(T/t_1)} c_{l,m}\mathrm{Y}^{l,m}(\theta,\phi)\,.
    \end{aligned}
    \label{sphf2}
\end{equation}

In the codimension-one $(4,6,12)$ limit we set $t_1/T=0$ on the l.h.s. side, thus (\ref{sphf2}) can be simplified to (\ref{splf_sphf}) as well, which means that their mass spectra are the same in codimension-one $(4,6,12)$ limit, insensitive to the details of the Weierstrass model.

The real and imaginary parts of $b^*$ are two solutions of the equation of motion. We can use the same method as four $D_4$ case to solve the spectrum. We find that the mass spectrum is similar to the four $D_4$ configuration with an $O(\sqrt{t_1/T})$ difference. But for each mass, the 2-form field only can be identified with $\operatorname{Re}(b^*)$ or $\operatorname{Im}(b^*)$, which means that each mass could correspond to two KK modes.  

We can also solve the KK mass spectrum of other codimension-one $(4,6,12)$ limits. For instance for $E_7$ approaching type III, we can assume that the $E_7$ bunches locate at $\xi_1 t_1$ and $\xi'_1 T$, and the type III bunches locate at $\xi_2 t_1$ and $\xi'_2 T$. When $t_1\rightarrow 0$ the codimension-one $(4,6,12)$ limit is reached. We can impose the monodromy condition on the complex solution of the equation of motion as
 \begin{equation}
     b^*=\beta \frac{[(t-\xi_1 t_1)(t-\xi'_1 T)]^{1/4}}{[(t-\xi_2 t_1)(t-\xi'_2 T)]^{1/4}} \left( \begin{array}{c}
        1/\sqrt{2}\\
        -i/\sqrt{2} 
     \end{array}
     \right)\\.
     \label{exp3}
 \end{equation}
$\beta$ is a function without monodromy behaviour. With a similar discussion, we can also achieve the equations (\ref{splf_sphf}) in the codimension-one $(4,6,12)$ limit, leading to the same mass spectrum.

The real and imaginary part of $b$ are two solutions of the equation of motion. The mass spectrum is the same as above cases with an $O(\sqrt{t_1/T})$ difference, and each mass corresponds to two states which are the real and imagine parts of $b^*$. 

For other cases such as $E_6$ approaching type IV, we get similar results. As a summary, any configuration with a constant $\tau$ background would have the mass spectrum of 2-form $c_{\mu\nu}$ KK-modes in the codimension-one $(4,6,12)$ limit. This is qualitatively different from the string junction spectrum which depends on the type of approaching 7-branes. 

Finally, we comment that the KK mass spectrum could get a large correction after higher order terms in the F-theory Lagrangian are taken into account. Nonetheless we still expect the qualitative behaviours w.r.t. $t_1$ and $V_{\mb{P}^1}$ to hold.

\section{Infinite distance limit of 8D F-theory}
\label{sec:infinite-distance}

\subsection{Distance conjecture}
\label{sec:infinite-conj}

In recent years, the swampland distance conjecture \cite{Vafa:2005ui,Ooguri:2006in,Heidenreich:2018kpg,Blumenhagen:2018nts,Lee:2018urn,Lee:2018spm,Grimm:2018cpv,Scalisi:2018eaz,Corvilain:2018lgw,Joshi:2019nzi,Marchesano:2019ifh,Font:2019cxq,Grimm:2019wtx,Erkinger:2019umg,Kehagias:2019akr,Lee:2019wij,Grimm:2019ixq,Bonnefoy:2019nzv,Gendler:2020dfp,Xu:2020nlh,Lanza:2020qmt,Heidenreich:2020ptx,Bastian:2020egp,Perlmutter:2020buo,Calderon-Infante:2020dhm,Ashmore:2021qdf,Brodie:2021ain,Lanza:2021udy,Castellano:2021yye,Stout:2021ubb,Buratti:2021fiv,Lee:2021qkx,Lee:2021usk,Basile:2022zee,Stout:2022phm,Shiu:2022oti,vandeHeisteeg:2023ubh,Baume:2023msm,Blumenhagen:2023yws,Calderon-Infante:2023ler,Basile:2023rvm,Etheredge:2024tok,Ibarra:2024hfm} is an important research direction, which mainly focuses on the asymptotic behaviour of the infinite distance limit of the (broadly defined) quantum gravity moduli space. Let $\mc{M}$ be the moduli space of a quantum gravity theory in $d\geq 4$ space-time dimensions, parameterized by the vacuum expectation value of all massless scalars. Then with one arbitrary reference point $p_0$ on $\mc{M}$, the theory at the point $p$ has an infinite light tower with mass scale
\begin{equation}
    m\sim \exp[-\alpha d(p,p_0)] \quad,\ d(p,p_0)\rightarrow \infty\, .
    \label{dis_conj_exp_behaviour}
\end{equation}
Here $\alpha$ is an order one constant, which depends on the type of the tower. $d(p,p_0)$ is the distance between the two points $p,p_0$ measured with the properly normalized (natural) metric of the moduli space. For a weak coupled quantum gravity theory, whose higher order corrections in the Lagrangian can be ignored, the natural metric is the metric from the kinetic term of the massless scalars.

Distance conjecture has two crucial implications. First, in the infinite distance limit, all quantum gravity theories are weakly coupled. Many theories with strong coupling behaviours, such as the $g_s\rightarrow\infty$ limit for 10D type IIB superstring theory, are equivalent to a weakly coupled theory via a duality (such as the S-duality). Hence it is expected that the strongly coupled  theory in the infinite distance limit is completely different from that in the bulk of the moduli space, where the latter may not have a weakly coupled dual description.

Second, in the weakly coupled theory in the infinite distance limit, there is an infinite tower with exponential asymptotic behaviour (\ref{dis_conj_exp_behaviour}). The numerical coefficient $\alpha$ for each type of tower can be computed. For the Kaluza-Klein tower from the compactification of a $D$-dimensional theory to $d$ space-time dimensions, the exponent is given by \cite{Agmon:2022thq} 
\begin{equation}
    \alpha_{\op{KK}}=\sqrt{\frac{D-2}{(D-d)(d-2)}}\,.
    \label{exponent_KK_tower}
\end{equation}
On the other hand, for the weakly coupled (tensionless) string tower in $d$ space-time dimensions, the exponent is
\begin{equation}
    \alpha_{\text{string}}=\frac{1}{\sqrt{d-2}}\,.
    \label{expoent_string_tower}
\end{equation}
According to the emergent string conjecture \cite{Lee:2019wij}, it was conjectured that the previously mentioned KK towers and string towers are the only two possible asymptotic towers.

As discussed in \cite{Lee:2021usk}, at the codimension-one $(4,6,12)$ limit of 8D F-theory, there should be an infinite light tower dual to the KK tower in 8D heterotic string theory. Below we will show that we can find such tower with the expected exponential behaviour. However, we encounter a subtlety of matching the numerical coefficient on the exponent due to the definition of Planck mass in F-theory.

\subsection{F/heterotic duality and the moduli space}
\label{sec:F-heterotic}

In order to discuss the infinite distance limit of the moduli space, we should first find the metric of the moduli space of F-theory. Here we only consider the complex structure moduli, corresponding to the coefficients of the Weierstrass model. However, it is hard to calculate the metric of these coefficients directly due to the difficulty of writing down the effective action for scalars in strongly coupled F-theory.

On the other hand, it is known that 8D F-theory on K3 surface is dual to 8D heterotic string theory on $T^2$ \cite{Vafa:1996xn}. And the metric of the moduli space of the 8D heterotic string description is much easier to compute. We can consider the case where there is no Wilson line, i.e. there are only two complex massless scalars $\hat{\tau},\hat{\rho}$ with non-zero vacuum expectation value \cite{Blumenhagen:2013fgp}. By the dimension reduction, which is shown in Appendix \ref{sec_metric_moduli_space}, we can get the kinetic term of such two scalars
\begin{equation}
    S_{\op{kin}}=\int\dd^8 x\;\frac{2}{3}\left(\frac{\d \hat{\tau} \overline{\d\hat{\tau}}}{\hat{\tau}_2^2}+\frac{\d \hat{\rho} \overline{\d\hat{\rho}}}{\hat{\rho}_2^2}\right)\,,
\end{equation}
and thus the metric of the moduli space of the 8D heterotic string is
\begin{equation}
    \dd s^2=\frac{2}{3}\left(\frac{\d \hat{\tau} \overline{\d\hat{\tau}}}{\hat{\tau}_2^2}+\frac{\d \hat{\rho} \overline{\d\hat{\rho}}}{\hat{\rho}_2^2}\right)\,.
    \label{metric_8D_heterotic}
\end{equation}
Here the two complex scalars are defined as \cite{Blumenhagen:2013fgp}
\begin{equation}
    \begin{aligned}
        \hat{\tau}&=B_{12}+i\sqrt{\op{det}g_{T_2}}=\hat{\tau}_1+i\hat{\tau}_2\,,\\
        \hat{\rho}&=\frac{g_{T_2,12}+i\sqrt{\op{det}g_{T_2}}}{g_{T_2,11}}=\hat{\rho}_1+i\hat{\rho}_2\,.
    \end{aligned}
    \label{complex_scalar_het_T2}
\end{equation}
$B$ denotes the NS-NS 2-form field and $g_{T_2}$ denotes the metric of the torus $T_2$.

By F/heterotic duality, we expect that a subspace of the moduli space of 8D F-theory is dual to the no Wilson line moduli space of the 8D heterotic string theory. Moreover, the 8D F-theory and 8D heterotic string descriptions should have the same moduli space metric, see e.g. \cite{LopesCardoso:1996hq}. The no Wilson line vacua of 8D heterotic string theory is dual to 8D F-theory with the Weierstrass model
\begin{equation}
    y^2=x^3-a t^4 x+b t^5+c t^6+d t^7\,.
\end{equation}
The relations between the moduli $\hat{\tau}$, $\hat{\rho}$ and the coefficients $a$, $b$, $c$, $d$ are
\begin{equation}
    \label{no_wilson_dual}
    \ba
        j(\hat{\tau}) j(\hat{\rho})&=1728^2 \frac{a^3}{27 b d}\,, \\
        (j(\hat{\tau})-1728)(j(\hat{\rho})-1728)&=1728^2 \frac{c^2}{4 b d}\,.
    \ea
\end{equation}
Obviously we require that $b$, $d$ are non-zero if there is no codimension-one $(4,6,12)$ singularity. 

\subsection{Codimension-one $(4,6,12)$ limit}
\label{sec:codimension-one-46-limit}

If the Weierstrass model has order of vanishing $\geq (4,6,12)$, it does not admit a conventional crepant resolution \cite{Weigand:2018rez}, which means that such Weierstrass model cannot describe 8D supergravity vacua at least in the bulk of the moduli space. Nonetheless from the discussions of \cite{Lee:2021usk}, such limits can correspond to the Kulikov degeneration limits of the K3 surface, corresponding to the infinite distance limit of 8D F-theory.

We summarize the following results from the F/heterotic duality. We can consider the decompactification limit from 8D to 10D of the 8D heterotic string, which corresponds to the point in the moduli space $\hat{\tau}\rightarrow i \infty,\hat{\rho}=\op{constant}$, see (\ref{complex_scalar_het_T2}). Then we can see that by (\ref{no_wilson_dual}), up to some rescaling factors, this limit can be obtained by
\begin{equation}
    b\sim C_1 j(\hat{\tau})^{-1}\quad c\sim C_2 j(\hat{\tau})^{-1}\quad a,d=\op{constant}\,.
    \label{cod_(4,6,12)_limit}
\end{equation}
This limit leads to the vanishing order $(4,6,12)$ when $\hat{\tau}\rightarrow i\infty$. For the simplification of calculations we can also tune the value of $\hat{\rho}$, which is transversal to the infinite distance direction of $\hat\tau$. For example if $\hat{\rho}=e^{i\pi/3}$, $j(\hat{\rho})=0$, we can set $a=0$ and $d\rightarrow 0$, which corresponds to the constant $\tau=e^{i\pi/3}$ mentioned above.

Recall the 8D heterotic string spectrum \cite{Blumenhagen:2013fgp}
\begin{equation}
    \label{heterotic_lat_T2_Ralph}
    \begin{aligned}
        & p_L=\sqrt{\frac{\alpha^{\prime}}{2}}\left(\mathbf{n}+\frac{1}{\alpha^{\prime}}(\mathbf{g}-\mathbf{b}) \mathbf{m}-\mathbf{a} \tilde{\mathbf{g}}^{-1} \mathbf{l}-\frac{1}{2} \mathbf{a} \tilde{\mathbf{g}}^{-1} \mathbf{a}^{\mathrm{T}} \mathbf{m}, \sqrt{\frac{2}{\alpha^{\prime}}}\left(\mathbf{l}+\mathbf{a}^{\mathrm{T}} \mathbf{m}\right)\right), \\
        & p_R=\sqrt{\frac{\alpha^{\prime}}{2}}\left(\mathbf{n}-\frac{1}{\alpha^{\prime}}(\mathbf{g}+\mathbf{b}) \mathbf{m}-\mathbf{a} \tilde{\mathbf{g}}^{-1} \mathbf{l}-\frac{1}{2} \mathbf{a} \tilde{\mathbf{g}}^{-1} \mathbf{a}^{\mathrm{T}} \mathbf{m}\right),\\
        & \mathbf{n},\mathbf{m},\mathbf{a}\in\mathbb{Z}^2 ,\  l_a E_A^{*a}\in \text{Lattice of } E_8\times E_8\text{ or }\operatorname{Spin}(32)/\mathbb{Z}_2\,,\\
        &\alpha' m^2=p_R^2+N_R=p_L^2+N_L\,.
    \end{aligned}
\end{equation}
The BPS condition is 
\begin{equation}
    N_R=0\,.
\end{equation}
When $\hat{\tau}\rightarrow i\infty$, the infinite tower is the KK tower from 8D to 10D parametrized by the 2-vector $\mathbf{n}$ with $\mathbf{m}=\mathbf{l}=0$. 

Then one can ask what is the dual tower in 8D F-theory. As discussed in \cite{Lee:2021usk}, one potential choice is the string junction tower of the $\hat{E}_9$ singularity in the codimension-one $(4,6,12)$ limit \cite{DeWolfe:1998pr}, which has a trivial total monodromy $I$.  Consider the simple case with constant $\tau=e^{i\pi/3}$, where one $E_8$ bunch approaching one type II bunch mentioned above. Up to an $\op{SL}(2,\mb{Z})$ transformation, it can be written as two bunches $E_8=\mbf{A^7BCC}$ and $\mbf{X_{-3,-1}A}$. As explicitly computed in Appendix \ref{sec_junction_spec}, we can find the BPS string junctions and the asymptotic $(P,Q)$ in (\ref{mass_string_junction_12brane}) as 
\begin{equation}
    \begin{aligned}
        \mbf{J}_{M,N,248}&=\bds{\lambda}_{248}+M (\bds{\omega}^p-\bds{\omega}'^p)+N (\bds{\omega}^q-\bds{\omega}'^q)\ \ ,\  (P,Q)=(M,N)\,,\\
        \mbf{J}_{M,N,1}&=M (\bds{\omega}^p-\bds{\omega}'^p)+N (\bds{\omega}^q-\bds{\omega}'^q)\ \ ,\  (P,Q)=(M,N)\,.
    \end{aligned}
\end{equation}
From (\ref{string_junction_mass_PQ}), we can see that the masses of these BPS string junctions are
\begin{equation}
    m_{M,N,248}=m_{M,N,1}=\left|\frac{M-Ne^{i\pi/3}}{\sqrt{\sqrt{3}\pi}}B\left(\frac{1}{6},\frac{5}{6}\right)\right|\sqrt{\left|\frac{V_{\mb{P}^1}}{\log t_1}\right|}\ ,\  M,N\in\mb{Z}\,.
    \label{E_8_type_II_mass}
\end{equation}
These are a part of the BPS spectrum of 8D heterotic string theory (\ref{heterotic_lat_T2_Ralph}).

In the vanishing limit of the discriminant $\Delta=0$, we find as $\hat\tau\rightarrow i\infty$, $t_1\rightarrow 0$,
\begin{equation}
    j(\hat\tau)=e^{2\pi\hat{\tau}_2}+\mc{O}(1)=-432\frac{(t_1+t_2)^2}{t_1t_2}=-432t_2\frac{1}{t_1}+\mc{O}(1)\,.
\end{equation}
\begin{equation}
    t_1=-432t_2 e^{-2\pi\hat{\tau}_2}+\mc{O}(e^{-4\pi\hat{\tau}_2})\,.
\end{equation}

From the metric of the moduli space of 8D heterotic string (\ref{metric_8D_heterotic}), we get the distance in the moduli space
\begin{equation}
    \Delta s=\sqrt{\frac{2}{3}}\log\hat{\tau}_2+\mc{O}(1)\,.
\end{equation}
Together with (\ref{E_8_type_II_mass}), we get the exponential asymptotic behaviour of the massive BPS string junction tower 
\begin{equation}
    m_{\op{junction}}\propto |\log t_1|^{-\frac{1}{2}}\sim |\hat{\tau}_2|^{-\frac{1}{2}}\sim e^{-\sqrt{\frac{3}{8}}\Delta s}
    \label{string_junction_distance_conjecture_8d_to_10d}\,,
\end{equation}
which should be consistent with the distance conjecture. Similarly, the KK tower  (\ref{KK_mode_mass_spectrum}) also has an exponential asymptotic behaviour
\begin{equation}
\label{8D-KK-behaviour}
    m_{\op{KK},F}\sim e^{\sqrt{\frac{3}{8}}\Delta s}
\end{equation}
It is interesting that the KK tower in 8D F-theory would become infinitely heavy as it approaches the infinite distance limit of the complex structure moduli space.

Now we encounter a problem: seemingly the exponent in (\ref{string_junction_distance_conjecture_8d_to_10d}) is incompatible with the F/heterotic duality. Because the only potential infinite light BPS tower on the heterotic side is the KK tower, which should have the exponent $\sqrt{\frac{2}{3}}$ according to (\ref{exponent_KK_tower}). The occurrence of such  incompatibility is due to the fact that all of the mass expressions (\ref{exponent_KK_tower}), (\ref{expoent_string_tower})  are computed with respect to the Planck mass in their respective frames. Thus the true meanings of these exponentials are
\begin{equation}
    \frac{M_{\op{KK}}}{M_{\op{P},d}}\sim \exp(-\alpha_{\op{KK}}\Delta s) \ ,\ \frac{M_{\op{string}}}{M_{\op{P},d}}\sim \exp(-\alpha_{\op{string}}\Delta s)\,.
\end{equation}

In a weakly coupled gravity theory, we can safely ignore  higher order corrections. The Einstein-Hilbert term plays the dominant role and we can compute the Planck mass from the prefactor of Einstein-Hilbert action
\begin{equation}
    S_{\op{EH}}= \frac{M_{\op{P},d}^{d-2}}{2}\int \dd^d x \sqrt{-g}R\,.
\end{equation}
The physical meaning of this definition of Planck mass is clear, because it corresponds to the Newton constant in the Newton potential
\begin{equation}
    U_{\op{Newton}}\sim M_{\op{P},d}^{d-2}\frac{m_1 m_2}{r^{d-3}}\,.
\end{equation}
At the infinite distance limit of the moduli space, the dual theory is weakly coupled and we can expect such value of $M_{\op{P},d}$ to be invariant under the duality. 

By F/heterotic duality, in 8D F-theory the volume $V_{\mb{P}^1}$ of the base is monotonically related to the string coupling constant $g_{\op{het}}$ of the 8D heterotic string theory \cite{Vafa:1996xn}. Because we only care about the variation over complex structure moduli, we can set $V_{\mb{P}^1}$ to some suitable value such that $g_{\op{het}}$ is small. Then on the heterotic side, we have a well-defined Planck mass from Einstein-Hilbert action. However on the F-theory side, we cannot directly derive the Planck mass from the supergravity action (\ref{type_IIB_sugra}), because $g_s=\frac{1}{\op{Im}\tau}\sim \mc{O}(1)$ at least in our case\footnote{Even for the configuration with only type $I_0^*$ branes, the naive IIB supergravity action is not a good approximation when these branes approaching each other to form codimension-one $(4,6,12)$ singularity.}. 

The discrepancy of the 8D Planck mass across the dual descriptions could explain why we get the exponent $\sqrt{\frac{3}{8}}$ instead of $\alpha_{\op{KK}}=\sqrt{\frac{2}{3}}$. 

We then conjecture that if the Planck mass of 8D F-theory has the asymptotic form of 
\begin{equation}
    M_{\op{P},F,8}\sim \left(M_{\op{P},10}^{8}V_{\mb{P}^1}\right)^{\frac{1}{6}}e^{\sqrt{\frac{1}{24}}\Delta s}\,.
    \label{possible_Planck_mass}
\end{equation}
Under this assumption, we can get the expected exponent $\alpha_{\op{junction}}=\sqrt{\frac{2}{3}}$ for the string junction tower. Furthermore, the exponent of KK modes in 8D F-theory becomes $\alpha_{\op{KK,F}}=-\sqrt{\frac{1}{6}}$ instead of $-\sqrt{\frac{3}{8}}$ in (\ref{8D-KK-behaviour}).

\section{Massive spectrum in lower dimensional F-theory}
\label{sec:massive-lower}

In the previous sections, we have studied the string junctions in 8D F-theory. In fact, such string junction states can also play an important role in lower dimensions, see e.g. \cite{Grassi:2013kha}, which will be the main focus of this section.

\subsection{Field configuration of lower dimension F-theory}

\label{subsec_field_config_lower_d_F_theory}

Similar to the 8D cases, we assume that there are no 3-branes and 5-branes in the type IIB supergravity background (\ref{type_IIB_sugra})\footnote{In 4D F-theory, generically there will be D3-branes from the tadpole cancellation condition $n_{D3}+\frac{1}{2}\int_{X_4}G_4\wedge G_4=\frac{\chi(X_4)}{24}$, where $X_4$ is the resolved Calabi-Yau fourfold and $G_4=dC_3$ is the flux in the dual M-theory description~\cite{Witten:1996md}.}, then the only non-trivial configuration is encoded in the complex scalar $\tau$ and the space-time metric $g$. In particular, the $\tau$ configuration is solely determined by the Weierstrass model, similar to the cases of 8D F-theory (\ref{j_tau_fg}).

The significant difficulty comes from the calculation of base metric, which we will provide a discussion as follows. In $(10-2n)$-dimensional F-theory setup, the base manifold $B_n$, on which the type IIB supergravity compactify, is a K\"ahler manifold. It has local holomorphic coordinates $t_i,i=1,\dots n$, and we can make an ansatz of base metric at least away from the 7-brane locus
\begin{equation}
    \dd s^2=g_{a\Bar{b}}\dd t_a\dd \Bar{t}_b + \sum_{i=0}^{9-2n} \dd x_i^2 .
\end{equation}
There is also a K\"ahler condition that the K\"ahler form $\omega=g_{a\Bar{b}}\dd t_a\wedge\dd \Bar{t}_b$ is closed $\dd \omega=0$.

From (\ref{type_IIB_sugra}), away from the 7-brane locus,  we can safely ignore the 7-brane action $S_{\text{7-branes}}$, and apply the Einstein equation
\begin{equation}
    \left[R_{a\bar{b}}-\frac{\d_{a}\tau\d_{\Bar{b}}\Bar{\tau}+\d_{\Bar{b}}\tau\d_{a}\Bar{\tau}}{2\tau_2^2}\right]-\frac{1}{2}g_{a\Bar{b}}\left[R_{c\bar{d}}g^{c\Bar{d}}+R_{\Bar{d}c}g^{\Bar{d}c} -\frac{\d_{c}\tau g^{c\Bar{d}}\d_{\Bar{d}}\Bar{\tau}}{2\tau_2^2}\right]=0
    \label{Einstein_equ_source_tau_hermitian}\,.
\end{equation}
For K\"ahler manifolds at least locally we can write down
\begin{equation}
    R_{a\Bar{b}}=-\d_a\d_{\Bar{b}}\log |g|\,.
\end{equation}
Here $|g|=\op{det}(g_{a\Bar{b}})$. There exists a simple class of solutions
\begin{equation}
    R_{a\Bar{b}}=-\d_a\d_{\Bar{b}}\log |g|=\frac{\d_{a}\tau\d_{\Bar{b}}\Bar{\tau}+\d_{\Bar{b}}\tau\d_{a}\Bar{\tau}}{2\tau_2^2}\,.
    \label{Einstein_equ_source_tau_hermitian_log} 
\end{equation}
Because $j(\tau)$ is holomorphic away from the 7-brane locus, $\tau$ should also be holomorphic, i.e. $\d_{\bar{a}}\tau=0$  away from the 7-brane locus. As a consequence, we can get a simple class of solution:
\begin{equation}
    |g|=\tau_2 F(t_a)\overline{F(t_a)}\,.
\end{equation}
Similar to the 8D cases in \cite{Greene:1989ya}, in order to keep $\op{SL}(2,\mb{Z})$ modular invariance of type IIB, we should have
\begin{equation}
    |g|=\tau_2\eta(\tau)^2 \bar{\eta}(\bar{\tau})^2 G(t_a)\overline{G(t_a)}\,.
    \label{g_solution_lower_d_G}
\end{equation}
$F(t_a)$ and $G(t_a)$ are holomorphic functions in $t_a$.

 Now the problem is that it is hard to impose more constraints on the form (\ref{g_solution_lower_d_G}) from the tree level supergravity action (\ref{type_IIB_sugra}) and determine all the components of the metric $g_{a\Bar{b}}$. To achieve this we need to fully determine the 7-brane actions along with the interaction terms between different 7-branes. Despite of the difficulties, we can still use the adiabatic approximation and some ansatz to obtain the possible form of the base  metric.

 With the adiabatic approximation, we can expect that when we are close to one bunch of 7-branes enough and far from other 7-branes enough, the metric should look like the 8D metric (\ref{metric_8d_F-theory}). So we can expect that in such region, where the nearby bunch locates at the divisor $D_u=\{u=0\}$, the determinant should have the form 
 \begin{equation}
     |g|=\tau_2\eta(\tau)^2 \bar{\eta}(\bar{\tau})^2 G'(t_a)\overline{G'(t_a)} u^{-k} \Bar{u}^{-k} .
 \end{equation}
 Here $G'$ is independent of $u$ locally. If the bunch at $u=0$ has $n_u$ 7-branes, we can expect that the global form of metric can be obtained by replacing $(t-t_i)$ in (\ref{metric_8d_F-theory}) with the defining equation $f_i$ of the divisor $D_i$. From this comparison we can also see that $k=\frac{n_u}{12}$. Consider the fact that all the singularities of the metric can only come from the 7-branes, globally the determinant of metric should have the form
\begin{equation}
    |g|=C_g^{2n}\tau_2\eta(\tau)^2 \bar{\eta}(\bar{\tau})^2 \prod f_i^{-\frac{n_i}{12}}\Bar{f}_i^{-\frac{n_i}{12}} .
    \label{g_solution_lower_d_det}
\end{equation}
Here there are $n_i$ 7-branes locating at the divisors $D_i=\{f_i=0\}$. Similar to the 8D cases, the numerical factor $C_g$ is only related to the volume of the base manifold $V_{B_n}$. Now the difficulty lies in the detailed metric ansatz with the determinant (\ref{g_solution_lower_d_det}), and we will discuss the local form of the metric in some special cases. For simplicity, below we consider the cases of 6D F-theory, while the cases of 4D F-theory can be generalized analogously. 

The simplest case is that mentioned above, where we consider the local region sufficiently near one bunch of 7-branes located at $D_u=\{u=0\}$ and far from the other bunches. The two local holomorphic coordinates of the 2D base are $u$ and $v$ s.t. the Jacobian determinant $\left|\frac{\d(t_1,t_2)}{\d(u,v)}\right|=1$ locally, such that we can write the metric in these coordinates without changing the determinant $|g|$. In addition, we assume that $\tau$ is almost constant everywhere at $D_u$, i.e. the singularity type at $D_u$ is not $I_0^*$, leading to $\d_v \tau=\d_v\eta=0$. In order to keep the K\"ahler condition, the simplest form of base metric is 
\begin{equation}
    \dd s^2=C_u\tau_2\eta^2\Bar{\eta}^2 u^{-\frac{n_u}{12}}\Bar{u}^{-\frac{n_u}{12}}\dd u \dd \bar{u}  +  C_v \dd v\dd \Bar{v}\,.
    \label{6d_metric_near_one_bunch}
\end{equation}
Here $C_u$, $C_v$ are nearly constant in this local region, and they get corrections away from $D_u$, from the other 7-branes.

A slightly different case is in the local region of two parallel bunches of 7-branes. Here parallel means that the supporting divisors do not intersect each other, which frequently occurs on a generic F-theory base~\cite{Morrison:2012np,Morrison:2012js,Taylor:2015isa}. Such two divisors can be put in the form of $D_{u_0}=\{u=0\}$, $D_{u_1}=\{u=u_1\}$. Then in this case, by adiabatic approximation and choosing proper local holomorphic coordinates $u,v$, in order to keep the K\"ahler condition, the simplest form of metric is 
\begin{equation}
    \dd s^2=C_u\tau_2\eta^2\Bar{\eta}^2 u^{-\frac{n_{u_0}}{12}}\Bar{u}^{-\frac{n_{u_0}}{12}}(u-u_1)^{-\frac{n_{u_1}}{12}}(\Bar{u}-\Bar{u}_1)^{-\frac{n_{u_1}}{12}}\dd u \dd \bar{u}  +  C_v \dd v\dd \Bar{v}\,.
    \label{6d_metric_near_two_parallel_bunches}
\end{equation}
Here $C_u$, $C_v$ are nearly constant in this local region, and they are affected by 7-branes other than $D_{u_0}$, $D_{u_1}$. 

Another simple case is when two bunches of 7-branes with the same $\tau$ value intersect. We can consider the local region  of the two divisors $D_u=\{u=0\}, D_v=\{v=0\}$, and solve from the equation of motion $\d_u \tau=\d_v\tau=0$. In order to keep the K\"ahler condition, the simplest form of base metric is
\begin{equation}
    \dd s^2=C_u u^{-\frac{n_u}{12}}\Bar{u}^{-\frac{n_u}{12}}\dd u \dd \bar{u}  +  C_v v^{-\frac{n_v}{12}}\Bar{v}^{-\frac{n_v}{12}} \dd v\dd \Bar{v} .
    \label{6d_metric_near_two_intersect_bunches}
\end{equation}
Here $C_u,C_v$ are nearly constant in this local region, and they are affected by 7-branes other than $D_u$ and the $\tau_2\eta^2\Bar{\eta}^2$ term. 

With above local forms (\ref{6d_metric_near_one_bunch}), (\ref{6d_metric_near_two_parallel_bunches}), (\ref{6d_metric_near_two_intersect_bunches}), we can calculate the global form of base metric in some simple cases. One example is 6D F-theory with base manifold $\mb{P}^1\times\mb{P}^1$, with the following special form of Weierstrass model
\begin{equation}
    f(t,t')=f_8(t)\prod_{i=1}^4 (t'-t'_i)^2 \ ,\ g(t,t')=g_{12}(t)\prod_{i=1}^4 (t'-t'_i)^3\,.
\end{equation}
The discriminant is
\begin{equation}
    \Delta(t,t')\propto\Delta_{24}(t)\prod_{i=1}^4(t'-t'_i)^6\propto \prod_{j=1}^{24}(t-t_j)\prod_{i=1}^4(t'-t'_i)^6 .
\end{equation}
Here $t$ and $t'$ denote the local coordinates of each $\mb{P}^1$, which we distinguish as $\mb{P}^1,\mb{P}^{1\prime}$. $f_8(t)$, $g_{12}(t)$ and $\Delta_{24}(t)$ are generic polynomials in $t$ of degree 8, 12 and 24 respectively. In this configuration, there are in total 48 7-branes. Among them, there are 24 7-branes wrapping the $\mb{P}^{1\prime}$, which locates at 24 different points on $\mb{P}^1$ (with possible gauge enhancements if some $t_j$ coincide). The other 24 7-branes wrapping the $\mb{P}^1$ locate 4 points on $\mb{P}^{1\prime}$, and there is type $I_0*$ $D_4$ enhancement at each of these points.

With this special configuration, it is natural to make an ansatz \cite{Asano:1997fu}
\begin{equation}
\label{P1P1-metric-ansatz}
    \dd s^2=e^{\phi(t,\Bar{t})}\dd t\dd \bar{t}+e^{\psi(t',\Bar{t}')}\dd t'\dd\Bar{t}'+ \sum_{i=0}^5 \dd x_i^2\,,
\end{equation}
\begin{equation}
    \begin{aligned}
        e^{\phi}=&C\tau_2 \eta(\tau)^2 \bar{\eta}(\bar{\tau})^2 \prod_{i=1}^{24}\left(t-t_i\right)^{-1 / 12}\left(\bar{t}-\bar{t}_i\right)^{-1 / 12}, \\
        e^{\psi}=&C'\prod_{i=1}^4\left(t'-t'_i\right)^{-1 / 2}\left(\bar{t}'-\bar{t}'_i\right)^{-1 / 2} . 
    \end{aligned}
\end{equation}
This ansatz is compatible with the above local form (\ref{6d_metric_near_one_bunch}), (\ref{6d_metric_near_two_parallel_bunches}), (\ref{6d_metric_near_two_intersect_bunches}). The numerical coefficient $C$ only depends on the volume $V_{\mb{P}_1}$ and $C'$ only depends on the volume $V_{\mb{P}_1'}$.

\subsection{Massive BPS string junctions in 6D F-theory}
\label{sec:massive-6D}

In a generic 6D F-theory configuration without a known global base metric, although we have some information about the local forms of the base metric, there are still some problems in the computation of BPS  string junctions. One underlying difficulty is how to precisely relate the prefactors $C_u,C_v$ in (\ref{6d_metric_near_one_bunch}), (\ref{6d_metric_near_two_intersect_bunches}), (\ref{6d_metric_near_two_parallel_bunches}) with the K\"ahler moduli of the base. The reason is that with the current local forms of metric, it is hard to compute the volume of all cycles on the base. We will leave the computation of base metric and string junction spectrum for a general 6D F-theory setup to the future work. 

On the other hand, the problem can be simplified when there are two approaching parallel bunches with total number of 7-branes $n_0+n_1=12$. As discussed around (\ref{volume_local_region}), the local volume will tend to infinite with a fixed prefactor, and the volume of the 2-cycle transversal to the 7-branes can be computed in a local patch. 

Let us consider the special case with two parallel bunches of 7-branes at $D_{u_0}=\{u=0\},D_{u_1}=\{u=u_1\}$. If in addition, $\tau$ is locally constant near the two bunches, then from the form (\ref{6d_metric_near_two_parallel_bunches}), we can compute the volume of the 2-cycle $D_v=\{v=0\}$ locally. As a consequence, we can apply a similar computation of BPS string junction mass as in the 8D case. More precisely, we only need to replace the $V_{\mb{P}^1}$ in (\ref{mass_string_junction_12brane}) with $V_v$ and $\log t_1$ with $\log u_1$, and obtain the mass
\begin{equation}
    m_{\text{junction}}=\left|B\left(1-\frac{n_0}{12},1-\frac{n_1}{12}\right)\frac{P-Q\tau}{\sqrt{2\pi\tau_2}}\left(V_v\log u_1\right)^{-\frac{1}{2}}\right| .
\end{equation}

If $D_v$ is also the 2-cycle with the largest volume among all possible 2-cycles, we can estimate the KK scale as 
\begin{equation}
    m_{\text{KK}}\sim V_v^{-\frac{1}{2}}\,.
\end{equation}
Under these assumptions, we get the ratio
\begin{equation}
    \frac{m_{\text{junction}}}{m_{\text{KK}}}\sim\left|B\left(1-\frac{n_0}{12},1-\frac{n_1}{12}\right)\frac{P-Q\tau}{\sqrt{2\pi\tau_2}}\left[\log u_1\right]^{-\frac{1}{2}}\right|\,,
\end{equation}
which is independent of the value $V_v$ and the numerical factors $C_u,C_v$ interestingly.

Of course, if we know the full global base metric, as in the special case in section \ref{subsec_field_config_lower_d_F_theory}, we can in principle compute the precise mass of BPS string junction. In the next section, we will apply this methodology to a special example of 4D F-theory and discuss the massive string junction spectrum.

\subsection{Example: 4D F-theory on $\mb{P}^1\times\mb{P}^1\times\mb{P}^1$}
\label{sec:4D}

In this section, we can construct a simple 4D F-theory example with potential phenomenology relevance. Consider the simple base manifold $B_3=\mb{P}^1\times\mb{P}^1\times\mb{P}^1$, with local coordinates $t,t',t''$. So the components can be denoted as $\mb{P}^1,\mb{P}^{1\prime},\mb{P}^{1\prime\prime}$.

\subsubsection{$E_7$ approaching type III}
\begin{figure}
    \centering
    \begin{tikzpicture}
        \draw[fill=gray!20] (0,0)--(1,0.5)--(1,2)--(0,1.5)--cycle;
        \draw[dashed] (0,0.5)--(1,1)
        (0,1)--(1,1.5);
        \draw[dashed] (0.33,0.167)--(0.33,1.667)
        (0.66,0.333)--(0.66,1.833);
        \draw[dashed,shift={(1.5,0)},fill=gray!20] (0,0)--(1,0.5)--(1,2)--(0,1.5)--cycle;
        \draw[dashed,shift={(4,0)},fill=gray!20] (0,0)--(1,0.5)--(1,2)--(0,1.5)--cycle;
        \draw [line width=2pt] plot [smooth] coordinates{(0.33,0.667) (0.67,0.777) (1.2,0.6) (1.83,0.667)};
        \node at (1.2,0.8) {$\mathbf{J}$};
        \node at (0.5,2.1) {$t_i'$};
        \node at (-0.3,0.75) {$t_i''$};
        \node at (0.2,-0.3) {$t_0=0$};
        \node at (1.7,-0.3) {$t_1$};
        \node at (4.2,-0.3) {$t_i,i\geq 2$};
        \draw (6.0,2)--(7,2);
        \node at (7.5,2) {$E_7$};
        \draw[dashed] (6.0,1)--(7,1);
        \node at (8,1) {type III};
        \draw[line width=2pt] (6.0,0)--(7,0);
        \node at (8.5,0) {String junction};
    \end{tikzpicture}
    \caption{7-branes and string junctions in the $E_7$ approaching type III configuration.}
    \label{fig:E7_III}
\end{figure}
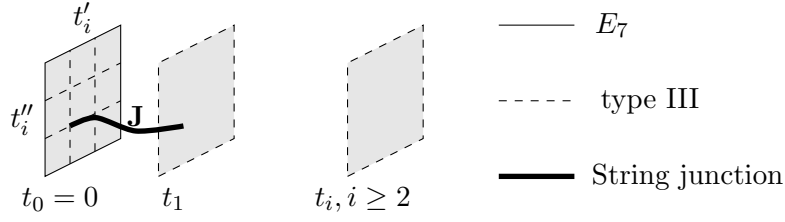

One simple example is the following constant $\tau=i$ configuration
\begin{equation}
    f=t^3 (t-t_1) \prod_{i=2}^5(t-t_i)\prod_{j=0}^7(t'-t'_j)\prod_{k=0}^7(t''-t''_k)\ ,\  g=0\,.
    \label{4D_E_7_type_III}
\end{equation}
Here $0<|t_1|\ll |t_i|,|t_i'|,|t_i''|$, which means that on $\mb{P}^1$, there is one $E_7$ bunch at $t=0$ approaching one type III $A_1$ bunch at $t=t_1$, and on $\mb{P}^{1\prime},\mb{P}^{1\prime\prime}$ there are a number of type III $SU(2)$ loci at $t_j'=0$ and $t_k''=0$ $(j,k=1,\dots,8)$ intersecting the $E_7$ locus. We depict the configuration in Figure \ref{fig:E7_III}.

In this case, there is a 4D $E_7$ gauge group at $t=0$ coupled to 16 strongly coupled conformal matter from the cod-2 $(4,6)$ loci at the matter curves $C_j:t=t_j'=0$ and $\widetilde{C}_k:t=t_k''=0$ $(j,k=1,\dots,8)$. Each of these can be thought as the dimensional reduction of a 6D rank-1 E-string theory (with $E_8$ global symmetry) on a matter curve $C_j$ or $\widetilde{C}_k$ with the topology of $\mb{P}^1$, and the subgroup $E_7\times SU(2)\subset E_8$ of the global symmetry is gauged~\cite{Kim:2017toz,Apruzzi:2018oge,Tian:2018icz}. 

On $\mb{P}^1$, the $E_7=\mbf{A^6BCC}$ bunch at $t=0$ approaches a type III ${H}_1=\mbf{A^2C}$ bunch and the singularity enhances to $\hat{E}_9$. As $|t_1|\ll |t_i|$ for $i\geq 2$, there are light massive BPS string junctions connecting the $E_7$ and the bunch at $t=t_1$, which may have impact on the low energy effective field theory, or even be a part of standard model.

As calculated in Appendix \ref{sec_junction_spec}, all BPS string junctions connecting these two bunches are
\begin{equation}
    \begin{aligned}
        \mbf{J}_{M+\frac{1}{2},N+\frac{1}{2},(0,0)}&=2(M+\frac{1}{2})(\bds{\omega}^p-\bds{\omega}^{'p})+2(N+\frac{1}{2})(\bds{\omega}^q-\bds{\omega}^{'q})\ ,\  (P,Q)=(2M+1,2N+1)\,,\\
        \mbf{J}_{M+\frac{1}{2},N+\frac{1}{2},(133,1)}&=2(M+\frac{1}{2})(\bds{\omega}^p-\bds{\omega}^{'p})+2(N+\frac{1}{2})(\bds{\omega}^q-\bds{\omega}^{'q})+\bds{\lambda}_{133}\ ,\  (P,Q)=(2M+1,2N+1)\,,\\
        \mbf{J}_{M+\frac{1}{2},N+\frac{1}{2},(1,3)}&=2(M+\frac{1}{2})(\bds{\omega}^p-\bds{\omega}^{'p})+2(N+\frac{1}{2})(\bds{\omega}^q-\bds{\omega}^{'q})+\bds{\lambda}'_3\ ,\  (P,Q)=(2M+1,2N+1)\,,\\
        \mbf{J}_{M,N,(0,0)}&=2M(\bds{\omega}^p-\bds{\omega}^{'p})+2N(\bds{\omega}^q-\bds{\omega}^{'q})\ ,\  (P,Q)=(2M,2N)\,,\\
        \mbf{J}_{M,N,(133,1)}&=2M(\bds{\omega}^p-\bds{\omega}^{'p})+2N(\bds{\omega}^q-\bds{\omega}^{'q})+\bds{\lambda}_{133}\ ,\  (P,Q)=(2M,2N)\,,\\
        \mbf{J}_{M,N,(1,3)}&=2M(\bds{\omega}^p-\bds{\omega}^{'p})+2N(\bds{\omega}^q-\bds{\omega}^{'q})+\bds{\lambda}'_{3}\ ,\  (P,Q)=(2M,2N)\,,\\
        \mbf{J}_{M+\frac{1}{2},N,(56,2)}&=2(M+\frac{1}{2}) \bds{\omega}^p+2N \bds{\omega}^q+\bds{\lambda}_{56}+\bds{\lambda}'_{2}\ ,\  (P,Q)=(2M+1,2N)\,,\\
        \mbf{J}_{M,N+\frac{1}{2},(56,2)}&=2M \bds{\omega}^p+2(N+\frac{1}{2}) \bds{\omega}^q+\bds{\lambda}_{56}+\bds{\lambda}'_{2}\ ,\  (P,Q)=(2M,2N+1)\ ,\  M,N\in\mb{Z}\,.
    \end{aligned}
\end{equation}
Here $\bds{\lambda}_{133}$ denotes the weights  of the adjoint representation $133$ of $E_7$, $\bds{\lambda}_{56}$ denotes the a weight of the representation $56$ of $E_7$, $\bds{\lambda}'_{3}$ denotes the a weight of the adjoint representation $3$ of $A_1$ and $\bds{\lambda}'_{2}$ denotes a weight of the fundamental representation $2$ of $A_1$. The BPS string junctions form a series of towers labeled by integers $M,N$. 
% The  asymptotic charge $(P,Q)$ of each state can be read out from $(M,N)$ as shown in Appendix \ref{sec_junction_spec}.

From (\ref{string_junction_mass_PQ}), treating the $\mb{P}^1$ with the local coordinate $t$ in the similar way as the $\mb{P}^1$ in the 8D case, we can obtain the mass of these towers 
\begin{equation}
    \begin{aligned}
        m_{M,N,(0,0)}&=m_{M,N,(133,1)}=m_{M,N,(1,3)}=m_{M+\frac{1}{2},N+\frac{1}{2},(0,0)}=m_{M+\frac{1}{2},N+\frac{1}{2},(133,1)}\\
        &=m_{M+\frac{1}{2},N+\frac{1}{2},(1,3)}=\left|\frac{2(M-iN)}{\sqrt{2\pi}} B\left(\frac{3}{4},\frac{1}{4}\right)  \right|\sqrt{\left|\frac{V_{t}}{\log t_1}\right|}\,,\\
        m_{M+\frac{1}{2},N,(56,2)}&=\left|\frac{2(M-iN)+1}{\sqrt{2\pi}} B\left(\frac{3}{4},\frac{1}{4}\right)  \right|\sqrt{\left|\frac{V_{t}}{\log t_1}\right|}\,,\\
        m_{M,N+\frac{1}{2},(56,2)}&=\left|\frac{2(M-iN)+i}{\sqrt{2\pi}} B\left(\frac{3}{4},\frac{1}{4}\right)  \right|\sqrt{\left|\frac{V_{t}}{\log t_1}\right|}\ ,\  M,N\in\mb{Z}\,.
    \end{aligned}    
\end{equation}
$V_t$ is the volume of the $\mb{P}^1$ with the local coordinate $t$. We list the lightest few massive states in Table \ref{tab:E7A1_light_state}.
\begin{table}[]
    \centering
    \begin{tabular}{c|c|c}
        mass $/\sqrt{\left|\frac{V_{t}}{\log t_1}\right|}$ &$(P,Q)$ & Representation $(E_7,A_1)$ \\\hline
        0&$(0,0)$&$(1,1)\oplus(133,1)\oplus (1,3)$\\\hline
        1.77&$(1,0)$ & $(56,2)$\\\hline
        1.77&$(-1,0)$ & $(56,2)$\\\hline
        1.77&$(0,1)$ & $(56,2)$\\\hline
        1.77&$(0,-1)$ & $(56,2)$\\\hline
        2.51&$(1,1)$ & $(1,1)\oplus (133,1) \oplus (1,3)$\\\hline
        2.51&$(1,-1)$ & $(1,1)\oplus (133,1) \oplus (1,3)$\\\hline
        2.51&$(-1,1)$ & $(1,1)\oplus (133,1) \oplus (1,3)$\\\hline
        2.51&$(-1,-1)$ & $(1,1)\oplus (133,1) \oplus (1,3)$\\\hline
        3.54&$(2,0)$ & $(1,1)\oplus (133,1) \oplus (1,3)$\\\hline
        3.54&$(-2,0)$ & $(1,1)\oplus (133,1) \oplus (1,3)$\\\hline
        3.54&$(0,2)$ & $(1,1)\oplus (133,1) \oplus (1,3)$\\\hline
        3.54&$(0,-2)$ & $(1,1)\oplus (133,1) \oplus (1,3)$\\\hline
    \end{tabular}
    \caption{The mass spectrum of light BPS string junctions in the $E_7$ approaching type III case.}
    \label{tab:E7A1_light_state}
\end{table}

\subsubsection{$E_6$ approaching type IV}
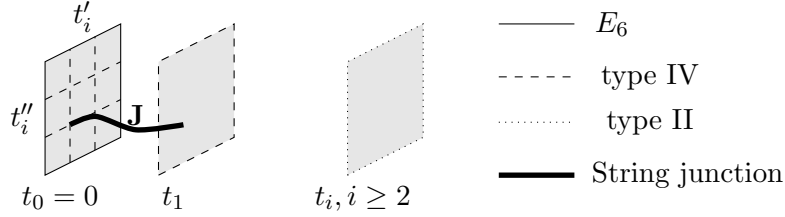
\begin{figure}
    \centering
    \begin{tikzpicture}
        \draw[fill=gray!20] (0,0)--(1,0.5)--(1,2)--(0,1.5)--cycle;
        \draw[dashed] (0,0.5)--(1,1)
        (0,1)--(1,1.5);
        \draw[dashed] (0.33,0.167)--(0.33,1.667)
        (0.66,0.333)--(0.66,1.833);
        \draw[dashed,shift={(1.5,0)},fill=gray!20] (0,0)--(1,0.5)--(1,2)--(0,1.5)--cycle;
        \draw[dotted,shift={(4,0)},fill=gray!20] (0,0)--(1,0.5)--(1,2)--(0,1.5)--cycle;
        \draw [line width=2pt] plot [smooth] coordinates{(0.33,0.667) (0.67,0.777) (1.2,0.6) (1.83,0.667)};
        \node at (1.2,0.8) {$\mathbf{J}$};
        \node at (0.5,2.1) {$t_i'$};
        \node at (-0.3,0.75) {$t_i''$};
        \node at (0.2,-0.3) {$t_0=0$};
        \node at (1.7,-0.3) {$t_1$};
        \node at (4.2,-0.3) {$t_i,i\geq 2$};
        \draw (6.0,2)--(7,2);
        \node at (7.5,2) {$E_6$};
        \draw[dashed] (6.0,1.3)--(7,1.3);
        \node at (8,1.3) {type IV};
        \draw[dotted] (6.0,0.7)--(7,0.7);
        \node at (8,0.7) {type II};
        \draw[line width=2pt] (6.0,0)--(7,0);
        \node at (8.5,0) {String junction};
    \end{tikzpicture}
    \caption{7-branes and string junctions in the $E_6$ approaching type IV configuration.}
    \label{fig:E6_IV}
\end{figure}

As another example, we can also consider the following model with a constant $\tau=e^{i\pi/3}$:
\begin{equation}
    f=0 \,\ ,\  g=t^4 (t-t_1)^2 \prod_{i=2}^7(t-t_i)\prod_{j=0}^5(t'-t'_j)^2\prod_{k=0}^5(t''-t''_k)^2\,.
    \label{4D_E_6_type_IV}
\end{equation}
This configuration is depicted in Figure \ref{fig:E6_IV}.

As before we have $0<|t_1|\ll |t_i|,|t_i'|,|t_i''|$, hence there is an $E_6$ bunch at $t=0$ approaching a type IV $SU(3)$ bunch at $t_1=0$. There are also 15 type IV $SU(3)$ gauge groups at $t_i=0$ $(i=2,3,4)$, $t'_j=0$ $(j=1,\dots,6)$ and $t''_k=0$ $(k=1,\dots,6)$.

In this case, the 4D $E_6$ gauge group at $t=0$ is also coupled to 12 strongly coupled conformal matter from the cod-2 $(4,6)$ loci at the matter curves $C_j:t=t_j'=0$ and $\widetilde{C}_k:t=t_k''=0$ $(j,k=1,\dots,6)$. Each of these can be thought as the dimensional reduction of a 6D rank-1 E-string theory on a matter curve $C_j$ or $\widetilde{C}_k$ with the topology of $\mb{P}^1$, and the subgroup $E_6\times SU(3)\subset E_8$ of the global symmetry is gauged. 

Similar to before, if one tries to realize a realistic $E_6$ GUT model, the chiral matter fields should arise from conformal matter after introducing $G_4$ flux.

On $\mb{P}^1$ the $E_6=\mbf{A^5BCC}$ bunch approaches a type IV $\tilde{H}_2=\mbf{A^2X_{0,-1}C}$ bunch and enhances to $\hat{E}_9$. With the calculation in Appendix \ref{sec_junction_spec}, we can find that all the BPS string junctions connecting these two bunches areThis configuration also  looks like Fi
\begin{equation}
    \begin{aligned}
        \mbf{J}_{M,N,(0,0)}&=3M(\bds{\omega}^p-\bds{\omega}^{'p})+N(\bds{\omega}^q-\bds{\omega}^{'q})\ ,\  (P,Q)=(3M,N)\,,\\
        \mbf{J}_{M,N,(78,1)}&=3M(\bds{\omega}^p-\bds{\omega}^{'p})+N(\bds{\omega}^q-\bds{\omega}^{'q})+\bds{\lambda}_{78}\ ,\  (P,Q)=(3M,N)\,,\\
        \mbf{J}_{M,N,(1,8)}&=3M(\bds{\omega}^p-\bds{\omega}^{'p})+N(\bds{\omega}^q-\bds{\omega}^{'q})+\bds{\lambda}'_8 \ ,\  (P,Q)=(3M,N)\,,\\
        \bds{J}_{M+\frac{1}{3},N,(27,\overline{3})}&=\bds{\lambda}_{27}+\bds{\lambda}'_{\overline{{3}}}+3(M+\frac{1}{3})(\bds{\omega}^p-\bds{\omega}'^p)+N(\bds{\omega}^q-\bds{\omega}'^q)\ ,\  (P,Q)=(3M+1,N)\,,\\
        \bds{J}_{M+\frac{2}{3},N,(\overline{27},{3})}&=\bds{\lambda}_{\overline{27}}+\bds{\lambda}'_{3}+3(M+\frac{2}{3})(\bds{\omega}^p-\bds{\omega}'^p)+N(\bds{\omega}^q-\bds{\omega}'^q)\ ,\  (P,Q)=(3M+2,N)\,,\\
        M,N\in\mb{Z}\,.
    \end{aligned}
\end{equation}
Here $\bds{\lambda}_{78}$ denotes the weights of the adjoint representation $78$ of $E_6$, $\bds{\lambda}_{27}$ denotes the weights of the representation $27$ of $E_6$, $\bds{\lambda}_{\bar{3}}$ denotes the weights of the adjoint representation $\bar{3}$ of $A_2$ and $\bds{\lambda}'_8$ denotes the weights of the adjoint representation $8$ of $A_2$. The BPS string junctions form a series of towers labeled by $(M,N)$. 

From (\ref{string_junction_mass_PQ}), treating the $\mb{P}^1$ with the local coordinate $t$ as the $\mb{P}^1$ in the 8D case, we can obtain the mass of these towers 
\begin{equation}
    \begin{aligned}
        m_{M,N,(0,0)}&=m_{M,N,(78,1)}=m_{M,N,(1,8)}=\left|\frac{3M-e^{i\pi/3}N}{\sqrt{\sqrt{3}\pi}} B\left(\frac{2}{3},\frac{1}{3}\right)  \right|\sqrt{\left|\frac{V_{t}}{\log t_1}\right|}\,,\\
        m_{M+\frac{1}{3},N,(27,\overline{3})}&=\left|\frac{3M+1-e^{i\pi/3}N}{\sqrt{\sqrt{3}\pi}} B\left(\frac{2}{3},\frac{1}{3}\right)  \right|\sqrt{\left|\frac{V_{t}}{\log t_1}\right|}\,, \\
        m_{M+\frac{2}{3},N,(\overline{27},{3})}&=\left|\frac{3M+2-e^{i\pi/3}N}{\sqrt{\sqrt{3}\pi}} B\left(\frac{2}{3},\frac{1}{3}\right)  \right|\sqrt{\left|\frac{V_{t}}{\log t_1}\right|}\ ,\  M,N\in\mb{Z}\,.
    \end{aligned} 
\end{equation}
$V_t$ is the volume of the $\mb{P}^1$ with the local coordinate $t$. We can find the lightest few massive states in Table \ref{tab:E6A2_light_state}.
\begin{table}[]
    \centering
    \begin{tabular}{c|c|c}
        mass $/\sqrt{\left|\frac{V_{t}}{\log t_1}\right|}$ &$(P,Q)$&  Representation $(E_6,A_2)$ \\\hline
        0 &$(0,0)$& $(1,1)\oplus (78,1)\oplus (1,8)$\\\hline
        1.56 &$(0,1)$& $(1,1)\oplus (78,1)\oplus (1,8)$\\\hline
        1.56 &$(0,-1)$& $(1,1)\oplus (78,1)\oplus (1,8)$\\\hline
        1.56 &$(1,0)$& $(27,\overline{3})$\\\hline
        1.56 &$(-1,0)$& $(\overline{27},3)$\\\hline
        3.11 &$(0,2)$& $(1,1)\oplus (78,1)\oplus (1,8)$\\\hline
        3.11 &$(0,-2)$& $(1,1)\oplus (78,1)\oplus (1,8)$\\\hline
        3.11 &$(-2,0)$& $(27,\overline{3})$\\\hline
        3.11 &$(2,0)$& $(\overline{27},3)$\\\hline
        4.11 &$(3,1)$& $(1,1)\oplus (78,1)\oplus (1,8)$\\\hline
        4.11 &$(-3,-1)$& $(1,1)\oplus (78,1)\oplus (1,8)$\\\hline
        4.11 &$(3,2)$& $(1,1)\oplus (78,1)\oplus (1,8)$\\\hline
        4.11 &$(-3,-2)$& $(1,1)\oplus (78,1)\oplus (1,8)$\\\hline
    \end{tabular}
    \caption{Mass spectrum of light BPS string junctions in the $E_6$ approaching type IV case.}
    \label{tab:E6A2_light_state}
\end{table}

\subsubsection{$E_6$ approaching type II}
\begin{figure}
    \centering
    \begin{tikzpicture}
        \draw[fill=gray!20] (0,0)--(1,0.5)--(1,2)--(0,1.5)--cycle;
        \draw[dashed] (0,0.5)--(1,1)
        (0,1)--(1,1.5);
        \draw[dashed] (0.33,0.167)--(0.33,1.667)
        (0.66,0.333)--(0.66,1.833);
        \draw[dotted,shift={(1.5,0)},fill=gray!20] (0,0)--(1,0.5)--(1,2)--(0,1.5)--cycle;
        \draw[dotted,shift={(4,0)},fill=gray!20] (0,0)--(1,0.5)--(1,2)--(0,1.5)--cycle;
        \draw [line width=2pt] plot [smooth] coordinates{(0.33,0.667) (0.67,0.777) (1.2,0.6) (1.83,0.667)};
        \node at (1.2,0.8) {$\mathbf{J}$};
        \node at (0.5,2.1) {$t_i'$};
        \node at (-0.3,0.75) {$t_i''$};
        \node at (0.2,-0.3) {$t_0=0$};
        \node at (1.7,-0.3) {$t_1$};
        \node at (4.2,-0.3) {$t_i,i\geq 2$};
        \draw (6.0,2)--(7,2);
        \node at (7.5,2) {$E_6$};
        \draw[dashed] (6.0,1.3)--(7,1.3);
        \node at (8,1.3) {type IV};
        \draw[dotted] (6.0,0.7)--(7,0.7);
        \node at (8,0.7) {type II};
        \draw[line width=2pt] (6.0,0)--(7,0);
        \node at (8.5,0) {String junction};
    \end{tikzpicture}
    \caption{7-branes and string junctions in the $E_6$ approaching type IV configuration.}
    \label{fig:E6_II}
\end{figure}
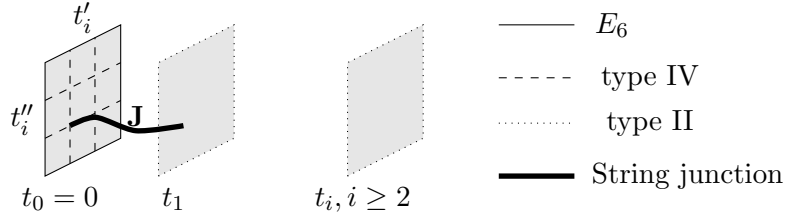

We can also consider an example with finite distance in the moduli space, for instance the following model with a constant $\tau=e^{i\pi/3}$:
\begin{equation}
    f=0 \,\ ,\  g=t^4 (t-t_1) \prod_{i=2}^8(t-t_i)\prod_{j=0}^5(t'-t'_j)^2\prod_{k=0}^5(t''-t''_k)^2\,.
    \label{4D_E_6_type_II}
\end{equation}
This configuration is depicted in Figure \ref{fig:E6_II}.

In the $E_6$ approaching type II cases (\ref{4D_E_6_type_II}), the $B_0$ is $E_6$, and as shown in \cite{DeWolfe:1998zf}, by putting all branch cuts downward, from left to right, it comprises of 7-branes $E_6=\mbf{A^5BCC}$. The $B_1$ is type II, and with similar requirement as $E_6$, it comprises of 7-branes $H_0=\mbf{AC}$. Then we can denote the basis of the strings end on the $B_0$ as $\mbf{a}_i (i=1,\dots 5)$, $\mbf{b}$, $\mbf{c}_i (i=1, 2)$, and $\mbf{J}_0$ only has components on these 7-branes. We denote the basis of the strings end on the $H_0$ bunch as $\mbf{a}',\mbf{c}'$, and $\mbf{J}_1$ only has components on these 7-branes.

% This case is simpler than the cases before. Because the two bunches $B_0,B_1$ can really collide. And when $t_1=0$, they enhance to $E_8$. As discussed in \cite{DeWolfe:1998zf}, when $t_1=0$, the BPS massless string junctions form the adjoint representation $248$ of $E_8$. When $t_1\neq 0$, some of the BPS string junctions in the $248$ representation become massive. And we can use the branching rule to find all such string junctions. When $E_8$ is broken to $E_6$, we have the branching rule $248\rightarrow 1^{\oplus 8}\oplus  27^{\oplus 3}\oplus  \overline{27}^{\oplus 3}\oplus 78$. As discussed in \cite{DeWolfe:1998zf}, only the adjoint representation $78$ of $E_6$ is massless. So the massive string junctions are $1^{\oplus 8}\oplus  27^{\oplus 3}\oplus  \overline{27}^{\oplus 3}$.

As calculated in Appendix \ref{sec_junction_spec}, the spectrum of BPS string junctions are shown in Table \ref{tab:E6II_light_state}. Here $C_t$ is the $C_g$ in (\ref{metric_8d_F-theory}). With fixed $V_t$, we can determine the $C_t$ with numerical calculation.

\begin{table}[]
    \centering
    \begin{tabular}{c|c|c}
        mass $/C_t|t_1|^{\frac{1}{6}}|\prod_{i=2}t_i^{-\frac{n_i}{12}}|$ &$(P,Q)$&  Representation $E_6$ \\\hline
        0&$(0,0)$&$78\oplus 1$\\\hline
        2.09&$(0,1)$&$1$\\\hline
        2.09&$(0,-1)$&$1$\\\hline
        2.09&$(1,0)$&$27$\\\hline
        2.09&$(1,1)$&$27$\\\hline
        2.09&$(-1,0)$&$\overline{27}$\\\hline
        2.09&$(-1,-1)$&$\overline{27}$\\\hline
        3.62&$(-2,-1)$&$27$\\\hline
        5.53&$(3,1)$&$1$\\\hline
        5.53&$(-3,-1)$&$1$\\\hline
        5.53&$(3,2)$&$1$\\\hline
        5.53&$(-3,-2)$&$1$\\\hline
    \end{tabular}
    \caption{Mass spectrum of BPS string junctions in the $E_6$ approaching type II case.}
    \label{tab:E6II_light_state}
\end{table}

\section{Discussions}
\label{sec:discussions}

In this paper we analyzed the massive spectrum of F-theory from massive BPS string junctions and KK modes. In particular, for 8D F-theory we have obtained a number of analytical result in a constant $\tau$ background and in particular, when two bunches of parallel 7-branes collide to form the codimension-one $(4,6,12)$ limit. Interestingly, we find that the light BPS string junctions connecting these bunches of 7-branes have a different asymptotic feature than the expected KK tower from the dual heterotic side. We conjecture that such discrepancy can be resolved by the correction in the definition of 8D Planck mass in the strongly coupled F-theory description. It would be interesting to further investigate this issue which could shed new light on the understanding of strongly coupled string theory. Such analysis should also be carried over to e.g. the 6D cases~\cite{Alvarez-Garcia:2023qqj}.

We have also discussed a few special cases of 6D and 4D F-theory, where there are sets of parallel 7-branes and one can write down the ansatz of base metric. It is an interesting future project to discover the base metric for a generic 6D or 4D F-theory setup, from which we can compute the mass of BPS string junctions as well as the KK modes.

Our study of KK modes in F-theory is also limited by the simplification of only taking the leading order kinetic terms of certain IIB fields, and only taking certain components after the dimensional reduction, which should be drastically improved in the future. Nonetheless, we found a qualitative result that the KK modes would become infinitely heavy in the infinite complex structure limit, if the volume of the base $\mb{P}^1$ is fixed.

It would also be interesting to further investigate the phenomenological implications of the massive states especially in the 4D cases, including the effects of D3 branes and other complications.

\acknowledgments
We thank Wei Cui, Wolfgang Lerche, Washington Taylor, Jiahua Tian, Cumrun Vafa, Timo Weigand, David Wu, Fengjun Xu, Yi Zhang for discussions. The work is supported by National Natural Science Foundation of China under Grant No. 12175004 and by Young Elite Scientists Sponsorship Program by CAST (2022QNRC001, 2023QNRC001, 2024QNRC001).

\appendix

\section{Special function}
\label{app:special-function}

Here we summarize the special functions used in this paper, see e.g. \cite{murty2016problems}.

\paragraph{Jacobi $j$-function}
The Jacobi $j$-function is an $\op{SL}(2,\mb{Z})$ modular invariant function defined with Eisenstein series
\begin{equation}
    j(\tau)=E_4(\tau)^3 / \Delta, \quad \Delta=\frac{E_4(\tau)^3-E_6(\tau)^2}{1728}\,.
\end{equation}
\begin{equation}
j\left(\frac{a \tau+b}{c \tau+d}\right)=j(\tau), \quad \forall\left(\begin{array}{ll}
a & b \\
c & d
\end{array}\right) \in \op{SL}(2,\mb{Z})\,.
\end{equation}
In the Weierstrass model
\begin{equation}
    j(\tau)=4 \frac{12^3 f(\tau)^3}{\Delta}, \quad \Delta=4 f^3(\tau)+27 g^2(\tau)\,.
\end{equation}
It has the following expansion around $i\infty$
\begin{equation}
    j(\tau)=\frac{1}{q}+744+196884 q+21493760 q^2+\ldots, \quad q=e^{2\pi i \tau}\,.
\end{equation}
And the $n$-th coefficient $c(n)$ has an asymptotic form as $n\rightarrow\infty$
\begin{equation}
    c(n) \sim \frac{e^{4 \pi \sqrt{n}}}{\sqrt{2} n^{3 / 4}} .
\end{equation}
There are also a number of special values, e.g. $j(i)=1728$, $j(e^{i\pi/3})=0$. 

\paragraph{Dedekind $\eta$-function}

\begin{equation}
\eta(\tau)=e^{\frac{\pi i \tau}{12}} \prod_{n=1}^{\infty}\left(1-e^{2 n \pi i \tau}\right)=q^{\frac{1}{24}} \prod_{n=1}^{\infty}\left(1-q^n\right)\,.
\end{equation}
\begin{equation}
    \eta\left(-\frac{1}{\tau}\right)=\sqrt{\frac{\tau}{i}} \eta(\tau)\,.
\end{equation}

\begin{equation}
\eta^2(-1 / \tau)=-i \tau \eta^2(\tau), \quad \eta^2(\tau+1)=e^{i \pi / 6} \eta^2(\tau)\,.
\end{equation}

\section{Mass of BPS string junctions in a constant $\tau$ background}
\label{sec_mass_string_junction}

We provide the details of computing the mass of BPS string junctions connecting two bunches in section \ref{subsec_massive_junction_two_bunches}. First, we assume that the background $\tau$ is constant. Second, only two bunches located at $t_0=0,t_1$ are close to each other, and all other bunches are far away, i.e. $0<|t_1|\ll |t_i|,i\geq 2$. We use the notation $t_i$ to denote the location of bunches instead of a single 7-brane.

We compute the analytical result of the mass of BPS string junctions with asymptotic charge $(P,Q)$
\begin{equation}
    \begin{aligned}
        m_{\text{junction}}&=C_g\left|\int_0^{t_1} h_{P,Q}\dd t\right|\\
        &=C_g\left|(P-Q\tau) \eta^2(\tau)\right|\left| \int_0^{t_1}\dd t\;t^{-\frac{n_0}{12}}(t-t_1)^{-\frac{n_1}{12}}\prod_{i=2}(t-t_i)^{-\frac{n_i}{12}} \right|\\
        &=C_g\left|(P-Q\tau) \eta^2(\tau)\right|\left| \int_0^{t_1}\dd t\;t^{-\frac{n_0}{12}} (t_1-t)^{-\frac{n_1}{12}}\prod_{i=2} t_i^{-\frac{n_i}{12}} \left(1-\frac{t}{t_i}\right)^{-\frac{n_i}{12}} \right|\\
        &=C_g\left|(P-Q\tau) \eta^2(\tau)\prod_{i=2} t_i^{-\frac{n_i}{12}}\right|\left|\int_0^{t_1}\dd t\; t^{-\frac{n_0}{12}} (t_1-t)^{-\frac{n_1}{12}}\left[1+t\sum_{i=2} \frac{n_i}{12t_i}+\mc{O}(t^2)\right]\right|\\
        &=C_g\left|(P-Q\tau) \eta^2(\tau)\prod_{i=2} t_i^{-\frac{n_i}{12}}B\left(1-\frac{n_0}{12},1-\frac{n_1}{12}\right)\right|\\
        &\quad\quad\times \left|1+\left(\frac{12-n_0}{24-n_0-n_1}\sum_{i=2} \frac{n_i}{12t_i}\right)t_1\right| |t_1|^{1-\frac{n_0+n_1}{12}}+\mc{O}(t_1^{3-\frac{n_0+n_1}{12}})\,.
    \end{aligned}
\end{equation}
Here $B(.,.)$ is the Beta function
\begin{equation}
    B(p, q)=\int_0^1 t^{p-1}(1-t)^{q-1} \dd t\,.
\end{equation}

As discussed in the main text, the numerical coefficient $C_g$ has dependence on the volume $V_{\mb{P}}^1$ of $\mb{P}^1$. Because we only change $t_1$, from (\ref{metric_8d_F-theory}), the metric far from the bunches at $t=0$ and $t=t_1$ are nearly constant. Hence if we want to discuss the variation of $V_{\mb{P}^1}$, we can only look at the neighborhood near the bunches $U=\left\{|t|<R\right\}$ with $|t_1|\ll R\ll |t_i|$ $(i\geq 2)$.
\begin{equation}
    \begin{aligned}
        V_U=&C_g^2\int_{|t|<R}\left| \tau_2 \eta(\tau)^2 \bar{\eta}(\bar{\tau})^2 \prod_{i=0}\left(t-t_i\right)^{-\frac{n_i}{12}}\left(\bar{t}-\bar{t}_i\right)^{-\frac{n_i}{12}}\right| \dd t \dd \bar{t}\\
        =&C_g^2\left|\tau_2 \eta(\tau)^2 \bar{\eta}(\bar{\tau})^2\prod_{i=2}|t_i|^{-\frac{n_i}{6}}\right|\left|t_1^{2-\frac{n_0+n_1}{6}}\right| \int_{|t|<R} \left|t^{-\frac{n_0}{6}}(t_1-t)^{-\frac{n_1}{6}}+\mc{O}(|t|^2)\right| \dd t\dd \bar{t}\,.
    \end{aligned}
\end{equation}
We can do a variable substitution $t=t_1 x$, and choose appropriate constant $x_1$ which is independent of $t_1$ such that $1\ll x_0\ll x_1=\left|\frac{R}{t_1}\right|$. For simplicity, we also define
\begin{equation}
    A(t_2,\dots)=\left|\tau_2 \eta(\tau)^2 \bar{\eta}(\bar{\tau})^2\prod_{i=2}|t_i|^{-\frac{n_i}{6}}\right|
\end{equation}
which is independent of $C_g$ and $t_1$. Then
\begin{equation}
    V_U= C_g^2 A\left|t_1^{2-\frac{n_0+n_1}{6}}\right|\left|\left(\int_{|x|\leq x_0}+\int_{x_0<|x|<x_1}\right)\left|x^{-\frac{n_0}{6}}(1-x)^{-\frac{n_1}{6}}\right| \dd x\dd \bar{x}+\mc{O}(1)\right|\,.
\end{equation}

It is easy to show that the integration over $|x|\leq x_0$ is a well-defined constant when $n_0,n_1<12$. Thus in the limit we only need to consider the second part
\begin{equation}
    \begin{aligned}
        V_U=& C_g^2 A \left|t_1^{2-\frac{n_0+n_1}{6}}\right| \left| \int_{x_0<|x|<x_1}\left|x^{-\frac{n_0}{6}}(x-1)^{-\frac{n_1}{6}}\right|\dd x\dd\Bar{x}+\mc{O}(1)  \right|\\
        =& C_g^2 A\left|t_1^{2-\frac{n_0+n_1}{6}}\right| \left| \int_{x_0<|x|<x_1}\left|x^{-\frac{n_0+n_1}{6}}\left(1+\mc{O}(x^{-1})\right)\right|\dd x\dd\Bar{x}+\mc{O}(1)  \right|\\
        =& C_g^2 A\left|t_1^{2-\frac{n_0+n_1}{6}}\right| \left| \int_{x_0}^{x_1}\left|x^{-\frac{n_0+n_1}{6}}\left(1+\mc{O}(x^{-1})\right)\right|2\pi |x| \dd |x|+\mc{O}(1)  \right|\,.
    \end{aligned}
    \label{local_volume_general}
\end{equation}
When $n_0+n_1< 12$, 
\begin{equation}
    \begin{aligned}
        V_U=& C_g^2 A \left|t_1^{2-\frac{n_0+n_1}{6}}\right| \left|\frac{12\pi}{12-n_0-n_1}\left(x_1^{2-\frac{n_0+n_1}{12}}+\mc{O}(x_1^{1-\frac{n_0+n_1}{12}})+\mc{O}(1)\right)\right|\\
        =&\frac{12\pi}{12-n_0-n_1} C_g^2 A\left|R^{2-\frac{n_0+n_1}{12}}+\mc{O}(|t_1|)+\mc{O}\left(\left|t_1^{2-\frac{n_0+n_1}{6}}\right|\right)\right|\,.
    \end{aligned}
\end{equation}
When $n_0+n_1=12$, 
\begin{equation}
        V_U= 2\pi C_g^2 A\left|\log x_1+\mc{O}(1)\right|= 2\pi C_g^2 A\left|\log t_1+\mc{O}(1)\right|\, .      
\end{equation}
When $n_0+n_1>12$
\begin{equation}
    V_U=C_g^2 A\times \mc{O}\left(\left|t_1^{2-\frac{n_0+n_1}{6}}\right|\right)\,.
    \label{ccl_V1}
\end{equation}

As we can see, in the limit $t_1\rightarrow 0$ while $C_g$ is fixed, the volume of the neighborhood can be ignored when $n_0+n_1<12$ and play the dominant role when $n_0+n_1\geq 12$. Hence when $n_0+n_1\geq 12$ and $|t_1|$ is small enough, we can use $V_U$ to estimate $V_{\mb{P}^1}$. 

When $n_0+n_1=12$, we can get 
\begin{equation}
    C_g=\left|\frac{V_{\mb{P}_1}}{2\pi A\log t_1}\right|^{\frac{1}{2}}+\mc{O}(|\log t_1|^{-\frac{3}{2}})\,.
    \label{ccl_V}
\end{equation}
The mass of BPS string junction is
\begin{equation}
    \begin{aligned}
        m_{\text{junction}}=&\left|\frac{P-Q\tau}{\sqrt{2\pi\tau_2}} B\left(1-\frac{n_0}{12},1-\frac{n_1}{12}\right)  \left[1+\left(\frac{12-n_0}{24-n_0-n_1}\sum_{i=2} \frac{n_i}{12t_i}\right)\right]\right|\sqrt{\left|\frac{V_{\mb{P}^1}}{\log t_1}\right|}\\&+\mc{O}(|t_1|^2|\log t_1|^{-\frac{1}{2}})\,.
    \end{aligned}    
\end{equation}

\section{The metric of the moduli space of 8D F-theory}
\label{sec_metric_moduli_space}
We review the derivation of the exponent of the KK tower from the dimension reduction of the $D$-dimensional Einstein-Hilbert action
\begin{equation}
    S_D=\int \dd^D x \sqrt{-\op{det}g_D}R_D\,
\end{equation}
to $d$ space-time dimensions. We choose the $D$-dimensional metric as 
\begin{equation}
    \dd s_D^2=g_{D,MN}\dd x^M \dd x^N=e^{-2\alpha \phi}\dd s_d^2+e^{-2\beta \phi}\dd x^i \dd x_i\,.
\end{equation}
Here $M,N=0,\dots, D-1$ denote the indices of the total space-time, $\mu,\nu=0,\dots, d-1$ denote the lower dimensional space-time and $i=d,\dots,D-1$ are the indices of the compact space. $g_{D,MN}$ denotes the metric of the total $D$-dim. space-time, $\dd s_d^2=g_{\mu\nu}\dd x^{\mu}\dd x^{\nu}$ is the metric of the compactified $d$-dimensional space-time. $\phi=\phi(x^i)$ is the dilaton of the total volume of the compact space with $V(X_{D-d})=e^{-(D-d)\beta\phi}$. $R_D$ is the Ricci scalar of the metric $g_{D,MN}$ and $R_d$ is the Ricci scalar of $g_{d,\mu\nu}$. Then we can find the relation
\begin{equation}
    \sqrt{-\op{det} g_D}R_D=e^{-[(D-d)\beta+(d-2)\alpha]\phi}\sqrt{-\op{det} g_d}R_d+\dots\,.
\end{equation}
In order to obtain the lower dimensional Einstein-Hilbert term in the Einstein frame, we have
\begin{equation}
    \alpha=-\frac{D-d}{d-2}\beta\,,
\end{equation}
\begin{equation}
    \sqrt{-\op{det} g_D}R_D=\sqrt{-\op{det} g_d}\left[R_d-\frac{(D-2)(d-2)}{D-d}\alpha^2\d^i\phi\d_i\phi\right]\,.
\end{equation}
To compare with the general result of distance conjecture \cite{Agmon:2022thq}, we need the normalized dilaton
\begin{equation}
    \sqrt{-\op{det} g_D}R_D=\sqrt{-\op{det} g_d}\left(R_d+\d^i\hat{\phi}\d_i\hat{\phi}\right)\ ,\  \hat{\phi}=\sqrt{\frac{(D-2)(d-2)}{D-d}}\alpha\phi\,.
\end{equation}
With the definition of the Planck mass in the Einstein-Hilbert action, we have 
\begin{equation}
    M_{\op{P},d}^{d-2}=M_{\op{P},D}^{D-2}V(X_{D-d})\,.
\end{equation}
For the KK tower of $T^n$, we can obtain the mass spectrum
\begin{equation}
    m_{\mbf{n}}= |\mbf{n}| V(X_{D-d})^{-\frac{1}{D-d}}\quad,\ \mbf{n}\in\mb{Z}^n\,.
\end{equation}
We can use this special case to estimate the scale of the KK tower in general
\begin{equation}
    M_{KK}\sim V(X_{D-d})^{-\frac{1}{D-d}}\,.
\end{equation}
Now we can calculate the exponent of KK tower in distance conjecture via
\begin{equation}
    \frac{M_{KK}}{M_{\op{P},d}}\sim V(X_{D-d})^{\frac{1}{D-d}+\frac{1}{d-2}}=e^{\frac{D-2}{d-2}\beta \phi}\sim e^{-\sqrt{\frac{D-d}{(D-2)(d-2)}\hat{\phi}}}\,,
\end{equation}
\begin{equation}
    \alpha_{\op{KK}}=\sqrt{\frac{D-d}{(D-2)(d-2)}}\,.
\end{equation}

For heterotic string compactified on $T^2$ with no Wilson lines, the additional complex scalar fields are $\hat{\tau},\hat{\rho}$ (\ref{complex_scalar_het_T2}). At the tree level, the supergravity action has the global symmetry $(\op{SL}(2)_{\hat{\tau}}\times\op{SL}(2)_{\hat{\rho}})\rtimes \mb{Z}_2$. Here $\mb{Z}_2$ denotes the exchange symmetry of $\hat{\tau}$ and $\hat{\rho}$. So the metric of the moduli space should have the form of
\begin{equation}
    \dd s^2=C\left(\frac{\d \hat{\tau}  \overline{\d\hat{\tau}}}{\hat{\tau}_2^2}+\frac{\d \hat{\rho}\overline{\d\hat{\rho}}}{\hat{\rho}^2}\right)\,.
\end{equation}
To fix the constant $C$, we can constrain the complex scalars in the form of $\hat{\rho}=i,\hat{\tau}=i\hat{\tau}_2$. By the definition of the complex scalar (\ref{complex_scalar_het_T2}), we have
\be
\hat{\tau}_2=\op{det} g_{T^2}=V(X_{D-d})=e^{-(D-d)\beta\phi}\ ,\ \phi=-\frac{1}{(D-d)\beta}\log \hat{\tau}_2\,.
\ee
Thus we find the metric of the moduli space for $\hat{\tau}_2$
\begin{equation}
    \dd s^2=\frac{(D-2)(d-2)}{D-d}\alpha^2\d_i\phi\d^i\phi=\frac{D-2}{(D-d)(d-2)}\frac{\d_i \hat{\tau}_2 \d^i \hat{\tau}_2}{\hat{\tau}_2^2}\,.
\end{equation}
We thus fix $C=\frac{2}{3}$ for 10D heterotic string compactified on $T^2$. 

\section{The spectrum of massive BPS string junctions}
\label{sec_junction_spec}
In this section we present the details of the computation of the spectrum of massive BPS string junctions.

\subsection{$E_8$ approaching type II}

First we need to identify the types of BPS string junctions. With the minimal length condition discussed in section \ref{subsec_massive_junction_two_bunches}, the BPS string junctions can be represented as the linear form (\ref{standard_rep_massless_junction}), with all coefficients $n_i(\mbf(J))$ to be integral. Then the string junctions connecting two bunches of 7-branes $B_0$ and $B_1$ can be divided into $\mbf{J}=\mbf{J}_0+\mbf{J}_1$, where $\mbf{J}_0$ only ends on $B_0$ and $\mbf{J}_1$ only ends on $B_1$. 

The bunch $B_0$ realizing $E_8$ can be described as $E_8=\mbf{A^7BCC}$ \cite{DeWolfe:1998zf} and the $B_1$ bunch realizing type II Kodaira singularity is $\op{II}=\mbf{X_{-3,-1}A}$, where the order of branes in from left to the right, all branch cuts pointing downward. The basis of the strings ending on the $E_8$ bunch $B_0$ are denoted as $\mbf{a}_i$ $(i=1,\dots 7)$, $\mbf{b}$, $\mbf{c}_i$ $(i=1, 2)$, and $\mbf{J}_0$ only has these types of components. The basis of the strings ending on the type II bunch $B_1$ are $\mbf{x_{-3,-1}}',\mbf{a'}$, and $\mbf{J}_1$ only has these types of components.

As discussed in \cite{DeWolfe:1998zf}, $\mbf{J}_0$ and $\mbf{J}_1$ can be expanded as
\begin{equation}
    \mbf{J}_0=\bds{\lambda}+P\bds{\omega}^p+Q\bds{\omega}^q, \quad \mbf{J}_1=\bds{\lambda}'-P\bds{\omega}'^p-Q\bds{\omega}'^q\,.
    \label{J_0_and_J_1_comp}
\end{equation}
Here $\bds{\lambda}$ has asymptotic charge zero with respect to $B_0$, and is a weight of a representation of gauge group at $B_0$. $\bds{\lambda}'$ has asymptotic charge zero with respect to $B_1$, and is a weight of a representation of the gauge group at $B_1$. The extended weights $\bds{\omega}^p$ and $\bds{\omega}^q$ are orthogonal to all possible $\bds{\lambda}$s, and have asymptotic charges $(1,0)$ and $(0,1)$. So all of the asymptotic charges connecting the two bunches $B_0$ and $B_1$ can contribute to $\bds{\omega}^p$ and $\bds{\omega}^q$. $P,Q$ in (\ref{J_0_and_J_1_comp}) are the asymptotic charges in the mass formula (\ref{string_junction_mass_PQ}), (\ref{mass_string_junction_12brane}). Similar statements hold for $\bds{\omega}'^p$ and $\bds{\omega}'^q$.

In (\ref{standard_rep_massless_junction}), all coefficients $n_i(\mbf{J})$ are integral. In the two bunches case, we can denote the basis of one string ending on one 7-brane in $B_0$ as $\mbf{s}_i$ and that in $B_1$ as $\mbf{s}'_i$. With this convention, we have
\begin{equation}
    \mbf{J}_0=\sum_i n_i(\mbf{J}_0) \mbf{s}_i\ ,\  \mbf{J}_1=\sum_i n'_i(\mbf{J}_1) \mbf{s}'_i \,.
    \label{J_0_and_J_1_standard}
\end{equation}
Here $n_i(\mbf{J}_0)$ and $n'_i(\mbf{J}_1)$ are all integral.

In the following calculation, in order to determine the representation of the string junctions, we will use the form (\ref{J_0_and_J_1_comp}) instead of (\ref{J_0_and_J_1_standard}). In $B_0$, for $\bds{\lambda}=\sum_i n_i(\bds{\lambda}) \mbf{s}_i$, $\bds{\omega}_p=\sum_i n_i(\bds{\omega}_p) \mbf{s}_i$, and $\bds{\omega}_q=\sum_i n_i(\bds{\omega}_q) \mbf{s}_i$. The coefficients $n_i(\bds{\lambda})$, $n_i(\bds{\omega}_p)$ and $n_i(\bds{\omega}_q)$ can be non-integral. In order to get integral coefficients in (\ref{J_0_and_J_1_standard}), we have a constraint
\begin{equation}
    n_i(\mbf{J}_0)=n_i(\bds{\lambda})+P n_i(\bds{\omega}_p)+ Q n_i(\bds{\omega}_q)\in\mb{Z}, \quad P,Q\in\mb{Z}\,.
    \label{integral_const_B0}
\end{equation}
In $B_1$, similarly, we have
\begin{equation}
    n'_i(\mbf{J}_1)=n'_i(\bds{\lambda'})-P n'_i(\bds{\omega}'_p)- Q n'_i(\bds{\omega}'_q)\in\mb{Z}, \quad P,Q\in\mb{Z}\,.
    \label{integral_const_B1}
\end{equation}
% However, the coefficients of $\bds{\lambda},\bds{\omega}^p,\bds{\omega}^q,\bds{\lambda}',\bds{\omega}',\bds{\omega}'^q$ can be non-integral. And this provide the constraints on the possible $(P,Q)$. 
With (\ref{integral_const_B0}), (\ref{integral_const_B1}) and the BPS condition (\ref{string_junction_BPS_condition}), we can determine all possible BPS string junctions connecting $B_0$ and $B_1$ by solving the integral solutions of these constraints.

In the codimension-one $(4,6,12)$ limit, the extended weights for $B_2$ are
\begin{equation}
    \bds{\omega}'^p=\mbf{a}'\ ,\  \bds{\omega}'^q=-3\mbf{a}'-\bds{x_{-3,-1}}'\,.
\end{equation}
Because there is no $\bds{\lambda}'$ in the above expansions, to keep $\mbf{J}_1$ integral, we must have
\begin{equation}
    \bds{J}_{1,M,N}=-M \bds{\omega}'^p-N \bds{\omega}'^q\,.
\end{equation}

As shown in \cite{DeWolfe:1998zf}, the simple roots of $E_8$ are
\begin{equation}
    \begin{aligned}
        & \boldsymbol{\alpha}_1=\mathbf{c}_1-\mathbf{c}_2 \,,\quad  \boldsymbol{\alpha}_2=-\mathbf{a}_1-\mathbf{a}_2+\mathbf{b}+\mathbf{c}_2\,, \\
        & \boldsymbol{\alpha}_3=\mathbf{a}_2-\mathbf{a}_3\,,\quad  \boldsymbol{\alpha}_4=\mathbf{a}_3-\mathbf{a}_4\,,\quad  \boldsymbol{\alpha}_5=\mathbf{a}_4-\mathbf{a}_5\,,\\
        & \boldsymbol{\alpha}_6=\mathbf{a}_5-\mathbf{a}_6\,,\quad  \boldsymbol{\alpha}_6=\mathbf{a}_7-\mathbf{a}_8\,,\quad  \boldsymbol{\alpha}_8=\mathbf{a}_1-\mathbf{a}_2\,, 
\end{aligned}
\end{equation}
with the Dynkin diagram
\begin{equation}
    \dynkin[labels={\bds{\alpha}_1,\bds{\alpha}_8,\bds{\alpha}_2,\bds{\alpha}_3,\bds{\alpha}_4,\bds{\alpha}_5,\bds{\alpha}_6,\bds{\alpha}_7},scale=2]E8
\end{equation}
The extended weights of $E_8$ are
\begin{equation}
\boldsymbol{\omega}^p  =-\sum_{i=1}^7 \mathbf{a}_i+4 \mathbf{b}+2 \sum_{i=1}^2 \mathbf{c}_i\,, \quad\boldsymbol{\omega}^q  =3 \sum_{i=1}^7 \mathbf{a}_i-11 \mathbf{b}-5 \sum_{i=1}^2 \mathbf{c}_i\,.
\end{equation}
The above equations have integral coefficients, hence  $\bds{\lambda}$ should also be integral. It implies that $\bds{\lambda}$ should can only be the weights $\bds{\lambda}_{248}$ of adjoint representation $248$ or the trivial representation $1$ of $E_8$:
\begin{equation}
    \begin{aligned}
        \mbf{J}_{0,M,N,248}&=\bds{\lambda}_{248}+M \bds{\omega}^p+N \bds{\omega}^q\,,\\
        \mbf{J}_{0,M,N,1}&=M \bds{\omega}^p+N \bds{\omega}^q\,.
    \end{aligned}
\end{equation}
As a result, the possible BPS string junctions and the asymptotic $(P,Q)$ in (\ref{mass_string_junction_12brane}) are 
\begin{equation}
    \begin{aligned}
        \mbf{J}_{M,N,248}&=\bds{\lambda}_{248}+M (\bds{\omega}^p-\bds{\omega}'^p)+N (\bds{\omega}^q-\bds{\omega}'^q)\ ,\  (P,Q)=(M,N)\,,\\
        \mbf{J}_{M,N,1}&=M (\bds{\omega}^p-\bds{\omega}'^p)+N (\bds{\omega}^q-\bds{\omega}'^q)\ ,\  (P,Q)=(M,N)\,.
    \end{aligned}
\end{equation}
Then we need to check if these string junctions are BPS. In Figure \ref{fig:string_junc_bet_bunch}, $\mbf{J}_0$ represents the part in the square $B_0$ and $\mbf{J}_1$ represents the part in the square $B_1$. Obviously we have 
\begin{equation}
    (\mbf{J}_0,\mbf{J}_1)=0\,.
\end{equation}
The following relations hold by the definition of the extended weights: 
\begin{equation}
    (\bds{\lambda},\bds{\omega}^{p})=(\bds{\lambda},\bds{\omega}^{q})=0\ ,\  (\bds{\lambda}',\bds{\omega}'^{p})=(\bds{\lambda}',\bds{\omega}'^{q})=0\,.
\end{equation}
Hence the total self-inner product of $\mbf{J}$ is 
\begin{equation}
    (\mbf{J},\mbf{J})=(\bds{\lambda},\bds{\lambda})+(P\bds{\omega}^p+Q\bds{\omega}^q,P\bds{\omega}^p+Q\bds{\omega}^q)+(\bds{\lambda}',\bds{\lambda}')+(P\bds{\omega}'^p+Q\bds{\omega}'^q,P\bds{\omega}'^p+Q\bds{\omega}'^q)\,.
\end{equation}
All these terms can be calculated through the rules in section \ref{subsec_basic_string_junction}. In the codimension-one (4,6,12) limit, the following (in)equalities hold
\begin{equation}
    \begin{aligned}
        (\bds{\lambda}_{248},\bds{\lambda}_{248})&\geq -2,\\
        (P\bds{\omega}^p+Q\bds{\omega}^q,P\bds{\omega}^p+Q\bds{\omega}^q)&=P^2-5PQ+7Q^2,\\
        (P\bds{\omega}'^p+Q\bds{\omega}'^q,P\bds{\omega}'^p+Q\bds{\omega}'^q)&=-P^2+5PQ-7Q^2.
    \end{aligned}
\end{equation}
Hence all of the string junctions $\mbf{J}_{M,N,248},\mbf{J}_{M,N,1}, M,N\in\mb{Z}$ are BPS, and from (\ref{mass_string_junction_12brane}) their mass takes the form of
\begin{equation}
    m_{M,N,248}=m_{M,N,1}=\left|\frac{M-Ne^{i\pi/3}}{\sqrt{\sqrt{3}\pi}}B\left(\frac{1}{6},\frac{5}{6}\right)\right|\sqrt{\left|\frac{V_{\mb{P}^1}}{\log t_1}\right|}\ ,\  M,N\in\mb{Z}\,.
\end{equation}

\subsection{$E_7$ approaching type III}

We consider the case of an $E_7=\mbf{A^6BCC}$ bunch $B_0$ approaching a type III ${H}_1=\mbf{A^2C}$ bunch $B_1$ (\ref{4D_E_7_type_III}), here the branes are listed from left to the right, putting all branch cuts downward. We  denote the basis of the strings ending on the $B_0$ as $\mbf{a}_i (i=1,\dots 6)$, $\mbf{b}$, $\mbf{c}_i (i=1, 2)$, and $\mbf{J}_0$ only have these components of 7-branes. The basis of the strings ending on the $H_1$ bunch as $\mbf{a}_i' (i=1,2),\mbf{c'}$, and $\mbf{J}_1$ only have these components of 7-branes.

As shown in \cite{DeWolfe:1998zf}, the simple roots of $E_7$ are
\begin{equation}
\begin{aligned}
& \boldsymbol{\alpha}_1=\mathbf{c}_1-\mathbf{c}_2, \quad \boldsymbol{\alpha}_2=-\mathbf{a}_1-\mathbf{a}_2+\mathbf{b}+\mathbf{c}_2\,, \\
& \boldsymbol{\alpha}_3=\mathbf{a}_2-\mathbf{a}_3, \quad \boldsymbol{\alpha}_4=\mathbf{a}_3-\mathbf{a}_4, \quad \boldsymbol{\alpha}_5=\mathbf{a}_4-\mathbf{a}_5\,, \\
& \boldsymbol{\alpha}_6=\mathbf{a}_5-\mathbf{a}_6, \quad \boldsymbol{\alpha}_7=\mathbf{a}_1-\mathbf{a}_2
\end{aligned}
\end{equation}
with the Dynkin diagram
\begin{equation}
    \dynkin[labels={\bds{\alpha}_1,\bds{\alpha}_7,\bds{\alpha}_2,\bds{\alpha}_3,\bds{\alpha}_4,\bds{\alpha}_5,\bds{\alpha}_6},scale=2]E7
\end{equation}
The extended weights of $E_7$ are
\begin{equation}
\boldsymbol{\omega}^p =-\frac{1}{2} \sum_{i=1}^6 \mathbf{a}_i+2 \mathbf{b}+\sum_{i=1}^2 \mathbf{c}_i\,,\quad \boldsymbol{\omega}^q =\frac{3}{2} \sum_{i=1}^6 \mathbf{a}_i-5 \mathbf{b}-2 \sum_{i=1}^2 \mathbf{c}_i\,.
\end{equation}
The simple root of $H_1$ is
\begin{equation}
    \bds{\alpha}'=\mbf{a}_1'-\mbf{a}_2'\,,
\end{equation}
and the extended weights of $H_1$ are
\begin{equation}
    \bds{\omega}'^{p}=\frac{1}{2}\sum_{i=1}^2\mbf{a}_i'\,, \quad \bds{\omega}'^{q}=-\frac{1}{2}\sum_{i=1}^2\mbf{a}_i'+\bds{c}'\,.
\end{equation}

There is a coincidence that the non-integral coefficients in $\bds{\omega}^{p,q}$ and $\bds{\omega}'^{p,q}$ are both $\frac{1}{2}$. This makes that the integral $\bds{\lambda}$ must be accompanied with the integral $\bds{\lambda}'$ and the non-integral $\bds{\lambda}$ must be accompanied with the non-integral $\bds{\lambda}'$, which constrains the possible form of BPS string junctions.

For the $H_1$ part, in order to keep $\mbf{J}_1$ integral, if $\bds{\lambda}'$ is integral, the possible $\mbf{J}_1$s are
\begin{equation}
    \begin{aligned}
        \mbf{J}_{1,M,N,3}&=\bds{\lambda}'_3+2M \bds{\omega}^{'p}+ 2N \bds{\omega}^{'q}\,,\\
        \mbf{J}_{1,M+\frac{1}{2},N+\frac{1}{2},3}&=\bds{\lambda}'_3+2(M+\frac{1}{2}) \bds{\omega}'^{p}+ 2(N+\frac{1}{2}) \bds{\omega}'^{q}\,,\\
        \mbf{J}_{1,M,N,1}&=2M \bds{\omega}'^{p}+ 2N \bds{\omega}'^{q}\,,\\
        \mbf{J}_{1,M+\frac{1}{2},N+\frac{1}{2},1}&=2(M+\frac{1}{2}) \bds{\omega}'^{p}+ 2(N+\frac{1}{2}) \bds{\omega}'^{q}\,.
    \end{aligned}
\end{equation}
If $\bds{\lambda}$ is integral, the possible $\mbf{J}_0$s are
\begin{equation}
    \begin{aligned}
        \mbf{J}_{0,M,N,133}&=\bds{\lambda}_{133}+2M \bds{\omega}^{p}+ 2N \bds{\omega}^{q}\,,\\
        \mbf{J}_{0,M+\frac{1}{2},N+\frac{1}{2},133}&=\bds{\lambda}_{133}+2(M+\frac{1}{2}) \bds{\omega}^{p}+ 2(N+\frac{1}{2}) \bds{\omega}^{q}\,,\\
        \mbf{J}_{0,M,N,1}&=2M \bds{\omega}^{p}+ 2N \bds{\omega}^{q}\,,\\
        \mbf{J}_{0,M+\frac{1}{2},N+\frac{1}{2},1}&=2(M+\frac{1}{2}) \bds{\omega}^{p}+ 2(N+\frac{1}{2}) \bds{\omega}^{q}\,.
    \end{aligned}
\end{equation}

In the $E_7$ approaching type II cases, we have the following inner product
\begin{equation}
    \begin{aligned}
        (\bds{\lambda}_{133},\bds{\lambda}_{133})&\geq-2\ ,\  (\bds{\lambda}'_{3},\bds{\lambda}'_{3})\geq-2\,,\\
        (P\bds{\omega}^p+Q\bds{\omega}^q,P\bds{\omega}^p+Q\bds{\omega}^q)&=\frac{1}{2}P^2-2PQ+\frac{5}{2}Q^2,\\
        (P\bds{\omega}'^p+Q\bds{\omega}'^q,P\bds{\omega}'^p+Q\bds{\omega}'^q)&=-\frac{1}{2}P^2+2PQ-\frac{5}{2}Q^2.
    \end{aligned}
\end{equation}

To satisfy the BPS condition and ensure that the total asymptotic charge of these two bunches equals to zero, all the possible BPS string junctions and their asymptotic $(P,Q)$ in (\ref{mass_string_junction_12brane}) are
\begin{equation}
    \begin{aligned}
        \mbf{J}_{M+\frac{1}{2},N+\frac{1}{2},(0,0)}&=2(M+\frac{1}{2})(\bds{\omega}^p-\bds{\omega}'^{p})+2(N+\frac{1}{2})(\bds{\omega}^q-\bds{\omega}'^{q})\ ,\  (P,Q)=(2M+1,2N+1)\,,\\
        \mbf{J}_{M+\frac{1}{2},N+\frac{1}{2},(133,1)}&=2(M+\frac{1}{2})(\bds{\omega}^p-\bds{\omega}'^{p})+2(N+\frac{1}{2})(\bds{\omega}^q-\bds{\omega}'^{q})+\bds{\lambda}_{133}\ ,\  (P,Q)=(2M+1,2N+1)\,,\\
        \mbf{J}_{M+\frac{1}{2},N+\frac{1}{2},(1,3)}&=2(M+\frac{1}{2})(\bds{\omega}^p-\bds{\omega}^{'p})+2(N+\frac{1}{2})(\bds{\omega}^q-\bds{\omega}^{'q})+\bds{\lambda}_3'\ ,\  (P,Q)=(2M+1,2N+1)\,,\\
        \mbf{J}_{M,N,(0,0)}&=2M(\bds{\omega}^p-\bds{\omega}'^{p})+2N(\bds{\omega}^q-\bds{\omega}'^{q})\ ,\  (P,Q)=(2M,2N)\,,\\
        \mbf{J}_{M,N,(133,1)}&=2M(\bds{\omega}^p-\bds{\omega}'^{p})+2N(\bds{\omega}^q-\bds{\omega}'^{q})+\bds{\lambda}_{133}\ ,\  (P,Q)=(2M,2N)\,,\\
        \mbf{J}_{M,N,(1,3)}&=2M(\bds{\omega}^p-\bds{\omega}'^{p})+2N(\bds{\omega}^q-\bds{\omega}'^{q})+\bds{\lambda}_3'\ ,\  (P,Q)=(2M,2N), \quad M,N\in\mb{Z}\,.
    \end{aligned}
\end{equation}
The mass of these string junctions is 
\begin{equation}
    \begin{aligned}
        &m_{M,N,(0,0)}=m_{M,N,(133,1)}=m_{M,N,(1,3)}\\
        &=m_{M+\frac{1}{2},N+\frac{1}{2},(0,0)}=m_{M+\frac{1}{2},N+\frac{1}{2},(133,1)}=m_{M+\frac{1}{2},N+\frac{1}{2},(1,3)}
        \\&=\left|\frac{2(M-iN)}{\sqrt{2\pi}} B\left(\frac{3}{4},\frac{1}{4}\right)  \right|\sqrt{\left|\frac{V_{t}}{\log t_1}\right|} , \quad M,N\in\mb{Z}\,.
    \end{aligned}
\end{equation}

If on the other hand $\bds{\lambda}'$ is non-integral, the possible $\mbf{J}_1$ are
\begin{equation}
    \begin{aligned}
        \mbf{J}_{1,M+\frac{1}{2},N,2}&=\bds{\lambda}'_2+2(M+\frac{1}{2}) \bds{\omega}'^p+2N \bds{\omega}'^q\,.\\
        \mbf{J}_{1,M,N+\frac{1}{2}}&=\bds{\lambda}'_2+2M \bds{\omega}'^p+2(N+\frac{1}{2}) \bds{\omega}'^q\,.\\
    \end{aligned}
\end{equation}
Here $\bds{\lambda}'_2$ is the weights of fundamental representation $2$ of $A_1$, with self-inner product $(\bds{\lambda}'_2,\bds{\lambda}'_2)=-\frac{1}{2}$. The highest weight is $\frac{1}{2}\bds{\alpha}'$.

To satisfy the BPS condition and the vanishing of the total asymptotic charge of the bunches, we find that the self-inner product of $\bds{\lambda}$ should have $(\bds{\lambda},\bds{\lambda})\geq -\frac{3}{2}$, and the non-integral part of $\bds{\lambda}$ is $\frac{1}{2}\sum_{i=1}^6\bds{a}_i$. Then we can find that all possible $\bds{\lambda}$ form the representation $56$ of $E_7$ with the highest weight 
\begin{equation}
    \bds{\lambda}_{56,h}=\bds{\alpha}_1+2\bds{\alpha}_2+3\bds{\alpha}_3+\frac{5}{2}\bds{\alpha}_4+2\bds{\alpha}_5+\frac{3}{2}\bds{\alpha}_6+\frac{3}{2}\bds{\alpha}_7\,.
\end{equation}
So the possible BPS string junctions and the asymptotic charges $(P,Q)$ in (\ref{mass_string_junction_12brane}) are
\begin{equation}
    \begin{aligned}
        \mbf{J}_{M+\frac{1}{2},N,(56,2)}=\bds{\lambda}'_{2}+\bds{\lambda}_{56}+2(M+\frac{1}{2}) (\bds{\omega}^p-\bds{\omega}'^p)+2N (\bds{\omega}^q-\bds{\omega}'^q)\ ,\  (P,Q)=(2M+1,2N)\,,\\
        \mbf{J}_{M,N+\frac{1}{2},(56,2)}=\bds{\lambda}'_{2}+\bds{\lambda}_{56}+2M (\bds{\omega}^p-\bds{\omega}'^p)+2(N+\frac{1}{2}) (\bds{\omega}^q-\bds{\omega}'^q)\ ,\  (P,Q)=(2M,2N+1)\,,\\
        \quad M,N\in\mb{Z}\,.
    \end{aligned}   
\end{equation}
The mass of these BPS string junctions is
\begin{equation}
    \begin{aligned}
        m_{M+\frac{1}{2},N,(56,2)}&=\left|\frac{2(M-iN)+1}{\sqrt{2\pi}} B\left(\frac{3}{4},\frac{1}{4}\right)  \right|\sqrt{\left|\frac{V_{t}}{\log t_1}\right|}\,,\\
        m_{M,N+\frac{1}{2},(56,2)}&=\left|\frac{2(M-iN)+i}{\sqrt{2\pi}} B\left(\frac{3}{4},\frac{1}{4}\right)  \right|\sqrt{\left|\frac{V_{t}}{\log t_1}\right|}\,,\quad M,N\in\mb{Z}\,.
    \end{aligned}
\end{equation}

\subsection{$E_6$ approaching type IV}

We consider the type $E_6=\mbf{A^5BCC}$ bunch $B_0$ approaching type IV bunch $\tilde{H}_2=\mbf{A^2X_{0,-1}C}$  (\ref{4D_E_6_type_IV}). We denote the basis of the strings ending on the $B_0$ as $\mbf{a}_i (i=1,\dots 5)$, $\mbf{b}$, $\mbf{c}_i (i=1, 2)$, and $\mbf{J}_0$ only has components on these 7-branes. The basis of the strings ending on the $\tilde{H}_2$ bunch as $\mbf{a}_i' (i=1,2),\mbf{x'_{0,-1}},\mbf{c'}$., and $\mbf{J}_1$ only has components on these 7-branes.

As shown in \cite{DeWolfe:1998zf}, the simple roots of $E_6$ are
\begin{equation}
\begin{aligned}
\boldsymbol{\alpha}_1 & =\mathbf{a}_1-\mathbf{a}_2, \quad \boldsymbol{\alpha}_2 =\mathbf{a}_2-\mathbf{a}_3, \quad \boldsymbol{\alpha}_3 =\mathbf{a}_3-\mathbf{a}_4\,, \\
\boldsymbol{\alpha}_4 & =\mathbf{a}_4+\mathbf{a}_5-\mathbf{b}-\mathbf{c}_1, \quad \boldsymbol{\alpha}_5  =\mathbf{c}_1-\mathbf{c}_2\,, \\
\boldsymbol{\alpha}_6 & =\mathbf{a}_4-\mathbf{a}_5\,,
\end{aligned}
\label{simple_roots_E6}
\end{equation}
with the dynkin diagram
\begin{equation}
    \dynkin[labels={\bds{\alpha}_1,\bds{\alpha}_6,\bds{\alpha}_2,\bds{\alpha}_3,\bds{\alpha}_4,\bds{\alpha}_5},scale=2]E6
\end{equation}
The extended weights of $E_6$ are
\begin{equation}
\boldsymbol{\omega}^p =-\frac{1}{3} \sum_{i=1}^5 \mathbf{a}_i+\frac{4}{3} \mathbf{b}+\frac{2}{3} \sum_{i=1}^2 \mathbf{c}_i\,, \quad\boldsymbol{\omega}^q =\sum_{i=1}^5 \mathbf{a}_i-3 \mathbf{b}-\sum_{i=1}^2 \mathbf{c}_i\,,
\label{extended_weights_E6}
\end{equation}
and the simple roots of $\tilde{H}_2$ ($A_2$) are
\begin{equation}
    \begin{aligned}
        \bds{\alpha}_1'&=-\mbf{a}_1'+\mbf{a}_2'\,,\\
        \bds{\alpha}_2'&=\mbf{a}_1'-\mbf{x_{0,-1}}'-\mbf{c}'\,.
    \end{aligned}
\end{equation}
The extended weights of $\tilde{H}_2$ ($A_2$) are
\begin{equation}
    \bds{\omega}^{'p}=\frac{1}{3}(\sum_{i=1}^2\mbf{a}'_i+\mbf{x'_{0,-1}}+\mbf{c'}), \quad \bds{\omega}^{'q}=-\mbf{x'_{0,-1}}\,.
\end{equation}

Similar to the $E_7$ approaching type III case, the non-integral factor of $E_6$ and type IV are both $\frac{1}{3}$, so $\bds{\lambda}$ and $\bds{\lambda}'$ must be both integral or non-integral. 

When they are both integral, the possible $\mbf{J}$ and the $(P,Q)$ charges are
\begin{equation}
    \begin{aligned}
        \mbf{J}_{M,N,(1,1)}&=3M(\bds{\omega}^p-\bds{\omega}^{'p})+N(\bds{\omega}^q-\bds{\omega}^{'q}), \quad (P,Q)=(3M,N)\,,\\
        \mbf{J}_{M,N,(78,1)}&=3M(\bds{\omega}^p-\bds{\omega}^{'p})+N(\bds{\omega}^q-\bds{\omega}^{'q})+\bds{\lambda}_{78}, \quad (P,Q)=(3M,N)\,,\\
        \mbf{J}_{M,N,(1,8)}&=3M(\bds{\omega}^p-\bds{\omega}^{'p})+N(\bds{\omega}^q-\bds{\omega}^{'q})+\bds{\lambda}'_{8}, \quad (P,Q)=(3M,N)\,,\\
        \quad M,N\in\mb{Z}\,.
    \end{aligned}
\end{equation}
Here $\bds{\lambda}_{78}$ denotes a weight of the adjoint representation $78$ of $E_6$ and $\bds{\lambda}'_{8}$ denotes a weight of the adjoint representation $8$ of $A_2$. 

The inner products are
\begin{equation}
    \begin{aligned}
        (\bds{\lambda}_{78},\bds{\lambda}_{78})&\geq -2\ ,\  (\bds{\lambda}'_{8},\bds{\lambda}'_{8})\geq -2\,,\\
        (P\bds{\omega}^p+Q\bds{\omega}^q,P\bds{\omega}^p+Q\bds{\omega}^q)&=\frac{1}{3}P^2-PQ+Q^2\,,\\
        (P\bds{\omega}'^p+Q\bds{\omega}'^q,P\bds{\omega}'^p+Q\bds{\omega}'^q)&=-\frac{1}{3}P^2+PQ-\frac{5}{2}Q^2\,.
    \end{aligned}
\end{equation}
So all these integral $\bds{J}$ are BPS, and their mass is
\begin{equation}
        m_{M,N,(1,1)}=m_{M,N,(78,1)}=m_{M,N,(1,8)}=\left|\frac{3M-e^{i\pi/3}N}{\sqrt{\sqrt{3}\pi}} B\left(\frac{2}{3},\frac{1}{3}\right)  \right|\sqrt{\left|\frac{V_{t}}{\log t_1}\right|}, \quad M,N\in\mb{Z}.
\end{equation}

When $\bds{\lambda}'$s are non-integral, we can find when $P=3M+1$ the non-integral parts are
\begin{equation}
    \begin{aligned}
        \bds{\lambda}_n=&\frac{1}{3}(\bds{\alpha}_1+2\bds{\alpha}_2+\bds{\alpha}_4-\bds{\alpha}_5)=\frac{1}{3}(\mbf{a}_1+\mbf{a}_2-2\mbf{a}_3+\mbf{a}_4+\mbf{a}_5-\mbf{b}-2\mbf{c}_1+\mbf{c}_2)\,,\\
        \bds{\lambda}'_n=&-\frac{1}{3}(2\bds{\alpha}'_1+\bds{\alpha}'_2)=-\frac{1}{3}(-\mbf{a}'_1+2\mbf{a}'_2-\mbf{x'_{[0,-1]}}-\mbf{c}')\,.
    \end{aligned} 
\end{equation}
Here $(\bds{\lambda}_n',\bds{\lambda}_n')=-\frac{2}{3}$ and $(\bds{\lambda}_n,\bds{\lambda}_n)=-\frac{4}{3}$, satisfying the BPS conditions. All the possible non-integral $\bds{\lambda}$s form the representation $27$ of $E_6$ with the highest weight 
\begin{equation}
\label{lambda27-h}
    \bds{\lambda}_{27,h}=\frac{1}{3}(4\bds{\alpha}_1+5\bds{\alpha}_2+6\bds{\alpha}_3+4\bds{\alpha}_4+2\bds{\alpha}_5+3\bds{\alpha}_6)\,.
\end{equation}
All the possible non-integral $\bds{\lambda}'$s form the fundamental representation $\Bar{3}$ of $A_2$ with the highest weight
\begin{equation}
    \bds{\lambda}'_{\Bar{3},h}=\frac{1}{3}(\bds{\alpha}'_1+2\bds{\alpha}'_2)\,.
\end{equation}
When $P=3M+2$ the non-integral parts are
\begin{equation}
    \begin{aligned}
        \bds{\lambda}_n=&-\frac{1}{3}(\bds{\alpha}_1+2\bds{\alpha}_2+\bds{\alpha}_4-\bds{\alpha}_5)=-\frac{1}{3}(\mbf{a}_1+\mbf{a}_2-2\mbf{a}_3+\mbf{a}_4+\mbf{a}_5-\mbf{b}-2\mbf{c}_1+\mbf{c}_2)\,,\\
        \bds{\lambda}'_n=&\frac{1}{3}(2\bds{\alpha}'_1+\bds{\alpha}'_2)=\frac{1}{3}(-\mbf{a}'_1+2\mbf{a}'_2-\mbf{x'_{[0,-1]}}-\mbf{c}')\,.
    \end{aligned} 
\end{equation}
Here $(\bds{\lambda}_n',\bds{\lambda}_n')=-\frac{2}{3}$ and $(\bds{\lambda}_n,\bds{\lambda}_n)=-\frac{4}{3}$, satisfying the BPS conditions. All the possible non-integral $\bds{\lambda}$s form the representation $\overline{27}$ of $E_6$ with the highest weight 
\begin{equation}
\label{lambda27-bar-h}
    \bds{\lambda}_{\overline{27},h}=\frac{1}{3}(2\bds{\alpha}_1+4\bds{\alpha}_2+6\bds{\alpha}_3+5\bds{\alpha}_4+4\bds{\alpha}_5+3\bds{\alpha}_6)\,.
\end{equation}
All possible non-integral $\bds{\lambda}'$s form the representation $3$ of $A_2$ with the highest weight
\begin{equation}
    \bds{\lambda}'_{3,h}=\frac{1}{3}(2\bds{\alpha}'_1+\bds{\alpha}'_2)\,.
\end{equation}
Hence the possible BPS string junctions and the asymptotic $(P,Q)$ in (\ref{mass_string_junction_12brane}) are
\begin{equation}
    \begin{aligned}
        \bds{J}_{M+\frac{1}{3},N,(27,\overline{3})}&=\bds{\lambda}_{27}+\bds{\lambda}'_{\overline{{3}}}+3(M+\frac{1}{3})(\bds{\omega}^p-\bds{\omega}'^p)+N(\bds{\omega}^q-\bds{\omega}'^q)\ ,\  (P,Q)=(3M+1,N)\,,\\
        \bds{J}_{M+\frac{2}{3},N,(\overline{27},{3})}&=\bds{\lambda}_{\overline{27}}+\bds{\lambda}'_{3}+3(M+\frac{2}{3})(\bds{\omega}^p-\bds{\omega}'^p)+N(\bds{\omega}^q-\bds{\omega}'^q)\ ,\  (P,Q)=(3M+2,N)\,,\\
        \quad M,N\in\mb{Z}.
    \end{aligned}
\end{equation}
Their mass is 
\begin{equation}
    \begin{aligned}
        m_{M+\frac{1}{3},N,(27,\overline{3})}&=\left|\frac{3M+1-e^{i\pi/3}N}{\sqrt{\sqrt{3}\pi}} B\left(\frac{2}{3},\frac{1}{3}\right)  \right|\sqrt{\left|\frac{V_{t}}{\log t_1}\right|}\,, \\
        m_{M+\frac{2}{3},N,(\overline{27},{3})}&=\left|\frac{3M+2-e^{i\pi/3}N}{\sqrt{\sqrt{3}\pi}} B\left(\frac{2}{3},\frac{1}{3}\right)  \right|\sqrt{\left|\frac{V_{t}}{\log t_1}\right|}\ ,\  M,N\in\mb{Z}\,.
    \end{aligned}
\end{equation}

\subsection{$E_6$ approaching type II}

We have a type $E_6=\mbf{A^5BCC}$ bunch $B_0$ approaching a type II $H_0=\mbf{AC}$ bunch $B_1$. Similar to before, we  denote the basis of the strings ending on $B_0$ as $\mbf{a}_i (i=1,\dots 5)$, $\mbf{b}$, $\mbf{c}_i (i=1, 2)$, and $\mbf{J}_0$ only has components of these 7-branes. The basis of the strings ending on the $H_0$ bunch is $\mbf{a}',\mbf{c}'$., and $\mbf{J}_1$ only has components of these 7-branes.

The extended weights of $B_1$ are
\begin{equation}
    \bds{\omega}'^p=\mbf{a}'\ ,\  \bds{\omega}'^q=-\mbf{a}'+\mbf{c}'\,,
\end{equation}
while extended weights and simple roots of $E_6$ are shown in (\ref{simple_roots_E6}), (\ref{extended_weights_E6}).

To fix the mass of the string junctions, we investigate the possible highest weights. For the representation $1$ of $E_6$, the highest weight is $\mbf{0}$, and the corresponding string junctions should have the form
\begin{equation}
    \mbf{J}_{M,N,1}=3M(\bds{\omega}^p-\bds{\omega}'^p)+N(\bds{\omega}^q-\bds{\omega}'^q)\,.
\end{equation}
Such form has self-inner product
\begin{equation}
    (\mbf{J}_{M,N,1},\mbf{J}_{M,N,1})=-6M^2+6MN+2N^2\,.
\end{equation}
For BPS string junctions in representation $1$, the possible $(M,N)$s are $(-1,-2)$, $(-1,-1)$, $(0,-1)$, $(0,0)$, $(0,1)$, $(1,1)$, $(1,2)$, and the corresponding asymptotic charge $(P,Q)$ are $(-3,-2)$, $(-3,-1)$, $(0,-1)$, $ (0,0)$, $(0,1)$, $(3,1)$, $(3,2)$.

For the representation $27$, the highest weight is $\bds{\lambda}_{27,h}$ (\ref{lambda27-h}), and the corresponding string junctions should have the form
\begin{equation}
    \mbf{J}_{M,N,27}=\bds{\lambda}_{27}+3(M+\frac{1}{3})(\bds{\omega}^p-\bds{\omega}'^p)+N(\bds{\omega}^q-\bds{\omega}'^q)
\end{equation}
with self-inner product
\begin{equation}
    (\mbf{J}_{M,N,27},\mbf{J}_{M,N,27})\geq -\frac{4}{3}-6(M+\frac{1}{3})^2-6(M+\frac{1}{3})N+2N^2\,.
\end{equation}
For BPS string junctions in representation $27$, the possible $(M,N)$s are $(-1,-1)$, $(0,0)$, $(0,1)$, and the corresponding asymptotic charges $(P,Q)$ are $(-2,-1)$, $(1,0)$, $(1,1)$.

For the representation $\overline{27}$, the highest weight is $\bds{\lambda}_{\overline{27},h}$ (\ref{lambda27-bar-h}). So these string junctions should have the form
\begin{equation}
    \mbf{J}_{M,N,\overline{27}}=\bds{\lambda}_{\overline{27}}-3(M+\frac{2}{3})(\bds{\omega}^p-\bds{\omega}'^p)+N(\bds{\omega}^q-\bds{\omega}'^q)\,.
\end{equation}
Such form has self-inner product
\begin{equation}
    (\mbf{J}_{M,N,\overline{27}},\mbf{J}_{M,N,\overline{27}})\geq -\frac{4}{3}-6(M+\frac{2}{3})^2+6(M+\frac{2}{3})N-2N^2\,.
\end{equation}
For BPS string junctions in representation $\overline{27}$, the possible $(M,N)$s are $(-1,-1)$, $(-1,0)$, $(0,1)$. The corresponding asymptotic charges $(P,Q)$ are $(-1,-1)$, $(-1,0)$, $(2,1)$.

With these asympototic charges, we can calculate the mass of these BPS string junctions by replacing $C_g$ with $C_t$ in (\ref{2-string_constant_tau})
\begin{equation}
        m_{P,Q}=C_t\left|(P-Qe^{i\pi/3}) \eta^2(e^{i\pi/3})\prod_{i=2} t_i^{-\frac{n_i}{12}}B\left(\frac{5}{6},\frac{1}{3}\right)\right||t_1|^{\frac{1}{6}}\,.
\end{equation}

\bibliographystyle{JHEP}
\bibliography{biblio.bib}

\end{document}